\documentclass[10pt,showkeys,twocolumn,amsmath,amssymb,aps,prb,showpacs,longbibliography,superscriptaddress]{revtex4-1}
\usepackage[latin9]{inputenc}
\usepackage{textcomp}
\usepackage{mathrsfs}
\usepackage{amsmath}
\usepackage{amssymb}
\usepackage{graphicx}
\usepackage{esint}
\usepackage{xcolor}
\usepackage{multirow}

\makeatletter

\usepackage{amsfonts}
\usepackage{graphics}

\newenvironment{dataavailability}{
  \section*{Data availability statement}
}{}

\makeatother

\begin{document}
\title{Altermagnetism and Superconductivity: A Short Historical Review}
\author{Zhao Liu}
\email{Corresponding author: zhaoliu@swin.edu.au}
\affiliation{Centre for Quantum Technology Theory, Swinburne University of Technology,
Melbourne 3122, Australia}
\author{Hui Hu}
\affiliation{Centre for Quantum Technology Theory, Swinburne University of Technology,
Melbourne 3122, Australia}
\author{Xia-Ji Liu}
\affiliation{Centre for Quantum Technology Theory, Swinburne University of Technology,
Melbourne 3122, Australia}
\date{\today}
\begin{abstract}
This review is organized into three parts. In the first part, we explore the deep interconnections among three seemingly unrelated concepts in condensed matter physics: electronic liquid crystal phases, multipole expansions, and altermagnetism. At the heart of these phenomena lies a shared foundation: spin-momentum locking in the nonrelativistic regime. Originally proposed in the context of electronic liquid crystal phases, nonrelativistic spin-momentum locking was later elegantly incorporated into the formalism of multipole expansions. This framework can be further extended across multiple atomic sites, making it particularly effective for describing altermagnets, which host localized magnetic moments with anisotropic magnetization densities distributed over sublattices. In the second part, we examine superconducting phenomena associated with altermagnetism from three complementary perspectives. First, we investigate superconductivity associated with nonrelativistic spin-momentum locked Fermi surfaces, the unifying theme of the first part, highlighting a rich variety of unconventional superconducting states. These include finite-momentum pairing, $d$-wave and spin-triplet superconductivity, and topological Bogoliubov Fermi surfaces, among others. We then review superconductivity emerging from either static altermagnetic order or altermagnetic fluctuations. Finally, we discuss the possible competition and intertwining between altermagnetic order and superconductivity, illustrated using the repulsive Hubbard model.
Additional related topics are addressed in the concluding part. Overall, this work offers both an accessible introduction to the newly identified magnetic order known as altermagnetism and a conceptual guide for researchers aiming to harness the ensuing unconventional superconductivity in the development of future quantum technologies. 
\end{abstract}


\maketitle

\section*{I. Introduction}
\begin{figure*}
\begin{centering}
\includegraphics[width=0.95\textwidth]{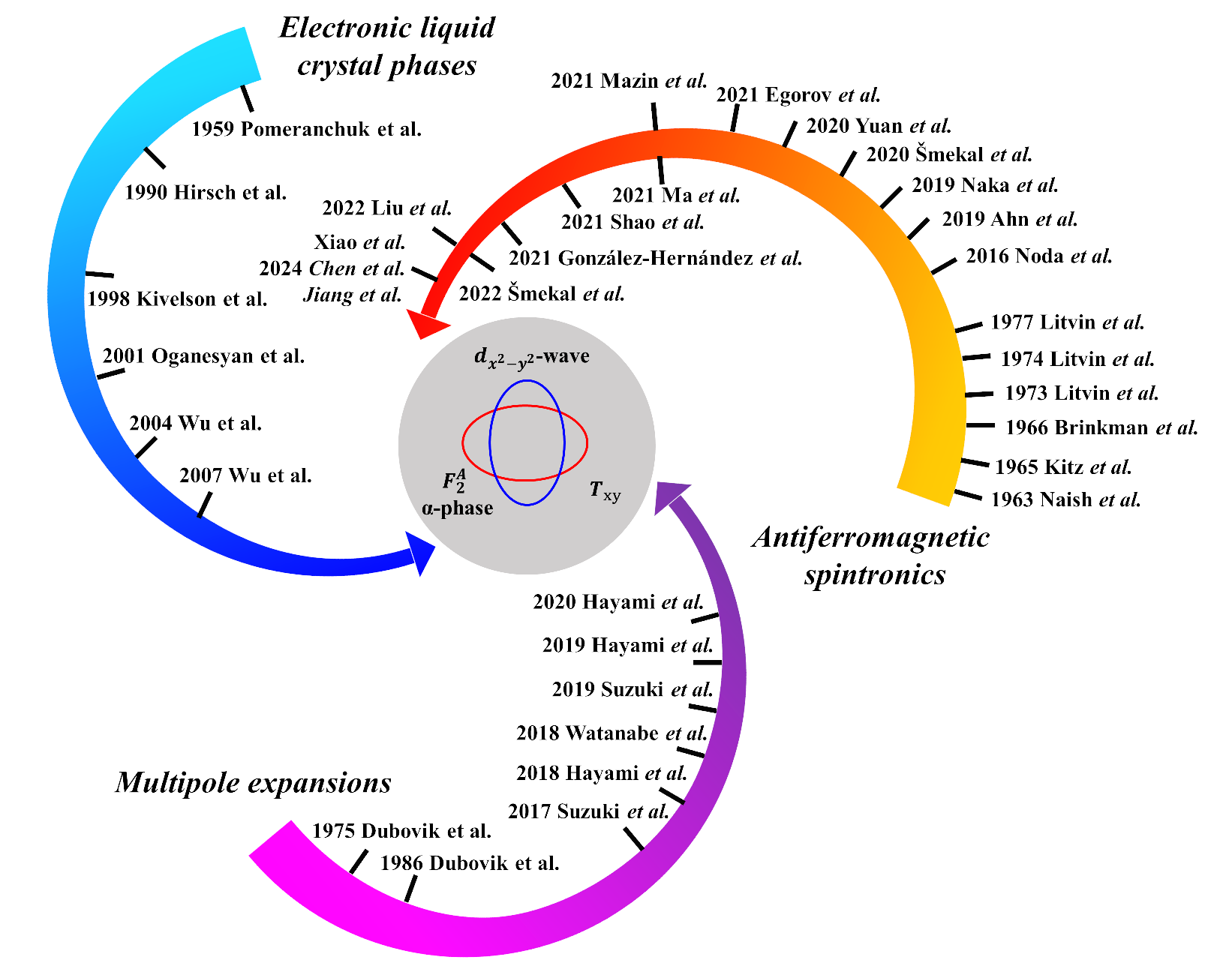}
\par\end{centering}
\caption{A tale of three, where significant progresses are chronologically displayed and labelled by year. While the idea of electronic liquid crystal phase is yet to be materialized, the multipole basis analysis of collinear AFMs \citep{Hayami2019}, spin-resolved transport in collinear AFMs \citep{Naka2019, Smejkal2020} and the symmetry classification of collinear magnets based on spin-group theory \citep{Smejkal2022-1} lead to the concept of altermagnetism based on spin-group theory.} 
\label{fig:Fig-1} 
\end{figure*}

In recent years, there has been a growing surge of interest in a newly discovered magnetic order known as altermagnetism (AM) \citep{Smejkal2020, Hayami2019, Smejkal2022-1, Smejkal2022-2, Mazin2022}. Although several excellent reviews have explored this rapidly evolving topic \citep{Bai2024, SongC2025, Guo2025, Jungwirth2025-1, Jungwirth2025-2, Jungwirth2025-3}, less attention has been given to the historical development of the underlying ideas, methodologies, and conceptual shifts that have defined the field. In this paper, we aim to shed light on key milestones and turning points that have shaped the study of AM (see, i.e. Fig.\ref{fig:Fig-1}). To this end, we briefly examine two closely related themes: (i) the interplay between electronic liquid crystal (ELC) phases \citep{Wu2007}, multipole expansions \citep{Hayami2018-PRB}, and the emergence of the AM framework \citep{Smejkal2020, Hayami2019, Naka2019, Smejkal2022-1, Smejkal2022-2}; and (ii) unconventional superconducting phenomena entangled with AM, by starting from this foundational interplay.

We begin in Sect. II by disentangling the similarities and differences among ELC phases, multipole expansions, and AM. This comparative analysis is structured chronologically to reflect their historical evolution, beginning with ELC phases, transitioning through multipole expansions, and finally arriving at AM. Despite their distinct origins, all three concepts share a core feature: nonrelativistic spin-momentum locking (NRSML) as shown at the center of Fig.\ref{fig:Fig-1}, although each is expressed in a different basis. In ELC phases, a single-band description suffices \citep{Wu2007}. This single-band picture is also shared by the language of multipole expansions \citep{Hayami2018-PRB}, which offer a unified scheme for classifying ELC phases across all the charge and spin channels $F^{S/A}_l$ (defined below). Moreover, multipole expansions can be systematically extended to multisite systems, enabling a nature description of diverse magnetic textures, including collinear, coplanar, and noncoplanar magnets \citep{Hayami2020}. These magnetic textures, when classified at the nonrelativisitic limit, fall within the framework of spin space groups \citep{Liu2022}. Among them, AM-a special type of collinear antiferromagnets (AFMs)-have attracted particular attention in modern spintronics \citep{Smejkal2022-1}. By integrating the  advantages of both ferromagnets (FMs) and N\'{e}el AFMs, AMs open a fertile arena for research, with rapid developments in recent years. Importantly, NRSML in AMs requires at least two sublattices with localized magnetic moments, meaning that a single-band description of NRSML in AMs should be understood as a low-energy effective approximation.

We then turn to the second theme: superconducting phenomena entangled with AMs. The marriage between magnetism and unconventional superconductivity (SC)-exemplified by cuprates \citep{Bednorz1986, Wu1987, Schilling1993}, heavy fermions \citep{Steglich1979, Ott1983, Stewart1984, Saxena2000, Petrovic2001}, iron-based materials \citep{Kamihara2008, Wang2012, Dai2015}, and more recently infinite-layer nickelates \citep{LiD2019, LiuZ2020, Lu2021}-stands among the most celebrated topics in condensed matter physics. It is therefore natural to add AMs into this narrative. To this end, we organize the discussion into three hierarchical classes of models, corresponding to the three complementary perspectives on AM. At the top level is the NRSML (Sect. III), which can be conveniently captured by a single-band model with anisotropic and spin-dependent hoppings. Upon introducing attractive interactions, this framework gives rise to a rich plethora of unconventional superconducting states. Although similar phenomenological model has already been investigated in the context of ELC phases over the past few decades \citep{Kivelson1998},  the emergence of AMs has renewed and sharpened interest in this class of models. At the intermediate level (Sect. IV), we treat anisotropic but spin-independent hoppings and magnetic order on equal footing, thereby correctly incorporating the  sublattice degree of freedom (dof).  Both linear spin-wave theory and spin-fluctuation mediated unconventional SC at weak to intermediate coupling limit are discussed, corresponding respectively to regimes deeply inside AM phase and in proximity to the AM-paramagnetism (PM) quantum critical point (QCP). At the bottom level lies the repulsive Hubbard model (Sect. V), formulated without the assuming pre-exist AM orders or a NRSML description. This unbiased approach is particularly valuable for exploring the competition and cooperation between AM and other ordered phases. In the strong coupling limit and in the presence of doping, the repulsive Hubbard model can be further downfolded to the $t-J$ model which has shown success in explaining the $d$-wave SC in cuprates \citep{Lee2006}. Here we will discuss a multiple sublattice $t-J$ model. 

In the concluding section (Sect. VI), we briefly discuss another type of unconventional magnetism, namely the $p$-wave magnet. We also briefly summarized the integration of multipole expansions and SC. In particular, we emphasize that Sect. III can be understood in the framework of multipole basis-fluctuation mediated unconventional SC.

This review is intended to serve both specialists and newcomers. For experts, it offers a conceptual consolidation of seemingly disparate ideas under a common framework; for a broader audience, it provides a heuristic introduction to the sequence of theoretical insights without dwelling on technical details. Our emphasis is on clarity of concepts, and we direct readers to excellent existing reviews when discussing topics that are not fully covered throughout the review. We apologize in advance for any omissions in the literatures.

In the following, $\hat{c}^{\dagger}_{i a \sigma}$ ($\hat{c}_{i a \sigma}$) represents a spin-1/2 fermion quasiparticle creation (annihilation) operator with $i$, $a$, and $\sigma$ representing unit cell, sublattice, and spin dof, respectively. Since we also use $\mathbf{\sigma}$ to label the three Pauli matrices, the label $\alpha$, $\beta$ are also used to represent spin dof to distinguish from Pauli matrices. Correspondingly, the creation (annihilation) operator in momentum space is denoted as  $\hat{c}^{\dagger}_{\mathbf{k} a \sigma}$ ($\hat{c}_{\mathbf{k} a \sigma}$). The occupation operator is $\hat{n}_{i a}$ and the localized magnetic moment and electron spin are denoted as $\mathbf{S}$ and $\mathbf{s}$ respectively. The reduced Planck constant $\hbar$, the Boltzmann constant $k_B$, and the light speed $c$ are set to 1. The relation between $\hat{\mathbf{s}}$ and electron creation and annihilation operators are $\hat{\mathbf{s}} = \sum_{\alpha \beta} \hat{c}^{\dagger}_{\alpha} \frac{\pmb{\sigma}_{\alpha \beta}}{2} \hat{c}_{\beta}$. $\mu$ is the chemical potential.

\section*{II. A tale of three}
High-$T_c$ cuprates \citep{Bednorz1986, Lee2006, Keimer2015} and $f$-electron systems \citep{Steglich1979, Stewart1984RMP} are two prototype strongly correlated systems. In cuprates, the parent compounds are layered AFM insulators. Upon electron or hole doping, the static AFM order in CuO$_2$ plane gradually melts, simultaneously the insulating phase evolves into a "bad" metal, then $d$-wave SC emerges. These intermediate "melted" phases, arising purely from electron-electron interactions, inspired the adoption of concepts in classical liquid crystals to classify the rich variety of electronic phases in cuprates and related materials \citep{Kivelson1998}. 

The $f$-electron systems, in contrast, are more naturally descried through the framework of multipole expansions \citep{Santini2009, Kuramoto2009}. This distinction originates from the different character of $d$ (the case in cuprates) and $f$ electrons. The $d$ electrons have much larger sensitivity to the crystal field produced by charges on neighboring ligands and a weaker spin-orbit coupling (SOC) when compared with $f$ electrons. This usually leads to an almost complete quenching of the orbital angular momentum of $d$ electrons. Conversely, the interplay of spin and unquenched orbital orbital angular momentum in $f$ electrons gives rise to a hierarchy of higher-order multipoles beyond the familiar isotropic monopole and dipole. This has lead to fruitful insights into ordering phenomena in $f$-electron systems. Not limited to $f$-electron systems, multipole basis has now been recognized as a unifying language for describing multiple degrees of freedom of electrons in solids, e.g. charge, spin, and orbital, and many other properties (see reviews \citep{Kusunose2022, Hayami2024}).  
 
Interestingly, the features of $d$ electrons-sensitive to crystal field and weak SOC-can spur a very intriguing phenomenon: crystal-symmetry-protected spin splitting in collinear AFMs even without SOC. This unusual spin splitting facilitates the signal read-out in antiferromagnetic spintronics, therefore holds immerse applications in energy-efficient quantum technologies. To capture the growing application list of such a phenomenon, a new class of collinear magnets--altermagnet--has recently been proposed, completing the traditional dichotomy of FMs and N\'{e}el AFMs. Today, AMs encompass both correlated metals and insulators, suggesting that their story is just beginning to unfold.  

\subsection*{A. Electronic liquid crystal phases}
Electronic liquid crystal (ELC) phases are states of correlated quantum electronic systems that spontaneously break either rotational or translational invariance \citep{Kivelson1998}. Following the symmetry-based scheme used in classical liquid crystals \citep{deGennes1993}, there are four typical ELC phases: 1, crystalline phases which break all continuous translation and rotation symmetries; 2, smectic phases, which break one translation and/or rotation symmetry, exhibiting electron liquid behavior along the remaining symmetric direction. 3, nematic phases which break rotation symmetry while preserving translational symmetry, resulting in an anisotropic electron liquid with a preferred orientation axis, and 4, isotropic (liquid) phases, which preserve all symmetries, allowing electrons to flow uniformly in all directions. These ELC phases are characterized by long-range order parameters of charge, spin, orbital, and phase coherence. Unlike their classical counterparts which are commonly melted by thermal fluctuations, ELC phases exhibit strong quantum mechanical effect, especially in the strongly correlation regime \citep{Kivelson1998}, therefore they can be quantum melted at low temperatures, which gives quantum phase transitions.   

Here, we focus primarily on the nematic phase of a Fermi liquid, using ELC phases as an instructive framework for organizing momentum-dependent spin textures. This phase is characterized by electronic uniformity combined with directional anisotropy. There are two main pathways to reach the nematic phase:  one involves a direct transition from the isotropic electronic fluid via a Pomeranchuk instability (PI) \citep{Pomeranchuk1958}, and the other occurs through the thermal or quantum melting a smectic phase \citep{Kivelson1998}. In the following discussion, we concentrate on the former - the PI of Landau Fermi liquid.

The central concept of the Landau Fermi liquid theory is the existence of quasiparticles near the Fermi surface (FS) which weakly interact. Such interactions are parameterized by the Landau interaction functions $F^{S/A}(\mathbf{k}, \mathbf{k}')$ quantifying the strength of the forward scattering amplitudes among quasiparticles at low energies with momenta $\mathbf{k}, \mathbf{k}'$ close to the FS in the charge channel (superscript "$S$") or the spin channel (superscript "$A$"). If the system has translation symmetry, the Landau interaction functions only depends on the difference of the two momenta: $F^{S/A}(\mathbf{k}, \mathbf{k}') = F^{S/A}(\mathbf{k}-\mathbf{k}') $. Moreover, if the system conserve rotation symmetry, the Landau interaction functions can be decomposed into orthogonal angular momentum basis (or partial waves) and characterized by dimensionless Landau parameters. In 3D, spherical harmonics is a natural basis and the Landau parameters takes the form $F^{S/A}_{l,m}$ where $l$ ($l \in \mathbb{N}$) and $m$ ($|m| \leq l$) is the angular momentum and magnetic quantum numbers. In 2D, circular harmonics is a natural basis and the Landau parameters is simply $F^{S/A}_{l}$. For a lattice model, the continuous rotation symmetry is broken into discrete point group symmetry of the lattice, then the Landau parameters $F^{S/A}_{l,m}$/$F^{S/A}_{l}$ are further split according to the irreducible representations (IRs) of the point group of the lattice. 

The thermodynamic stability of a Fermi liquid state relies on the Pauli pressure from the Pauli exclusion principle. In 1958, Pomeranchuk \citep{Pomeranchuk1958} argued that if in one channel, the forward scattering interaction becomes sufficiently negative to overcome the stabilizing effects of the Pauli pressure, the Fermi liquid becomes unstable and undergeos distortion compatible with the symmetry of the unstable channel. Expressed in Landau parameters, we have the condition for PI:
\begin{equation}
\begin{split}
F^{S/A}_{l,m} &\leq -(2l+1) \mbox{ in } 3D,  \\
F^{S/A}_{l \neq 0} &\leq -2 \mbox{ in } 2D, \\
\end{split}
\label{eq:Wu-2007-A5}
\end{equation}
We note that the $l=1$ and $l=2$ basis in 2D are $\{x, y\}$ and $\{2xy, x^2-y^2\}$ in the real representation, and $\{x \pm iy\}$ and $\{(x \pm iy)^2\}$ in the complex representation. 

In 2001, Oganesyan \textit{et al.} \cite{Oganesyan2001} firstly studied the PI in the $F^S_2$ channel (charge nematic state) with continuum models. Using a 2D single-band \textit{spinless} system, they introduced the quadrupole density operator as a symmetric traceless tensor
\begin{equation}
\hat{\mathbf{Q}}(\mathbf{r}) \equiv -\frac{1}{k^2_F} \Psi^{\dagger}(\mathbf{r}) 
\left(\begin{array}{cc}
\partial^2_x - \partial^2_y & 2\partial_x\partial_y \\
2\partial_x\partial_y & \partial^2_y - \partial^2_x
\end{array}\right)
\Psi(\mathbf{r})
\label{eq:Oganesyan-2001-1}
\end{equation} 
where $k_F$ is the Fermi wave vector and $\Psi^{\dagger}(\mathbf{r}), \Psi(\mathbf{r})$ are the field operators in real space. Such a tensor is equivalent to a complex operator $\hat{Q}_2 (\mathbf{r}) \equiv -\frac{1}{k^2_F} \Psi^{\dagger}(\mathbf{r})(\partial_x + i \partial_y)^2 \Psi(\mathbf{r}) $ (where the subscript "2" represent the angular momentum $l=2$). 

To obtain nonvanishing $\mathbf{Q} \equiv \left\langle \hat{\mathbf{Q}}(\mathbf{r}) \right\rangle$, Oganesyan \textit{et al.} considered the following model Hamiltonian:
\begin{equation}
\begin{split}
H &= \int d\mathbf{r} \Psi^{\dagger}(\mathbf{r}) \epsilon(\overrightarrow{\nabla})\Psi(\mathbf{r}) \\ 
&+ \frac{1}{4} \int d\mathbf{r} \int d\mathbf{r}' F_2(\mathbf{r}-\mathbf{r}') \mathrm{Tr}[\hat{\mathbf{Q}}(\mathbf{r})\hat{\mathbf{Q}}(\mathbf{r}')] \\
\end{split}
\label{eq:Oganesyan-2001-4}
\end{equation}

where the free-fermion dispersion (near the FS) is $\epsilon(\mathbf{k}) = v_F q[1+a(q/k_F)^2]$ with $v_F$ the Fermi velocity and $q \equiv |\mathbf{k}| - k_F$, and the interparticle interaction is given in the form $F_2(\mathbf{r})=  \int \frac{d\mathbf{q}}{(2\pi)^2} e^{i\mathbf{q} \cdot \mathbf{r}}\frac{F_2}{1+\kappa F_2\mathbf{q}^2}$, where $F_2$ is an appropriate parameter related to the $l=2$ Landau parameter $F^S_2$, and $\kappa$ measures the range of the two-body interactions.

\begin{figure*}
\begin{centering}
\includegraphics[width=0.95\textwidth]{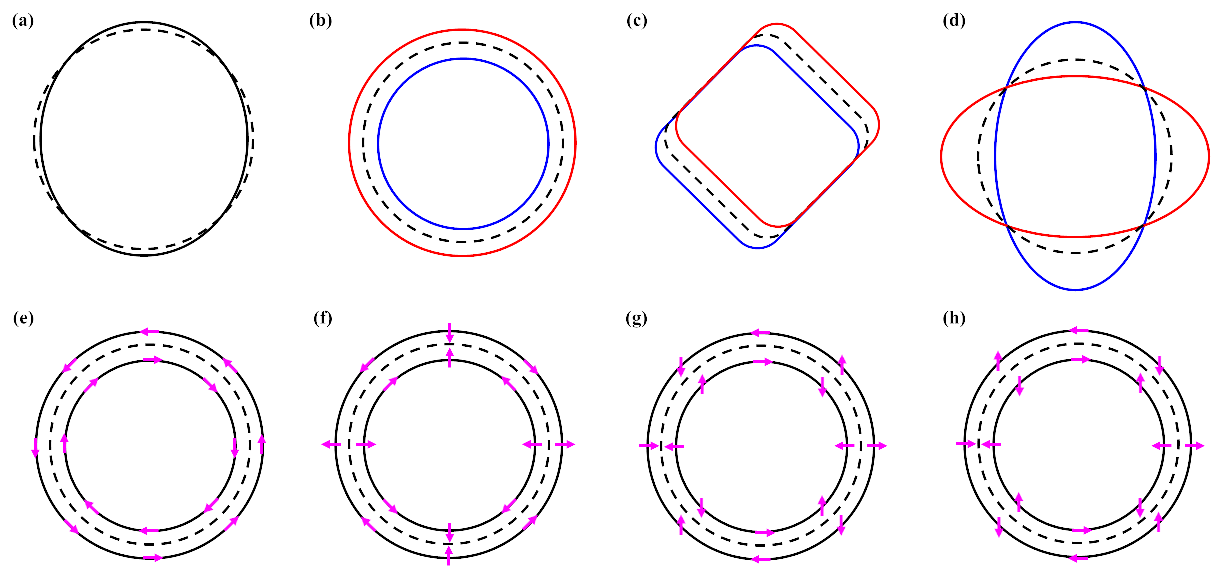}
\par\end{centering}
\caption{Fermi surfaces of different electronic liquid crystal phases. (a) charge nematic state, (b) itinerant FM state, (c) $l=1$ or $p$-wave $\alpha$ state, (d) $l=2$ or $d$-wave $\alpha$ state, which bears similarity to the $d$-wave AM, (e)-(h) various $\beta$ phases. 
The dashed line denotes the isotropic FSs. When spin is a good quantum number, spin-up and spin-down FS are marked by red and blue color; otherwise spin projections are labelled by pink arrows.}
\label{fig:Fig-ELC} 
\end{figure*}

By utilizing a Hubbard-Stratonovich decoupling, the Landau energy density functional for this model can be written as  \citep{Oganesyan2001}:
\begin{equation}
F[\mathbf{Q}] = E[\mathbf{Q}] - \frac{\tilde{\kappa}}{4}\mathrm{Tr}[\mathbf{Q} \pmb{D} \mathbf{Q}] - \frac{\tilde{\kappa}'}{4}\mathrm{Tr}[\mathbf{Q}^2 \pmb{D} \mathbf{Q}] + ...
\label{eq:Oganesyan-2001-3}
\end{equation}
where $D_{ij} \equiv \partial_i \partial_j$, and $\tilde{\kappa}$ and $\tilde{\kappa}'$ are the two effective Franck constants. The uniform part of the energy functional $E[\mathbf{Q}]$ is given by
\begin{equation}
E[\mathbf{Q}] = E(0) + \frac{A}{4}\mathrm{Tr}[\mathbf{Q}^2] + \frac{B}{8}\mathrm{Tr}[\mathbf{Q}^4] + ...
\label{eq:Oganesyan-2001-2}
\end{equation}
where $A = 1/(2N_F)+F_2$ with $N_F$ the density of state at the FS, and $B = (3aN_F|F_2|^3)/(8v^2_Fk^2_F)$. The isotropic Fermi liquid phase becomes unstable provided $A < 0$, or equivalently, $2 N_F F_2 < -1$, which is the PI condition in this spinless model. The result of such an instability is a spontaneously distorted FS characterized by the nonzero order parameter $\textbf{Q}$ (see Fig.\ref{fig:Fig-ELC}(a)), which has the name nematic phase in analogy to classical electronic liquid phase. 
Results on lattice models can be found in Kee \textit{et al.} \citep{Kee2003}, Khavkine \textit{et al.} \citep{Khavkine2004}, Yamase \textit{et al.} \citep{Yamase2005}, Lamas \textit{et al.} \citep{Lamas2008},  Quintanilla \textit{et al.} \citep{Quintanilla2008}.

Now we discuss the spin channel ($F^A_l$), which is more relevant. The simplest case the is the $l = 0$ channel, the PI $F^A_0 < -1$ gives just the well-known Stoner criterion $N_F U >1$ for itinerant FMs if we consider the Hubbard model with on-site Hubbard interaction strength $U$ \citep{Stoner1938}. Obviously, the spatial inversion ($ \mathcal{P}$) and rotation symmetries are conserved but time reversal ($\mathcal{T}$) invariance is broken in this channel, as shown in Fig.\ref{fig:Fig-ELC}(b). 

In 1990, Hirsch \textit{et al.} \cite{Hirsch1990} first proposed a spin-split state, i.e. a $F^A_1$ channel PI, by considering the following Hamiltonian:
\begin{equation}
\hat{H} =  \sum_{<i, j>} \sum_{\sigma} (\hat{c}^{\dagger}_{i \sigma} \hat{c}_{j \sigma} + h.c.) + J \hat{c}^{\dagger}_{i \sigma} \hat{c}_{j \sigma} \hat{c}^{\dagger}_{j \overline{\sigma}} \hat{c}_{i \overline{\sigma}}, J > 0
\label{eq:Hirsch-1990-2}
\end{equation}
Actually the second term resembles the XY-model $\sum_{<i, j>} -J (\hat{\mathrm{s}}^{+}_{i} \hat{\mathrm{s}}^{-}_{j} + \hat{\mathrm{s}}^{-}_{i} \hat{\mathrm{s}}^{+}_{j} )$, where $\hat{\mathrm{s}}^+$ ($\hat{\mathrm{s}}^-$) is the spin raising (lowering) operator $\hat{\mathrm{s}}^{\pm}=\hat{\mathrm{s}}^{x} \pm i \hat{\mathrm{s}}^{y}$. Such a model hosts three phases: PM phase preserving both  $\mathcal{P}$ and  $\mathcal{T}$, FM phase preserving $\mathcal{P}$ but breaking $\mathcal{T}$, and spin-split state preserving $\mathcal{T}$ but breaking $\mathcal{P}$. For a 2D square lattice near half-filling, such a spin-split state is shown in Fig.\ref{fig:Fig-ELC}(c), where the spin-up  and spin-down FSs shift along opposite direction without changing the number of spin-up and spin-down electrons. In 2006, Varma \textit{et al.} \citep{Varma2006} considered a continuum model in 3D and obtained a similar spectrum shown in Fig.\ref{fig:Fig-ELC}(c), such states have been proposed to be helicity-ordered states, which is believed to be the "hidden order parameter" in URu$_2$Si$_2$ \citep{Palsta1985, Maple1986}.

A systematic investigation of $F^{A}_l$ channel PI was laid out by Wu \textit{et al.} \cite{Wu2004, Wu2007}, generalizing the earlier idea by Oganesyan \textit{et al.} \cite{Oganesyan2001} to the spin channel. Wu \textit{et al.} \cite{Wu2004, Wu2007} discussed both 2D and 3D cases. Here, for simplicity we focus on the 2D case. In 2D, the multipole spin density operator in complex representation is given by:
\begin{equation}
\hat{Q}^{a}_{l} = \sum_{\alpha \beta } \Psi^{\dagger}_{\alpha}(\mathbf{r}) \sigma^{a}_{\alpha \beta}(\partial_x + i \partial_y)^l \Psi_{\beta}(\mathbf{r})
\label{eq:Wu-2007-2.1-2.2}
\end{equation}
where $\sigma^{a} (a = x, y, z)$ represent the three Pauli matrices. The order parameter $Q^a_l \equiv \left\langle \hat{Q}^{a}_{l} \right\rangle$ obeys the following transformation laws: (1) $\mathcal{P} Q^a_l \mathcal{P} ^{-1} = (-1)^{l} Q^a_l$; (2) $\mathcal{T} Q^a_l \mathcal{T} ^{-1} = (-1)^{l+1} Q^a_l$; and (3) $Q^a_l$ is invariant under a rotation by $\pi/l$ followed by a spin flip. The first two properties indicate that $Q^a_l$ transforms like electronic and magnetic toroidal multipoles for odd and even $l$ (see Tab.\ref{tab:four-multipole-symmetry}). The third transformation law, under a combined effect of independent spatial and spin rotations, is a key property of spin group \citep{Liu2022}, which plays an important role in AM, as we shall see. 

The order parameter $Q^a_l$ can be decomposed into the real and complex part, $Q^a_l \equiv n^a_1 + in^a_2$ and hence can be represented by two vectors $\mathbf{n}_1 = (n^x_1, n^y_1, n^z_1)$ and  $\mathbf{n}_2 = (n^x_2, n^y_2, n^z_2)$. Wu \textit{et al.} \citep{Wu2004, Wu2007} have shown that the Landau-Ginzburg free energys takes the form:
\begin{equation}
F[\mathbf{n}_1, \mathbf{n}_1] = r(\mathbf{n}_1^2+\mathbf{n}_2^2) + v_1(\mathbf{n}_1^2+\mathbf{n}_2^2)^2 + v_2 |\mathbf{n}_1 \times \mathbf{n}_2|^2
\label{eq:Wu-2007-3.4}
\end{equation}
where $r$, $v_1$, and $v_2$ are three parameters (or coupling constants) to be determined by the microscopic model Hamiltonian. The PI occurs at $r < 0$ for $F^A_{l\neq 0} < -2$. There are two symmetry broken phases: (1) if $v_2 > 0$, then $\mathbf{n}_{1} \parallel \mathbf{n}_{2}$. This is dubbed as "$\alpha$" phase, a name comes from the $A$ phase in liquid $^3$He. In the $\alpha$ phase, the spin-up and spin-down FSs are distorted but are rotated from each other by $\pi/l$. The spin-split state (see Fig.\ref{fig:Fig-ELC}(c)) proposed by Hirsch \textit{et al.} and the helicity-ordered state proposed by Varma \textit{et al.} are the $l =1$ $\alpha$ phase.  Fig.\ref{fig:Fig-ELC}(d) shows the $l =2$ $\alpha$ phase. This is also the "nematic-spin-nematic" state discussed briefly by Kivelson \textit{et al.} in 2003 \cite{Kivelson2003}. (2) If $v_2 < 0$, then $\mathbf{n}_{1} \perp \mathbf{n}_{2}$ and $|\mathbf{n}_{1}| = |\mathbf{n}_{2}|$. This is dubbed as "$\beta$" phases, in analogy to the $B$ phase in liquid $^3$He. In the $\beta$ phases, the FSs split into two parts with different areas, while each one keeps the circle shape as shown in Fig.\ref{fig:Fig-ELC}(e)-(h). The spin texture in $\beta$ phase exhibit the vortex structure in $\mathbf{k}$ space, so spin is no longer a good quantum number. To see the spin texture, Wu \textit{et al.} defined the $\mathbf{d}$ vector  \citep{Wu2007} 
\begin{equation}
\mathbf{d}(\mathbf{k}) = (\cos(l\theta_{\mathbf{k}}), \sin(l\theta_{\mathbf{k}}))
\label{eq:Wu-2007-4.6}
\end{equation}
where $\theta_{\mathbf{k}}$ is the azimuthal angle of $\mathbf{k}$ in the 2D plane. For the same $l$, there are two different spin configurations characterized by opposite winding numbers. For example, Fig.\ref{fig:Fig-ELC}(e)-(f) both have $l =1$, but the spin circulates counterclockwisely when enclosing the outer FS in Fig.\ref{fig:Fig-ELC}(e), clockwisely in Fig.\ref{fig:Fig-ELC}(f). The same applies for Fig.\ref{fig:Fig-ELC}(g)-(h). The spin texture in Fig.\ref{fig:Fig-ELC}(e) and Fig.\ref{fig:Fig-ELC}(f) are the well-known Rashba-type SOC and Dresselhaus-type SOC, but Fig.\ref{fig:Fig-ELC}(g) and Fig.\ref{fig:Fig-ELC}(h) are totally undiscovered at that time \citep{Wu2004}. 

Based on the above analysis, Wu \citep{Wu2007-2} termed the $F^A_{l} (l >0)$ channel PI as "unconventional magnetism", a name following the spirit of unconventional SC to emphasize the non-trivial representations of the rotation group. Although the term "magnetism" appears in this definition, it should be noted that there can be no localized magnetic moments in unconventional magnetism. Unconventional magnetism with magnetic dipolar interactions, which is a form of SOC, can be found in Fu \textit{et al.} \citep{Fu2015}, Norman \textit{et al.} \citep{Norman2015}, and Yuan \textit{et al.}  \citep{Yuan2025}. 

From above classification, it is readily seen that to identify an ELC phase, four types of indexes are required: the channel index $l$ and the charge/spin index of the Landau parameter $F^{S/A}_l$, the $\alpha$ or $\beta$ phase for isotropic or anisotropic FSs, and a winding number to distinguish the spin texture with the same $l$. In the framework of multipole basis, we will see that these four indexes will be unified.

\subsection*{B. Multipole expansions}

\begin{figure*}
\begin{centering}
\includegraphics[width=0.83\textwidth]{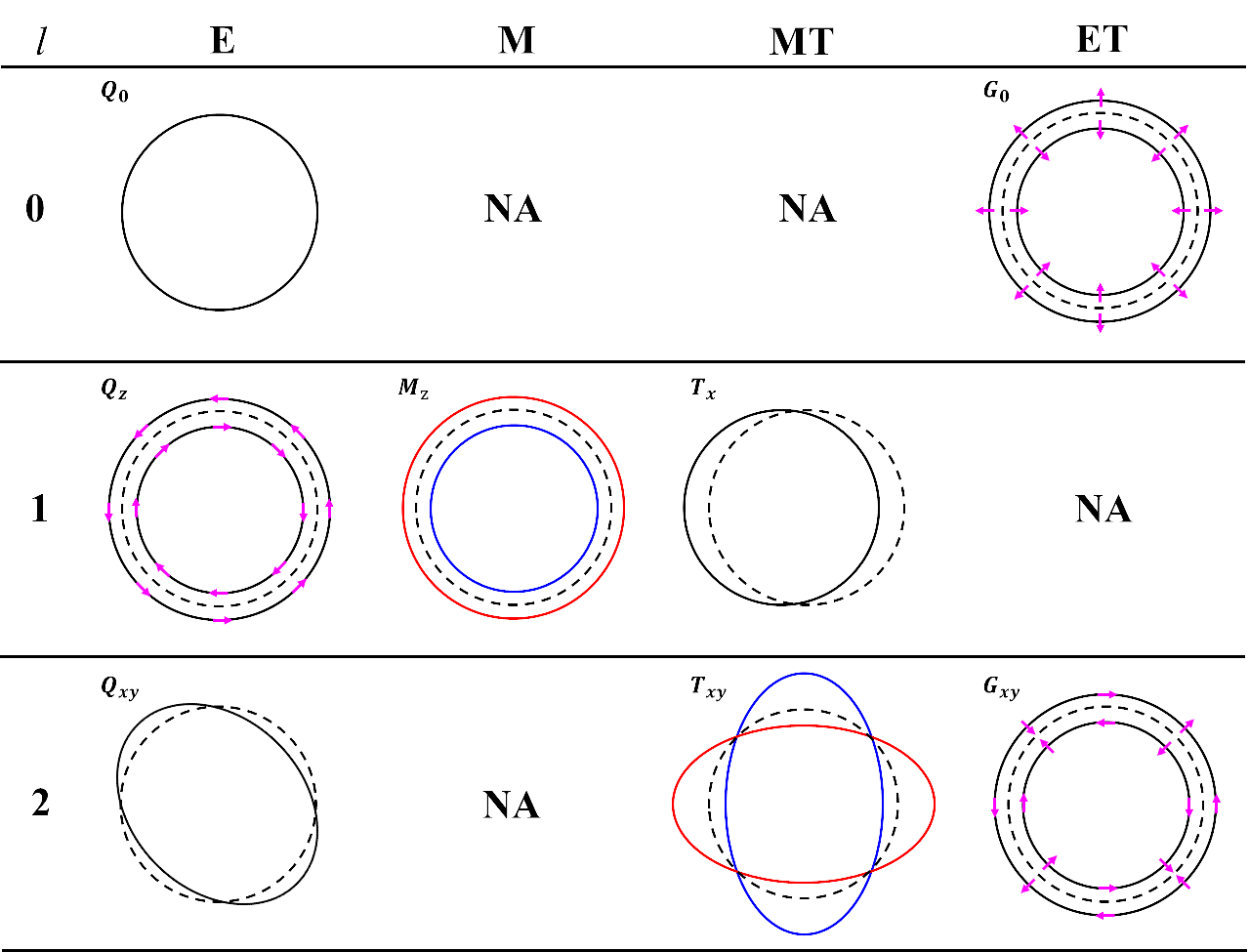}
\par\end{centering}
\caption{Multipole basis up to $l=2$. The dashed line denotes the Fermi surfaces with vanishing multipoles. "NA" means not allowed by symmetry. It is useful to compare the various phases in this figure with the ones in Fig.\ref{fig:Fig-ELC}. In particular, the dispersion relation of the $\textbf{MT}$ quadrupole $T_{xy}$ should be contrasted with that of $F^{A}_{2}$ $\alpha$-phase shown in Fig.\ref{fig:Fig-ELC}(d), both of which exhibit a similar feature as the $d$-wave AMs. }
\label{fig:Fig-3} 
\end{figure*}

Textbooks in classical electrodynamics introduce two types of multipoles, electric ($\textbf{E}$) and magnetic ($\textbf{M}$) multipoles \citep{Jackson1999}.  In 1975, Dubovik \textit{et al.} \citep{Dubovik1975} discovered the magnetic toroidal ($\textbf{MT}$) multipoles "hidden" in the classical Maxwell's equation. In condensed matter physics, $\textbf{MT}$ multipoles have been extensively investigated owing to its potential role in exotic phenomena. Interested readers can refer to reviews \citep{Dubovik1990, Spaldin2008}.  

\begin{table}[b]
	\caption{Transformation properties of $Q_{lm}$, $M_{lm}$, $G_{lm}$, and $T_{lm}$ multipoles under $\mathcal{P}$ and $\mathcal{T}$.}
	\label{tab:four-multipole-symmetry}
	\begin{tabular}{c|c|c|c}
		\hline\hline
		Multipole types & Multipole symbol  & $\mathcal{P}$ & $\mathcal{T}$  \\
		\hline
		$\textbf{E}$         & $Q_{lm}$  & $(-1)^l$       & + \\
		$\textbf{M}$        & $M_{lm}$  & $(-1)^{l+1}$ & - \\
		$\textbf{MT}$      & $T_{lm}$  & $(-1)^l$       & - \\
		$\textbf{ET}$      & $G_{lm}$   & $(-1)^{l+1}$ & + \\
		\hline\hline
	\end{tabular}
\end{table}

It seems that there should only be three multipoles $\textbf{E}$, $\textbf{M}$, and $\textbf{MT}$ in nature. However, they are insufficient to form a complete vector basis of multipole representations under $\mathcal{P}$ and $\mathcal{T}$. Therefore, in 1986 Dubovik \textit{et al.} \citep{Dubovik1986} introduced a forth multipole called electric toroidal ($\textbf{ET}$) multipole. A phenomenological introduction about this history is given by Nanz \textit{et al.} \citep{Nanz2016}. $\textbf{E}$, $\textbf{M}$, $\textbf{MT}$, and $\textbf{ET}$ transform as a polar tensor with time-reversal even, axial tensor with time-reversal odd, polar tensor with time-reversal odd, and axial tensor with time-reversal even, therefore exhausting all possibilities under $\mathcal{P}$ and $\mathcal{T}$. Under rotational symmetry, $\textbf{E}$, $\textbf{M}$, $\textbf{MT}$, and $\textbf{ET}$ can be represented in different partial waves as $Q_{lm}$, $M_{lm}$, $G_{lm}$, and $T_{lm}$ in 3D.
The transformation property of $Q_{lm}$, $M_{lm}$, $G_{lm}$, and $T_{lm}$ multipoles under $\mathcal{P}$ and $\mathcal{T}$ is shown in Tab.\ref{tab:four-multipole-symmetry}.   

The concept of multipole expansions is also applied to describe atomic-scale electromagnetic distribution of the wave-function of electrons bound to a single-centered atom, leading to the so called (atomic) multipole basis. Such a terminology emphasizes that the multipoles form a complete basis set and each multipole basis is equally important for describing physical properties, to be distinguished from the conventional multipole expansion where higher-order multipoles are less important and contribute only weakly. Due to the highly localized $f$ electrons, higher-order multipole basis has a long history in studying $f$-electron systems,
 especially the "hidden order" whose order parameters are hard to identify experimentally. For more details, we refer to the excellent reviews \citep{Santini2009, Kuramoto2009}.

Although the expressions of quantum operators for $\textbf{E}$, $\textbf{M}$, and $\textbf{MT}$ multipole can be direct derived from their classical definitions \citep{Kusunose2008}, the expressions for $\textbf{ET}$ multipoles are known only recently by Hayami \textit{et al.} in 2018 \citep{Hayami2018-JPSJ} as it does not appear in the multipole expansions. For a single-band model, these operators in $\mathbf{k}$ space are given by \citep{Hayami2018-PRB}:
\begin{subequations}
\begin{align}
Q_{lm}(\mathbf{k}) &\equiv 
\begin{cases}
\sigma_0 O_{lm}(\mathbf{k})  &(l=0, 2, ...)\\
(\mathbf{k} \times \pmb{\sigma}) \cdot \nabla_{\mathbf{k}} O_{lm}(\mathbf{k}) &(l=1, 3, ...) \\
\end{cases}
, \label{eq:Hayami-2018-PRB-21} \\
M_{lm}(\mathbf{k}) &\equiv 
\begin{cases}
0  &(l=0, 2, ...)\\
\pmb{\sigma} \cdot \nabla_{\mathbf{k}} O_{lm}(\mathbf{k}) &(l=1, 3, ...) \\
\end{cases} 
, \label{eq:Hayami-2018-PRB-23} \\
T_{lm}(\mathbf{k}) &\equiv 
\begin{cases}
0  &(l=0)\\
(\mathbf{k} \times \pmb{\sigma}) \cdot \nabla_{\mathbf{k}} O_{lm}(\mathbf{k}) &(l=2, 4, ...) \\
\sigma_0 O_{lm}(\mathbf{k})  &(l=1, 3, ...)\\
\end{cases} 
, \label{eq:Hayami-2018-PRB-22} \\
G_{lm}(\mathbf{k}) &\equiv 
\begin{cases}
\mathbf{k} \cdot \pmb{\sigma} &(l=0) \\
\pmb{\sigma} \cdot \nabla_{\mathbf{k}} O_{lm}(\mathbf{k})  &(l=2, 4, ...)\\
0 &(l=1, 3, ...) \\
\end{cases} 
, \label{eq:Hayami-2018-PRB-24} 
\end{align}
\end{subequations} 
where $\pmb{\sigma} = (\sigma^x, \sigma^y, \sigma^z)$ and the harmonics $O_{lm}(\mathbf{k})$ is defined as
\begin{equation}
O_{lm}(\mathbf{k}) = \sqrt{\frac{4\pi}{2l+1}}k^l Y^*_{lm}(\hat{\mathbf{k}})
\end{equation}
with $Y_{lm}(\hat{\mathbf{k}})$ the spherical harmonics of the angles $\hat{\mathbf{k}} = \mathbf{k}/k$. The index $l$ represents the order (also known as rank) of multipole basis: $l$ = 0 (monopole), 1 (dipole), 2 (quadrupole), 3 (octupole), 4 (hexadecapole), and so on. At first glance, Eq.\ref{eq:Hayami-2018-PRB-21} and Eq.\ref{eq:Hayami-2018-PRB-22} appear to share the same formal structure, differing only in parity of $l$. To demonstrate why this occurs, here
we elaborate in detail how Eq.\ref{eq:Hayami-2018-PRB-21} -Eq.\ref{eq:Hayami-2018-PRB-24} are systematically constructed from the basis function $O_{lm}(\mathbf{k})$. We first notice that under $\mathcal{P}$ and $\mathcal{T}$, $O_{lm}(\mathbf{k})$ acquires the same factor $(-1)^l$. According to Tab.\ref{tab:four-multipole-symmetry}, the even-order $\sigma_0 O_{lm}(\mathbf{k})$ thus represents the $\textbf{E}$ multipoles (first line of Eq.\ref{eq:Hayami-2018-PRB-21}), whereas the odd-order $\sigma_0 O_{lm}(\mathbf{k})$ corresponds to the $\textbf{MT}$ multipoles (third line of Eq.\ref{eq:Hayami-2018-PRB-22}). Next the operator $(\mathbf{k} \times \pmb{\sigma}) \cdot \nabla_{\mathbf{k}} $ is $l$-conserving, $\mathcal{P}$-even, and $\mathcal{T}$-odd, therefore $(\mathbf{k} \times \pmb{\sigma}) \cdot \nabla_{\mathbf{k}} O_{lm}(\mathbf{k})$ still has order $l$, but now transforms with parity $(-1)^l$ under $\mathcal{P}$ and $(-1)^{l+1}$ under $\mathcal{T}$. As a result, the odd-order $(\mathbf{k} \times \pmb{\sigma}) \cdot \nabla_{\mathbf{k}} O_{lm}(\mathbf{k})$ represents the $\textbf{E}$ multipoles (second line of Eq.\ref{eq:Hayami-2018-PRB-21}), while the even-order $(\mathbf{k} \times \pmb{\sigma}) \cdot \nabla_{\mathbf{k}} O_{lm}(\mathbf{k})$ gives the $\textbf{MT}$ multipoles (second line of Eq.\ref{eq:Hayami-2018-PRB-22}). Note that $T_0$ vanishes as the gradient of $O_{00}$ is either 0 or proportional to $\mathbf{k}$. Finally, the operator $\pmb{\sigma} \cdot \nabla_{\mathbf{k}}$ is $l$-conserving, $\mathcal{P}$-odd, and $\mathcal{T}$-even, therefore $\pmb{\sigma} \cdot \nabla_{\mathbf{k}} O_{lm}(\mathbf{k})$ still has order $l$, but now gathers the parity $(-1)^{l+1}$ under $\mathcal{P}$ and $(-1)^{l}$ under $\mathcal{T}$. According to Tab.\ref{tab:four-multipole-symmetry}, the odd-order $\pmb{\sigma} \cdot \nabla_{\mathbf{k}} O_{lm}(\mathbf{k})$ represents the $\textbf{M}$ multipoles (second line of Eq.\ref{eq:Hayami-2018-PRB-23}), while the even-order $\pmb{\sigma} \cdot \nabla_{\mathbf{k}} O_{lm}(\mathbf{k})$ gives the $\textbf{ET}$ multipoles (second line of Eq.\ref{eq:Hayami-2018-PRB-24}). On the other hand, it is impossible to construct the even-order $\textbf{M}$  and the order $\textbf{ET}$ multipoles within the single-band systems, they require the multiorbital or sublattice dof, see Appendix B of Hayami \textit{et al.} \citep{Hayami2018-PRB} for more information.
To make a specific connection with the previous section, we note that the electric quadrupole $O_{22}(\mathbf{k})=(k_x+ik_y)^2$. In the second quantization , it takes the form $\Psi^{\dagger}(\mathbf{r})(\partial_x + i\partial_y)^2\Psi(\mathbf{r})$, which is an equivalent representation of Eq.\ref{eq:Oganesyan-2001-1}.

The one-electron Hamiltonian is a $\mathcal{P}$- and $\mathcal{T}$-invariant scalar \citep{Hayami2018-PRB}:
\begin{equation}
\hat{H}_0 = \sum^{Q, M, T, G}_{X} \sum_{\mathbf{k} \alpha \beta} \sum_{l m} X^{ext}_{lm} X^{\alpha \beta}_{l m}(\mathbf{k}) \hat{c}^{\dagger}_{\mathbf{k} \alpha} \hat{c}_{\mathbf{k} \beta}
\label{eq:Hayami-2018-PRB-31}
\end{equation}
Here we have assume $X_{l m}$ in real representation (which can be achieved by taking linear combination of complex $X_{l m}$) so that the Hermite conjugation is omitted in the scalar product. $X^{ext}_{lm}$ represent "symmetry breaking" fields (also known as conjugate fields), which cause symmetry-breaking for certain multipoles. In order to examine the effect of multipoles on the band structure, in the single-band systems, the Hamiltonian in Eq.\ref{eq:Hayami-2018-PRB-31} can be divided as
\begin{equation}
\hat{H}_0 = \sum_{\mathbf{k} \alpha \beta} [\varepsilon^{E}(\mathbf{k})\delta_{\alpha \beta} + \varepsilon^{O}(\mathbf{k})\delta_{\alpha \beta} + f^{E}_{\alpha \beta}(\mathbf{k}) + f^{O}_{\alpha \beta}(\mathbf{k})] \hat{c}^{\dagger}_{\mathbf{k} \alpha} \hat{c}_{\mathbf{k} \beta}
\label{eq:Hayami-2018-PRB-33}
\end{equation}
where $\varepsilon$ ($f_{\alpha \beta}$) denotes the charge (spin) sector and the superscript $E(O)$ represents symmetric (antisymmetric) contribution with respect to $\mathbf{k}$. As the even-rank $Q_{lm}(\mathbf{k})$, $T_{lm}(\mathbf{k})$, and the odd-rank $M_{lm}(\mathbf{k})$ are even function of $\mathbf{k}$, and other multipoles are odd function of $\mathbf{k}$, each coefficient in Eq.\ref{eq:Hayami-2018-PRB-33} is identified as:
\begin{subequations}
\begin{align}
\varepsilon^{E}(\mathbf{k}) &= \sum^{even}_{l m}  Q^{ext}_{lm} Q_{lm}(\mathbf{k})
, \label{eq:Hayami-2018-PRB-34} \\
\varepsilon^{O}(\mathbf{k}) &= 
\sum^{odd}_{l m}  T^{ext}_{lm} T_{lm}(\mathbf{k})
, \label{eq:Hayami-2018-PRB-35} \\
f^{E}_{\alpha \beta}(\mathbf{k}) &= \sum^{odd}_{l m} M^{ext}_{lm} M^{\alpha \beta} _{lm}(\mathbf{k})
+ \sum^{even}_{l m} T^{ext}_{lm} T^{\alpha \beta}_{lm}(\mathbf{k})
, \label{eq:Hayami-2018-PRB-36} \\
f^{O}_{\alpha \beta}(\mathbf{k}) &= \sum^{odd}_{l m} Q^{ext}_{lm} Q^{\alpha \beta} _{lm}(\mathbf{k})
+ \sum^{even}_{l m} G^{ext}_{lm} G^{\alpha \beta}_{lm}(\mathbf{k})
. \label{eq:Hayami-2018-PRB-37} 
\end{align}
\end{subequations} 

By diagonalizing Eq.\ref{eq:Hayami-2018-PRB-33} with  constants $X^{ext}_{lm}$, we can obtain different types of band structures. In the following, typical cases are discussed and are contrasted with the 2D results of the $F^{S/A}_l$ channels in the previous subsection:  

i) $\varepsilon^{E}(\mathbf{k})$ in Eq.\ref{eq:Hayami-2018-PRB-34} represents a symmetric band dispersion without spin splitting. This band structure is present when both $\mathcal{P}$ and $\mathcal{T}$ exist. The $\textbf{E}$ monopole $Q_{0}$ gives the kinetic energy of free electron $\varepsilon(\mathbf{k}) \propto Q^{ext}_{0} \mathbf{k}^2$. This is shown in the first row (i.e., $l = 0$) and first column (i.e., the column $\textbf{E}$) of Fig.\ref{fig:Fig-3}. The $\textbf{E}$ quadrupole $Q_{xy}$, accompanying the $\textbf{E}$ monopole $Q_{0}$,  gives the band dispersion $\varepsilon(\mathbf{k}) \propto Q^{ext}_{0} \mathbf{k}^2 + Q^{ext}_{xy} k_x k_y$ which describes an ellipse as shown in the third row and first column of Fig.\ref{fig:Fig-3}. This quadrupole-type deformation corresponds to the orbital nematic order $F^S_{2}$ described by Oganesyan \textit{et al.} \citep{Oganesyan2001} (see Fig.\ref{fig:Fig-ELC}(a)). 

ii) $\varepsilon^{O}(\mathbf{k})$ in Eq.\ref{eq:Hayami-2018-PRB-35} represents the asymmetric-type band dispersions with spin degeneracy. For example, the $\textbf{MT}$ dipole $T_x$, accompanying the $\textbf{E}$ monopole $Q_{0}$, leads to a shift along $k_x$ direction in the dispersion relation $\varepsilon(\mathbf{k}) \propto Q^{ext}_{0} \mathbf{k}^2 + T^{ext}_{x} k_x$, as shown in the second row and third column of Fig.\ref{fig:Fig-3}. Obviously, this band structure occurs when both $\mathcal{P}$ and $\mathcal{T}$ are broken. Actually it corresponds to the $F^S_{1}$ channel instability.

iii) $f^{E}_{\alpha \beta}(\mathbf{k})$ in Eq.\ref{eq:Hayami-2018-PRB-36}  represents a symmetric-type band dispersion with spin splitting, which appears for system with conserved $\mathcal{P}$ but broken $\mathcal{T}$. The simplest case is considering the $\textbf{M}$ monopole $M_z$, accompanying the $\textbf{E}$ monopole $Q_0$,  gives the band dispersion $\varepsilon_{\sigma}(\mathbf{k}) \propto Q^{ext}_{0} \mathbf{k}^2 + M^{ext}_{z} \sigma$ with $\sigma = \pm 1$. With $M^{ext}_{z}$ contributed by either Weiss molecular field of the spontaneous ferromagnetic ordering or a Zeeman field, this term give an isotropic spin splitting as shown in the second row and second column of Fig.\ref{fig:Fig-3}, corresponding to the $F^A_0$ channel PI. Another non-trivial case is the $\textbf{MT}$ quadrupole $T_{xy}$, accompanying the $\textbf{E}$ monopole $Q_{0}$, it leads to the band dispersion $\varepsilon_{\sigma}(\mathbf{k}) \propto Q^{ext}_{0} \mathbf{k}^2 + \sigma T^{ext}_{xy} (k^2_x - k^2_y) $ (where we have set $k_z = 0$ for simplicity). This is an anisotropic spin splitting as shown in the third row and third column of Fig.\ref{fig:Fig-3}, corresponding to the $F^A_2$  channel PI in the $\alpha$ phase, or corresponding to the spin splitting in the $d_{x^2-y^2}-$wave AM, as we shall see.   

iv) $f^{O}_{\alpha \beta}(\mathbf{k})$ in Eq.\ref{eq:Hayami-2018-PRB-37} represents the asymmetric-type band dispersion with spin splitting. For example, the $\textbf{E}$ dipole $Q_z$ gives the Rashba-type SOC $k_y \sigma_x - k_x \sigma_y$, and the $\textbf{E}$ octupole $Q_{xyz}$ gives the Dresselhuas-type SOC $k_x(k^2_y-k^2_z)\sigma_x + k_y(k^2_z-k^2_x)\sigma_y + k_z(k^2_x-k_y^2)\sigma_z$. In 2D, the $\textbf{E}$ dipole $Q_x$ becomes $k_y \sigma_z$, accompanying $\textbf{E}$ monopole $Q_0$, gives the band dispersion $\varepsilon_{\sigma}(\mathbf{k}) = Q^{ext}_{0} \mathbf{k}^2 + \sigma Q^{ext}_{x} k_y$, which gives the spin-split state in Fig.\ref{fig:Fig-ELC}(c), and the $\textbf{E}$ octupole $Q_{3a}$ ($Q_{3b}$) gives the Ising-type SOC $(3k^2_x k_y - k^3_y) \sigma_z$ ($(3k_x k^2_y - k^3_x) \sigma_z$) near a valley in hexagonal systems (see the third line of Tab.\ref{tab:four-FFLO}).
The $\textbf{ET}$ monopole $G_0$ gives the hedgedog-type SOC $\mathbf{k} \cdot \mathbf{\sigma}$, and the $\textbf{ET}$ quadrupole $G_{xy}$ gives the SOC $k_x \sigma_y + k_y \sigma_x$. 

From the above discussion, we can readily see that multipole basis can be understood as a unified description of the charge and spin channel PI for a single-band model Hamiltonian. 

Multipole basis can also be employed to describe magnetic structures which will be discussed in the following subsection. To achieve this, it is necessary to generalize single-site multipole basis to multi-site systems, just as the way cluster orbitals are introduced on the top of atomic orbitals. 
In 2017, Suzuki \textit{et al.} \citep{Suzuki2017} introduced the concept of cluster multipoles to provide a unified framework for describing the intrinsic anomalous Hall effect (AHE) in magnetic materials. By taking linear combination of  atomic $\textbf{M}$ multipoles at symmetry-equivalent Wyckoff positions, they constructed cluster $\textbf{M}$ multipoles, which can be 
further classified according to the point group symmetry of the cluster. This type of multipole expansions now is referred to as the symmetry-adapted multipole basis (SAMB) \citep{Kusunose2022}. Suzuki \textit{et al.} demonstrated that a necessary condition for the AHE is a nonvanishing $\textbf{M}$ SAMB that transforms like a magnetic dipole moment \citep{Suzuki2017}. Since $\textbf{M}$ and $\textbf{MT}$ multipoles are odd under $\mathcal{T}$, their corresponding SAMBs can be used to construct arbitrary magnetic structures \citep{Suzuki2019}. In other words, multi-site multipoles \citep{Yatsushiro2021} provide an alternative method to scrutinize spin space group (see, i.e., the next subsection). Furthermore, the two polar $\textbf{E}$ and $\textbf{MT}$ SAMBs can be used to describe SML in AFMs \citep{Watanabe2018, Bhowal2024, Fernandes2024, Hayami2019, Hayami2020}.

\begin{figure*}
\begin{centering}
\includegraphics[width=0.85\textwidth]{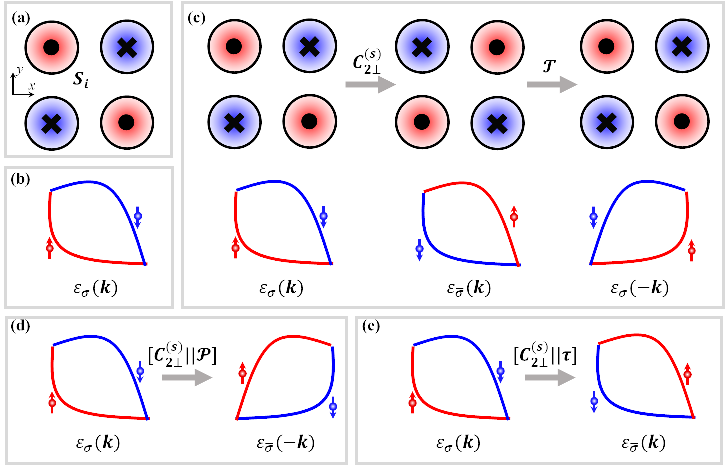}
\par\end{centering}
\caption{(a) Schematic representation of a collinear magnet, here the local magnetic moment $S_i$ is perpendicular to the paper plane ($xoy$ plane). (b) Spin-polarized band structure of collinear magnets. (c) Top: both $C^{(s)}_{2\perp}$ and $\mathcal{T}$ flips local magnetic moment $S_i$ in real space. Bottom: the combination of $C^{(s)}_{2\perp}$ and $\mathcal{T}$ functions as $\mathcal{P}$ for the energy spectrum. (d) $[C^{(s)}_{2\perp}||\mathcal{P}]$ works as $\mathcal{T}$ for the energy spectrum. (e) $[C^{(s)}_{2\perp}||\boldsymbol{\tau}]$ protects spin degeneracy. }
\label{fig:Fig-AM} 
\end{figure*}

\subsection*{C. Altermagnetism}
AFM could represent next-generation spintronic applications \citep{Jungwirth2016, Baltz2018}, owing to their unique combination of properties: i) they are inherently robust against perturbation against external magnetic field disturbances; ii) they generate zero stray fields, allowing for high-density device integration; and iii) they exhibit ultrafast dynamics reaching the terahertz range, enabling exceptionally high-speed response. However, the absence of net magnetization in AFMs poses a significant challenge for signal readout, as traditional electromagnetic methods do not couple with the AFM order parameter. 
  
The discovery of the AHE in the noncollinear AFM Mn$_3$Sn \citep{Nakatsuji2015} has partially addressed the signal readout challenge in AFM systems. This breakthrough has lead to the emergence of a new subfield known as noncollinear AFM spintronics (see reviews \citep{Chen2022, Rimmler2025}). However, the intricate magnetic texture inherent to noncollinear AFMs can always lead to rapid decoherence of spin excitations. This limitation has driven interest in \textit{collinear} AFM spintronics, where the focus is on achieving longer spin coherence lengths.

\subsubsection*{1, Symmetry aspects of collinear AFMs}
Collinear magnets possess two intrinsic on-site symmetries, regardless of their specific magnetic configuration \citep{Smejkal2022-1, Liu2022}. 
For a collinear magnet illustrated in Fig.\ref{fig:Fig-AM}(a), all local magnetic moments S$_i$ at site "$i$" are aligned perpendicular to the paper plane, i.e. along the $z$-direction. Because the spin configuration remains invariant under arbitrary spin rotations about the $z$-axis, thus collinear magnet exhibit a U$_s$(1) symmetry, where the subscript "$s$" highlights that the rotation occurs in spin space. As a result, spin remains a good quantum number, and each Bloch state maintains a definite spin projection of $\pm \frac{1}{2}$, as shown by the red/blue color in Fig.\ref{fig:Fig-AM}(b). In addition to this continuous spin rotational symmetry, collinear magnets also exhibit a second discrete symmetry, shown in Fig.\ref{fig:Fig-AM}(c): an 180$^\circ$ spin rotation about an axis perpendicular to the $z$-axis (denoted as $C^{(s)}_{2\perp}$ and corresponding to the spin flip in discussing symmetries of Eq.\ref{eq:Wu-2007-2.1-2.2}), followed by the time reversal operation $\mathcal{T}$.  Such a combined symmetry imposes the constraint $\varepsilon_{\sigma}(\mathbf{k}) = \varepsilon_{\sigma}(-\mathbf{k})$ on the band structure. While the system may lack real-space inversion symmetry $\mathcal{P}$, this combined symmetry effectively plays the role of $\mathcal{P}$ in determining the energy spectrum when SOC is absent. 

At this point, it is helpful to clarify why conventional (or N\'{e}el) AFMs typically exhibit
Kramers degeneracy $\varepsilon_{\sigma}(\mathbf{k}) = \varepsilon_{\overline{\sigma}}(\mathbf{k})$ ($\overline{\sigma}$ represent the opposite of $\sigma$). This spin degeneracy arises from two fundamental symmetries \citep{Smejkal2022-1}. The first one is $C^{(s)}_{2\perp}$ followed by $\mathcal{P}$ as shown in Fig.\ref{fig:Fig-AM}(d). Together, this combined symmetry effectively mimic $\mathcal{T}$, as the Bloch state now satisfies $\varepsilon_{\sigma}(\mathbf{k}) = \varepsilon_{\overline{\sigma}}(-\mathbf{k})$. Since collinear AFMs inherently posses an effective $\mathcal{P}$ as we just mentioned, in the absence of SOC, both effective $\mathcal{P}$ and effective $\mathcal{T}$ guarantees that each Bloch state is doubly degenerated. The second symmetry involves the same $C^{(s)}_{2\perp}$ spin rotation, but followed by a fractional lattice translation $\pmb{\tau}$ connecting the two magnetic sublattices. Because $\pmb{\tau}$ introduces only a phase factor to the Bloch state without changing the energy spectrum, this combined symmetry operation leads to  Kramers degeneracy as shown in Fig.\ref{fig:Fig-AM}(e). Therefore, to realize nonrelativistic spin splitting, a collinear AFM must not possess $[C^{(s)}_{2\perp}||\mathcal{P}]$ or $[C^{(s)}_{2\perp}||\pmb{\tau}]$, where the transformation on the left and right sides of the double vertical bar acts in spin and spatial space, respectively. Since $C^{(s)}_{2\perp}$ flips the magnetic moments on all sublattices, this condition is equivalently stated as requiring that $\mathcal{P}$ or $\pmb{\tau}$ (if they exist) must belong to the subgroup interchanging the same sublattice. This rule serves as the symmetry criterion in screening AMs from collinear AFMs \citep{Smolyanyuk2024}.

\subsubsection*{2, Prelude to altermagnetism}
Breaking Kramers degeneracy, even just over a small portion of the first Brillouin zone (FBZ), provides the first step toward collinear AFM spintronics. First-principles calculations has played important roles in searching for lifted Kramers degeneracy in collinear AFM. 
By employing DFT+$U$ calculation for different MnO$_2$ phases,  Noda \textit{et al.} \citep{Noda2016} firstly reported NRSML in collinear AFM. It should be noted that before the work by Nada \textit{et al.} \cite{Noda2016},  Franchini \textit{et al.} \citep{Franchini2007} have studied the ground state of $\beta$-MnO$_2$ and Cockayne \textit{et al.} \citep{Cockayne2012} have studied the ground state of $\alpha$-MnO$_2$. However, no band structures are explicitly displayed in those studies. 

Noda \textit{et al.} \citep{Noda2016} also demonstrated a symmetry analysis to unveil the spin splitting and spin degeneracy at a given $\mathbf{k}$-point in the FBZ. In the Kohn-Sham framework, the single-electron potential $V_{\sigma}$ is spin dependent for a collinear magnet, so the Kohn-Sham equation for spin-up and spin-down electrons are:
\begin{subequations}
\begin{align}
[\frac{1}{2}(\mathbf{k}-i\vec{\nabla})^2 +V_{\uparrow}]u_{\uparrow}(\mathbf{k}) = \varepsilon_{\uparrow}(\mathbf{k})u_{\uparrow}(\mathbf{k})
, \label{eq:Noda2016-1a} \\
[\frac{1}{2}(\mathbf{k}-i\vec{\nabla})^2 +V_{\downarrow}]u_{\downarrow}(\mathbf{k}) = \varepsilon_{\downarrow}(\mathbf{k})u_{\downarrow}(\mathbf{k})
, \label{eq:Noda2016-1b}
\end{align}
\end{subequations} 
where $u_{\sigma}(\mathbf{k})$ denotes the periodic eigenfunctions at each $\mathbf{k}$. Now suppose a space-group operation $\mathcal{R}$ that maps the single-electron potential as $\mathcal{R}V_{\sigma}\mathcal{R}^{-1} = V_{\overline{\sigma}}$. In the case that a certain wave vector $\mathbf{k}$  is invariant under $\mathcal{R}$, i.e. $\mathcal{R}\mathbf{k} = \mathbf{k}$ (equivalence up to a reciprocal lattice vector), the Hamiltonian of the spin-up and spin-down electron are exchanged under $\mathcal{R}$, which gives Kramers degeneracy $\varepsilon_{\uparrow}(\mathbf{k}) = \varepsilon_{\downarrow}(\mathbf{k})$. However, away from these high symmetric points, for general wave vectors $\mathbf{k}$ and $\mathbf{k}'=\mathcal{R}\mathbf{k}$, then we have SML $\varepsilon_{\uparrow}(\mathbf{k}) = \varepsilon_{\downarrow}(\mathbf{k}')$. 

It is also worth noting that Noda \textit{et al.} \citep{Noda2016} highlighted the critical role of oriented MnO$_6$ octahedron in modifying the magnetization densities around Mn atoms, leading to the emergence of a staggered potential $V_{\sigma}$. A common mechanism for introducing such oriented MO$_6$ octahedron in transition metal compounds is the Jahn-Teller distortion. In particular, this effect in LaMO$_3$ (M = Cr, Mn, Fe) has been shown to give rise to spin splitting \citep{Okugawa2018}. LaMO$_3$ (M = Cr, Mn, Fe) is only one member of the perovskite structure where lattice distortion is commonly seen, therefore perovskites should be a zoo for AMs \citep{Naka2021, Autieri2025}, as reviewed by Naka \textit{et al.} \citep{Naka2025}.

After the work by Noda \textit{et al.} \citep{Noda2016}, theoretical studies of momentum-dependent spin splitting protected by crystal symmetry in collinear AFM have significantly advanced. Notable reports include Naka \textit{et al.} on 2D orangic $\kappa$-Cl \citep{Naka2019}, Ahn \textit{et al.} \citep{Ahn2019} and \v{S}mejkal \textit{et al.} \citep{Smejkal2020} on RuO$_2$,  Yuan \textit{et al.} \citep{Yuan2020} and Egorov \textit{et al.} \citep{Egorov2021} on MnF$_2$, Ma \textit{et al.} \citep{Ma2021} on V$_2$Se$_2$O and Mazin \textit{et al.} \citep{Mazin2021} on doped FeSb$_2$, among others. Beyond NRSML, \v{S}mejkal \textit{et al.} \citep{Smejkal2020} demonstrated that collinear AFMs can exhibit AHE, a phenomenon traditionally associated only with FMs and noncollinear AFMs. This property is particularly important for  spintronics, as it enables the detection of  magnetic states through electrical signals. Also in this work, \v{S}mejkal \textit{et al.} realized that the characterization of certain magnets requires the detailed shape of magnetization density.  Simultaneously, works by Naka \textit{et al.} \citep{Naka2019}, Shao \textit{et al.} \citep{Shao2021}, and Rafael \textit{et al.} \citep{Rafael2021} reported that spin currents along with NRSML can be generated in collinear AFMs, further highlighting their potentials in spintronics applications. And \v{S}mejkal \textit{et al.}  \citep{Smejkal2022-3} illustrated the tunneling magnetoresistance in junctions formed by collinear AFMs. These discoveries indicate the possibility of converting a collinear AFM state into measurable electric signals, thus making collinear AFM spintronics possible.
These fascinating phenomena has also been systematically explored within the framework of SAMB \citep{Watanabe2018, Hayami2019, Hayami2020}.

\begin{figure*}
\begin{centering}
\includegraphics[width=0.95\textwidth]{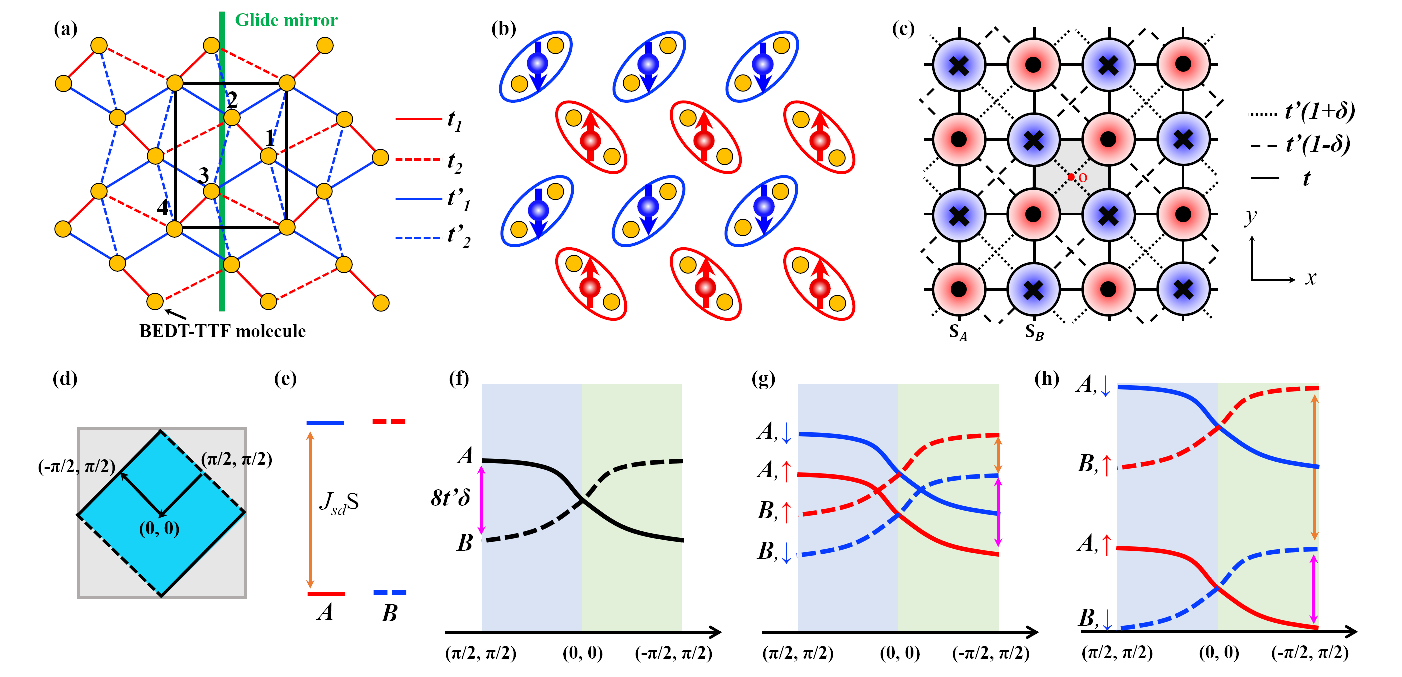}
\par\end{centering}
\caption{(a) Schematic illustration of the lattice structure of $\kappa$-Cl conducting layer. Each yellow dot represents one BEDT-TTF molecule. The black rectangle marks the unit cell. (b) AFM order driven by on-site Hubbard interaction. Each BEDT-TTF dimer (red and blue ellipses) contributes one localized magnetic moment. (c) A N\'{e}el state on an alternating anisotropic square lattice is a $d$-wave AM. It is invariant under $[C^{(s)}_{2\perp}||C^+_{4z}]$ symmetry, where the red dot labels the rotation center. The grey shaded region is the PM unit cell. (d) The grey (cyan) shaded region corresponds to the Brillouin zone of PM unit cell (N\'{e}el AFM supercell). For Hamiltonian $h_0(\mathbf{k})$ in Eq.\ref{eq:s-d-model-2}, at the N\'{e}el AFM supercell Brillouin zone boundary ($cos k_x +  cos k_y = 0$), the Bloch states exhibit an alternating sublattice composition, as indicated by the solid and dashed black lines, which correspond to sublattice $A$ and $B$, respectively. (e) Collinear AFM order introduces homogeneous spin-splitting, as described by $\hat{H}_{sd}$ in Eq.\ref{eq:s-d-model-2}. Here the blue/yellow color represents spin down/up, respectively. (f) Anisotropic NNN hopping introduces momentum-dependent sublattice-staggered splitting, as described by the last term in $h_0(\mathbf{k})$, Eq.\ref{eq:s-d-model-2}. (g)-(h) Collinear AFM order gives rise to $d_{xy}$-wave NRSML when the momentum-dependent sublattice-staggered potential is nonzero. Here the inter-sublattice hybridization $t$ is not considered.}
\label{fig:projection} 
\end{figure*}

Remarkably, the microscopic origin of staggered $V_{\sigma}$ in the AFM phase has been explored using a \textit{microscopic} repulsive Hubbard model by Naka \textit{et al.} \citep{Naka2019}. In $\kappa$-Cl the conducting layers are formed by BEDT-TTF molecules where four molecules in a unit cell arrange into two dimers (sublattice 1, 2 and sublattice 3, 4 in Fig.\ref{fig:projection}(a)). These two dimers are connected to each other via glide mirror symmetry (see green line in Fig.\ref{fig:projection}(a)). This class of organic materials is known to have a simple electronic structure composed of frontier molecular orbitals \citep{Kino1996, Liu2018}. In Fig.\ref{fig:projection}(a), the frontier orbitals in each BEDT-TTF dimer become strongly hybridized by the intra-dimer hopping $t_1$ and constitute both bonding and antibonding orbitals. With each BEDT-TTF molecule contributes one frontier molecular orbital, the low-energy effective model contains four bands (or eight-band including spin dof) in the unit cell:  the two lower-energy and two higher-energy bands formed by the bonding and antibonding orbitals, respectively. The system has three electrons per dimer on average, and hence, the four bands are 3/4 filled. Therefore, the following four-band Hubbard model can be constructed \citep{Naka2019}:
\begin{equation}
\begin{split}
\hat{H} &= \hat{H}_0 + \hat{H}_{int} \\
\hat{H}_0 &=  \sum_{i j a b} \sum_{\sigma} (t_{ia,jb} \hat{c}^{\dagger}_{i a \sigma} \hat{c}_{j b \sigma} + h.c.) -  \sum_{i a} \mu \hat{n}_{i a} \\
\hat{H}_{int} &= U \sum_{i a} \hat{n}_{i a \uparrow} \hat{n}_{i a \downarrow}  \\
\end{split}
\label{eq:Naka2019-1}
\end{equation}
where $t_{ia,jb}$ represents the four hoppings $(t_1, t_2, t'_1, t'_2) = (-0.207, -0.067, -0.102, 0.043)$ eV labelled in Fig.\ref{fig:projection}(a) and $a, b = 1-4$ is the index for the four sublattices.  Within the Hartree approximation, at 3/4 filling, increasing the on-site repulsion $U$ first drives a transition from a PM metal to an AFM metal (see Fig.\ref{fig:projection}(b)), and then a metal-insulator transition from an AFM metal to an AFM insulator. Once the AFM order emerges, the glide-mirror-symmetric hopping present in the PM phase becomes broken in either the spin-up or spin-down channel, but preserved in intra-spin channels. These resulting spin-dependent hoppings lead to NRSML.

\subsubsection*{3, Emergence of altermagnetism}
As more exotic physical properties discovered in such type of collinear AFM (see recent reviews \citep{Smejkal2022-2, Bai2024,  SongC2025, Guo2025, Jungwirth2025-1, Jungwirth2025-2, Jungwirth2025-3}), in 2022, \v{S}mejkal \textit{et al.} proposed the idea of type-III collinear magnets, to be distinguished from the well-established known FM and N\'{e}el AFM. Type-III collinear magnets specify collinear magnets which have compensating magnetic moments in real space and NRSML protected by crystal symmetry in the reciprocal space. Type-III collinear magnet was also termed \textit{altermagnetism} \citep{Smejkal2022-1}. Here, the prefix "alter" carries two distinct meanings. First, it refers to alternating orders in both real and reciprocal space: in real space, this reflects the compensating collinear magnetic moments; in reciprocal space, it signifies the alternating NRSML protected by crystal symmetry. In addition to this, "alter" also conveys the idea of alternating physical properties shared by FM and N\'{e}el AFM. Specifically, an "alter-magnet" can exhibit characteristics typical of FM in one context, such as NRSML, AHE, nonrelativistic spin current etc, while simultaneously displaying behaviours associated with AFM in the others including zero stray field, ultrafast dynamics up to terahertz and so on, as summarized by Mazin \textit{et al.} \citep{Mazin2022}. 

Following this thought, Liu \textit{et al.} \citep{LiuQ2025} proposed a second definition of unconventional magnetism: magnets adopt antiferromagnetic configurations yet display properties reminiscent of FMs. Within this framework, altermagnets (AMs) are considered a specific subset of  unconventional magnets. Cheong \textit{et al.} \citep{Cheong2024, Cheong2025} defines altermagnetism more broadly as magnetism characterized by broken $\mathcal{PT}$ symmetry, with a ground state exhibiting full spin compensation in the nonrelativistic limit. This definition also includes noncollinear antiferromagnets and aligns with the conception of unconventional magnetism by Liu \textit{et al.} \citep{LiuQ2025}.

To grasp the key characteristics of AMs, we begin with the standard $s-d$ model in 2D square lattices with two sublattices (see Fig.\ref{fig:projection}(c)) to see the role of static AFM order on spin splitting. Such a model analysis serves to introduce two ground-breaking insights --- one by Hayami \textit{et al.}  \citep{Hayami2019, Hayami2020} using the SAMB framework, and the other by \v{S}mejkal \textit{et al.} \citep{Smejkal2022-1, Smejkal2022-2} based on spin-group theory. Both perspectives are essential for developing a comprehensive understanding of AMs.

The Hamiltonian of Fig.\ref{fig:projection}(c) is rewritten as:
\begin{equation}
\begin{split}
\hat{H} &= \hat{H}_{0} + \hat{H}_{sd} \\
\hat{H}_{0} &=  - \sum_{i,j, \sigma} t_{ij}(\hat{c}^{\dagger}_{i \sigma} \hat{c}_{j \sigma} + h.c.) - \mu \sum_{i } \hat{n}_{i } \\
\hat{H}_{sd} &= - J_{sd} \sum_{i} \hat{\mathbf{S}}_i \cdot \hat{\mathbf{s}}_i  
\end{split}
\label{eq:s-d-model}
\end{equation}
where the hopping $t_{ij}$ extends up to next-nearest-neighbour (NNN) range. As illustrated in Fig.\ref{fig:projection}(c), it features an isotropic nearest-neighbour (NN) hopping $t$ and anisotropic NNN hoppings; for sublattice $A$, the latter take values of $t'(1+\delta)$ or $t'(1-\delta)$ along the $(1, 1)$-direction or $(1, -1)$-direction, respectively, where $\delta $ is a dimensionless parameter measuring NNN hopping anisotropy. For sublattice $B$, the NNN hopping is $t'(1-\delta)$ along the $(1, 1)$-direction and $t'(1+\delta)$ along the $(1, -1)$-direction, respectively. In real materials, these hoppings are obtained by integrating out the contributions of nonmagnetic ligand sites, and $t'$ can be larger than $t$. In $\hat{H}_{sd}$, the negative sign indicates FM coupling between spin-1/2 fermions and localized magnetic moments. Since we consider collinear AFM order, without loss of generality, $\hat{\mathbf{S}}_i$ can be chosen to +S and -S along the $z$-direction for sublattice $A$ and $B$, respectively. 
Such a collinear AFM order can be driven by on-site Hubbard repulsions, as explored by Das \textit{et al.} \citep{DasP2024}. At half-filling and zero temperature, for parameters $t'/t = 0.3$ and $\delta = 0.2$, a relatively weak Hubbard repulsion $U/t \approx 2.5$ is sufficient to induce a phase transition from a PM metal to an AM metal within the Hartree-Fock approximation \citep{DasP2024}. 

Hayami \textit{et al.} first realized that the existence of $C^{(s)}_{2\perp} \mathcal{T}$ as we discussed earlier inspires a multipole expansion of the dispersion in terms of some conjugate fields, similar to $X^{ext}_{lm}$ in Eq.\ref{eq:Hayami-2018-PRB-31}. The most crucial observation by Hayami \textit{et al.} is that the collinear AFM ordering can \textit{activate} particular conjugate fields, and hence can cause symmetry-breaking for the related multipoles. In other words, the N\'{e}el vector plays the role of a primary order parameter, and it induces  multipole basis like $T_{xy}$ as secondary order parameter depending on the detailed crystal structure. This idea has been recently highlighted and further developed by McClarty \textit{et al.} to formulate a Landau theory for AMs \citep{McClarty2024}. 

To better understand such a symmetry-breaking concept, it is useful to analyze the symmetry of 2D square lattice in Fig.\ref{fig:projection}(c). At the PM phase, Fig.\ref{fig:projection}(c) has a high symmetry described by the point group $C_{4v} =  \{E, C^+_{4z}, \mathcal{P} = C_{2z}, C^{-}_{4z}, \mathcal{M}_x, \mathcal{M}_y, \mathcal{M}_{11}, \mathcal{M}_{1 \bar{1}} \}$, where $C^+_{4z}$/$C^-_{4z}$ is the four-fold counterclockwise/clockwise rotation along the $z$-axis, and $\mathcal{M}_x$/$\mathcal{M}_y$/$\mathcal{M}_{11}$/$\mathcal{M}_{1 \bar{1}}$ denotes the mirror reflection perpendicular to $x$/$y$/$(1,1)$/$(1,-1)$ direction. Since the horizontal mirror plane is redundant, the two-fold rotation along the $z$-axis $C_{2z}$ is identical to the spatial inversion $\mathcal{P}$. Now we consider symmetry breaking where collinear AFM order is established as depicted in Fig.\ref{fig:projection}(c). As the collinear AFM order does not change under the invariant subgroup $C_{2v} = \{E, C_{2z}, \mathcal{M}_x, \mathcal{M}_y\}$ and acquires a "-1" under the coset $C^+_{4z}C_{2v}$, it belongs to the 1D IR $B_1$. Consequently, this collinear AFM order induces a conjugate field associated with the multipole basis $T_{xy}$ that transforms according to $B_1$. This conjugate field leads to $d$-wave NRSML. For a more detailed and rigorous treatment, see Sect. V of Ref. \citep{Hayami2020}.

To explicitly see the multipole basis $T_{xy}$ in Eq.\ref{eq:s-d-model}, we now perform Fourier transformation and introduce the spin-sublattice basis $\Psi_{\mathbf{k}} = \left(\hat{c}_{\mathbf{k}A \uparrow}, \hat{c}_{\mathbf{k}B \uparrow}, \hat{c}_{\mathbf{k}A \downarrow}, \hat{c}_{\mathbf{k}B \downarrow} \right)^T$, then Eq.\ref{eq:s-d-model} becomes
\begin{equation}
\hat{H} = \sum_{\mathbf{k}} \Psi^{\dagger}_{\mathbf{k}} h(\mathbf{k}) \Psi_{\mathbf{k}}
\end{equation}
with
\begin{equation}
\begin{split}
h(\mathbf{k})&= h_0(\mathbf{k}) + h_{sd} \\
h_0(\mathbf{k}) &= -(4t' \cos k_x \cos k_y + \mu) \sigma_0 \tau_0 \\
&-2t(\cos k_x + \cos k_y )\sigma_0 \tau_x 
+4t' \delta \sin k_x \sin k_y \sigma_0 \tau_z  \\
h_{sd} &= - \frac{J_{sd} \mathrm{S}}{2} \sigma_z \tau_z  \\
\end{split} 
\label{eq:s-d-model-2}
\end{equation}
where the lattice constants are set to unity and the Brillouin zone is given by $|k_x \pm k_y| \leq \pi$ as shown by the cyan shaded region on the right bottom of Fig.\ref{fig:projection}(d). The Pauli matrices $\sigma_{x/y/z}$ ($\tau_{x/y/z}$) stand for the spin (sublattice) dof. $\hat{H}_{sd}$ is an on-site interaction so it generates homogeneous spin splitting proportional to $J_{sd}$S. And due to the collinear AFM order, such spin splitting has opposite signs among these two sublattices, as shown in Fig.\ref{fig:projection}(e). The last term in $h_0(\mathbf{k})$ represents a momentum-dependent sublattice-staggered potential. Analogous to the Stark effect in spatial space, it induces band splitting in momentum space with opposite signs on the two sublattices, as shown in Fig.\ref{fig:projection}(f).  In the limit $t' \delta = 0$, $h_0(\mathbf{k})$ reduces to a trivial Brillouin zone folding without introducing new physics, and Eq.\ref{eq:s-d-model-2} exhibits Kramers degeneracy protected by $[C^{(s)}_{2\perp}||\pmb{\tau}]$. When $t' \delta \neq 0$, the collinear AFM order gives rise to NRSML as illustrated in Fig.\ref{fig:projection}(g)-(h), corresponding to the regimes where the exchange spin splitting $J_{sd}$S is much smaller and much larger than the Stark-like splitting $8t' \delta$, respectively. In the present discussion, we focus on the low-filling limit when $J_{sd}$S is the dominant energy scale (as shown in Fig.\ref{fig:projection}(h)), to remain consistent with the single-band treatment in Sect. I-B. Under such circumstances, we can safely retain $\hat{c}_{\mathbf{k} A \uparrow}$, $\hat{c}_{\mathbf{k} B \downarrow}$, and ignore $\hat{c}_{\mathbf{k} A \downarrow}$, $\hat{c}_{\mathbf{k} B \uparrow}$, as the latter have higher energies. In the new basis $\Psi_{\mathbf{k}} = \left(\hat{c}_{\mathbf{k}A \uparrow},  \hat{c}_{\mathbf{k}B \downarrow} \right)^T$, the effective Hamiltonian is:
\begin{equation}
\begin{split}
\varepsilon_{\sigma}(\mathbf{k}) &\approx  - 4t' \cos k_x \cos k_y  - \mu - \frac{J_{sd} \mathrm{S}}{2} \\
 & +4t' \delta \sin k_x \sin k_y  \sigma \\
\end{split}
\label{eq:s-d-model-3}
\end{equation}
You can find that anisotropic spin-independent hoppings in Eq.\ref{eq:s-d-model-2} now becomes NRSML in Eq.\ref{eq:s-d-model-3}. To contrast Eq.\ref{eq:s-d-model-3} to the known expressions of multipole basis discussed earlier, we use $\cos k \approx 1 - k^2/2$ and $\sin k \approx k$ to expand the spectrum Eq.\ref{eq:s-d-model-3} around $\Gamma$ point:
\begin{equation}
\varepsilon_{\sigma}(\mathbf{k}) \approx E_0 + 2t' \mathbf{k}^2 + 4t'\delta k_x k_y \sigma
\label{eq:s-d-model-4}
\end{equation}
with the constant $E_0 = -4t' - \mu -\frac{J_{sd}\mathrm{S}}{2}$. Eq.\ref{eq:s-d-model-4} clearly shows the symmetry-breaking conjugate field associated with the \textbf{MT} quadrupole $T^{ext}_{xy} = t' \delta$ (after transforming the coordinate system from PM unit cell to the magnetic supercell by $k_x \rightarrow \frac{k_x - k_y}{2}$, $k_y \rightarrow \frac{k_x + k_y}{2}$), in additional to the trivial conjugate field associated with the \textbf{E} monopole $Q^{ext}_0 = 2t' $.

From above analysis, it is not difficult to construct general minimal models for AM from a space group $\mathbf{G}$ containing a halving group $\mathbf{H}$ ($\mathcal{P} \in \mathbf{H}$ and/or $\pmb{\tau} \in \mathbf{H}$) with corresponding group division $\mathbf{G} = \mathbf{H} \cup (\mathbf{G} -  \mathbf{H}) = \mathbf{H} \cup \mathcal{A} \mathbf{H}$ (where $\mathcal{A} \in \mathbf{G}$ is chosen that $\mathcal{A} \mathbf{H} \equiv \mathbf{G} -  \mathbf{H} $), here $\mathcal{A} \mathbf{H}$ can contain both symmorphic and nonsymmorphic operations (see Fig.\ref{fig:projection}(a) for an example). By considering a 1D IR of the site symmetry group for sublattice degrees of freedom, Roig \textit{et al.} developed forty such models for 3D AMs in the form \citep{Roig2024}:
\begin{equation}
h(\mathbf{k}) = \varepsilon_0(\mathbf{k}) + t_x(\mathbf{k}) \tau_x + t_z(\mathbf{k}) \tau_z +  \pmb{J} \cdot \pmb{\sigma}  \tau_z + \mathbf{\lambda}(\mathbf{k}) \cdot \pmb{\sigma} \tau_y
\label{eq:Roig-2024-1}
\end{equation}
with a sublattice independent dispersion $ \varepsilon_0(\mathbf{k})$, inter- and intrasublattice hopping coefficients $t_x(\mathbf{k})$ and $t_z(\mathbf{k})$, a primary order parameter $\pmb{J}$, and a SOC term $\pmb{\lambda}(\mathbf{k})$. The form of $\pmb{\lambda}(\mathbf{k})$ can be determined by considering the transformation properties of the sublattice operator $\tau_y$ (which follows the same 1D IR at $\tau_z$) and of the spin operator $\pmb{\sigma}$, under the space group $\mathbf{G}$. 

Finally, let us briefly mention the spin-group analysis by \v{S}mejkal \textit{et al.} \citep{Smejkal2022-1}, which was already well-documented by several reviews \citep{Bai2024, Jungwirth2025-2}. The key observation by \v{S}mejkal \textit{et al.} \citep{Smejkal2022-1} is the existence of the following symmetry operations,
\begin{equation}
[E||\mathbf{H}] + [C^{(s)}_{2\perp}||\mathcal{A}\mathbf{H}]
\label{eq:Smejkal-2022-1}
\end{equation}
in the spin group the lattice structure. In the 2D square lattice shown in Fig.\ref{fig:projection}(c), we have $\mathbf{G} = C_{4v}$, $\mathbf{H} = C_{2v}$, and $\mathcal{A} = C^+_{4z}$. These symmetries are crucial in determining the characteristic anisotropy of the spin density distribution on each sublattice. For example, the symmetry operations $[C^{(s)}_{2\perp}||\mathcal{A}\mathbf{H}]$ guarantees the exchange between sublattices with opposite spin orientations, thereby enforcing a vanishing net magnetization - similar to conventional AFM. However, the coset $\mathcal{A}\mathbf{H}$ does not contain $\mathcal{P}$ (which is already one element of $\mathbf{H}$), indicating the broken of $\mathcal{T}$. For a generic wave-vector $\mathbf{k}$, its little group also lacks the element of $\mathcal{A}\mathbf{H}$, implying the spin-split band structure, reminiscent of FM. The symmetry operation in Eq.\ref{eq:Smejkal-2022-1} thereby define the magnetic order of AM, different from either AFM or FM. They also provide a useful alternative description of the collinear AFM ordering induced symmetry breaking, from a symmetry point of view based on the powerful tool of spin-group theory \citep{Naish1963, Kitz1965, Brinkman1966-1, Brinkman1966-2, Litvin1973, Litvin1974, Litvin1977}. 

In this context, it is worth highlighting several recent developments of spin-group theory as applied to AMs and the related phenomena. In 2022, Liu \textit{et al.} \citep{Liu2022} classified spin point groups for collinear, coplanar, and non-coplanar configurations. More recently, in 2024, three independent groups systematically enumerated the spin space groups and their representation \citep{Xiao2024, Chen2024, JiangY2024}. It is also noteworthy that the concept of spin groups has been widely employed in describing quantum phase transition in spin systems, particularly in the limit where SOC vanishes. For instance, in the Heisenberg model 
\begin{equation}
\hat{H} = \sum_{<i, j>} J_{i j} \hat{\mathbf{S}}_i \cdot \hat{\mathbf{S}}_j
\label{eq:Heisenberg}
\end{equation}
where $J_{ij}$ is the exchange coupling strength, the Hamiltonian remains invariant under a global rotation of all spins by an arbitrary angle.

\subsubsection*{4, Relationship to the previous subsections}
We now establish the connection between AMs, the ELC phases, and multipole basis discussed in previous subsections. For AM metals, the spin-momentum locked FSs resembles those of $F^A_{l >1}$ channel $\alpha$ phase (see Fig.\ref{fig:Fig-ELC}) or corresponding multipole basis (see Fig.\ref{fig:Fig-3}). Accordingly, the types of NRSML in AM metals can be classified by the angular momentum number $l$; for $l = 0, 1, 2, 3, ...$, these corresponds to $s$-, $p$-, $d$-, $g$-, and higher-wave SML \citep{Smejkal2022-1}. As concrete examples, the $d_{xy}$- and $d_{x^2-y^2}$-wave SML, which are frequently encountered in the next section, can be expressed as $\sin k_x \sin k_y \sigma_z$ and $(\cos k_x - \cos k_y)\sigma_z$, respectively. Near the $\Gamma$ point, these terms linearize to $k_x k_y \sigma_z$ and $k^2_x - k^2_y \sigma_z$, which directly correspond to $T_{x^2-y^2}$ and $T_{xy}$ multipoles. Therefore, $F^A_{2}$ channel $\alpha$-phase, $T_{xy}$ multipole, and $d_{x^2-y^2}$-wave AMs can manifest the same type of NRSML, as shown in the centre of Fig.\ref{fig:Fig-1}. However, it is important to note that these $F^A_{l}$ channel FSs - or more generally, atomic multiple basis - are constructed within a single-band framework, which does not generally apply to AMs. In AMs, the compensated magnetic moments requires at least two sublattices (the exact number determined by the underlying spin space group). Consequently, a multiband description is necessary to capture the FS topology. For the same reason, multipole basis description are insufficient; instead, SAMB is required to fully describe AMs, as seen in the works by Hayami \textit{et al.} \citep{Hayami2019, Hayami2020}.

The anisotropic kinetic energy plays an essential role in generating NRSML in AMs. However, this is not required for ELC phases, where SML is instead driven by strong correlations, either acting alone or in combination with SOC. Since the N\'{e}el order serves as the primary order parameter, NRSML in AMs can also be regarded as a dynamic phenomenon: it vanishes upon crossing either an AM-PM or an AM-N\'{e}el AFM quantum phase transition. Importantly, this dynamic behavior provides a clear experimental route for identifying AM, as the appearance of spin splitting coincides with the onset of AFM order \citep{Krempasky2024, Reimers2024, Zeng2024, Lu2025, Jiang2025, ZhangF2025}.  
We finally note that, the origin of spin-momentum locked FSs from PI remains poorly understood. Theoretically, determining Landau parameters in strongly correlated electronic systems is notoriously challenging, especially when a sublattice dof is involved. Investigating the potential for PI-induced AMs in future studies would be highly intriguing.

\subsubsection*{5, AM as a quantum AFM}
In the strong interaction regime, an AM metal undergoes metal-insulator transition to a Mott-insulating AM. With AFM exchange coupling $J_{ij} \approx 4t^2_{ij}/U$, the anisotropic NNN hoppings $t'(1 \pm \delta)$ now leads to anisotropic NNN exchange couplings $J'(1 \pm \delta)$. Therefore, insulating AM can be described by an anisotropic $J-J'-\delta$ spin-S Heisenberg model in general. For the 2D square lattice shown in Fig.\ref{fig:projection}(c), the Hamiltonian reads:
\begin{equation}
\hat{H}_{AM} = \sum_{ i, j } J_{ij} \hat{\mathbf{S}}_{i} \cdot \hat{\mathbf{S}}_{j} 
\label{eq:Heisenberg-AM}
\end{equation}
where the exchange coupling $J_{ij}$ follow the same spatial pattern as $t_{ij}$ in Eq.\ref{eq:s-d-model}. In particular, $J_{ij}$ equals to $J$ for NN, while is $J'(1+\delta)$ ($J'(1-\delta)$) and $J'(1-\delta)$ ($J'(1+\delta)$) along the (1, 1)-direction and (1, -1)-direction for sublattice $A$ ($B$), respectively. For concreteness, we consider positive $J$ (AFM coupling) and small positive (to avoiding frustration) or negative $J'$ (FM coupling) so that Eq.\ref{eq:Heisenberg-AM} stabilizes an AM ground state when the anisotropy $\delta$ is small.

Using a Holstein-Primakoff transformation \citep{Holstein1940}, we can now examine the spin-wave excitations of this anisotropic Heisenberg model. Let $\hat{a}_i$ and $\hat{b}_j$ denote bosonic operators on sublattice $A$ and $B$ respectively, such that
$\hat{\mathrm{S}}^z_{iA} = S - \hat{a}^{\dagger}_i \hat{a}_i$ and $\hat{\mathrm{S}}^z_{iB} = -S + \hat{b}^{\dagger}_i \hat{b}_i$. Expanding Eq.\ref{eq:Heisenberg-AM} in these bosons and retaining only bilinear terms (linear spin-wave approximation), one obtains a quadratic spin-wave Hamiltonian which can be diagonalized by a Bogoliubov transformation. The resulting magnon dispersion consists of two branches \citep{Cui2023}
\begin{equation}
\omega_{\pm}(\mathbf{k}) = \frac{1}{2} \sqrt{(A(\mathbf{k})+B(\mathbf{k}))^2 - (2C(\mathbf{k}))^2} \pm \frac{A(\mathbf{k}) - B(\mathbf{k})}{2}
\label{eq:Heisenberg-AM-LSW-spectrum}
\end{equation}
where
\begin{equation}
\begin{split}
A(\mathbf{k}) &= 4J' S (\cos k_x \cos k_y - \delta \sin k_x \sin k_y) + 4(J-J')S \\
B(\mathbf{k}) &= 4J' S (\cos k_x \cos k_y + \delta \sin k_x \sin k_y) + 4(J-J')S \\
C(\mathbf{k}) &= 2JS(\cos k_x + \cos k_y) \\
\end{split}
\end{equation}
The anisotropic exchange coupling $J' \delta \neq 0$ lifts the degeneracy between these magnon modes. In particular, according to Eq.\ref{eq:Heisenberg-AM-LSW-spectrum}, their frequency splitting reads $\omega_{+}(\mathbf{k}) - \omega_{-}(\mathbf{k})= -8J' \delta \rm{S} \sin k_x \sin k_y$, which have the same momentum dependence as the last term in Eq.\ref{eq:s-d-model-3} (with $t'$ replaced by $J'$S). In other words, the magnon band splitting and NRSML occur along the same directions in the Brillouin zone. Such magnon band splitting has been observed in atermagnetic MnTe \citep{LiuZY2024}, $\alpha$-Fe$_2$O$_3$ \citep{Hoyer2025, SunQY2025}, and CrSb \citep{Singh2025}. Moreover, the two magnon branches carry opposite chiralities \citep{Cui2023, Smejkal2023}, and such magnon band splitting and opposite chirality persist even in the generalized SU($N$) altermagnet \citep{Consoli2025}.

Finally, we consider the continuum description of the insulating AM. It is well known that the long-wavelength fluctuations of a quantum AFM can by captured by a nonlinear sigma model (NLSM) \citep{Haldane1983-PLA, Haldane1983-PRL, Chakravarty1989}. Since the insulating AM considered here is essentially a collinear AFM, Eq.\ref{eq:Heisenberg-AM} should admit a corresponding NLSM description, supplemented by a correction term proportional to $J' \delta$. Indeed, Lundemo \textit{et al.} \citep{Lundemo2025} derived a NLSM for Eq.\ref{eq:Heisenberg-AM}, with the following Euclidean Lagrangian density:
\begin{equation}
\mathcal{L} = \frac{(J - 2J') \rm{S}^2}{2} (\partial_{\mu} \mathbf{n})^2 - i \frac{J' \delta}{J} \rm{S} \epsilon^{abc} n^a \partial_{\tau} n^b \partial_x \partial_y n^c
\label{eq:Lundemo-2025-PRB-1}
\end{equation}
together with the unit-vector constraint $\mathbf{n}^2 = 1$. Here $\partial_{\mu} \equiv (\frac{1}{\rm{S}\sqrt{2J(J-2J')}} \partial_{\tau}, \mathbf{\nabla})$, $\tau$ is the imaginary time, and $\epsilon^{abc}$ is the Levi-Civita symbol. The first term in Eq.\ref{eq:Lundemo-2025-PRB-1} is the familiar O$(3)$ sigma-model stiffness term for a N\'{e}el AFM, proportional to $(J-2J')\rm{S}^2$ (the effective spin stiffness). The second term is a higher-order Berry phase term, its proportionality to $J'\delta$ is in agreement with prior expectations. The mixed spatial derivatives term again indicates that AM order involves simultaneous ordering of the N\'{e}el order and some high-order multipole moment of the magnetization \citep{Bhowal2024, McClarty2024}. We note that, although this term is superficially reminiscent of a Wess-Zumino-Witten term in 1+1 dimensions \citep{Witten1984, Affleck1989}, it is not topological here and does not alter the one-loop renormalization-group flow of the NLSM coupling \citep{Lundemo2025}.

\begin{figure*}
\begin{centering}
\includegraphics[width=0.95\textwidth]{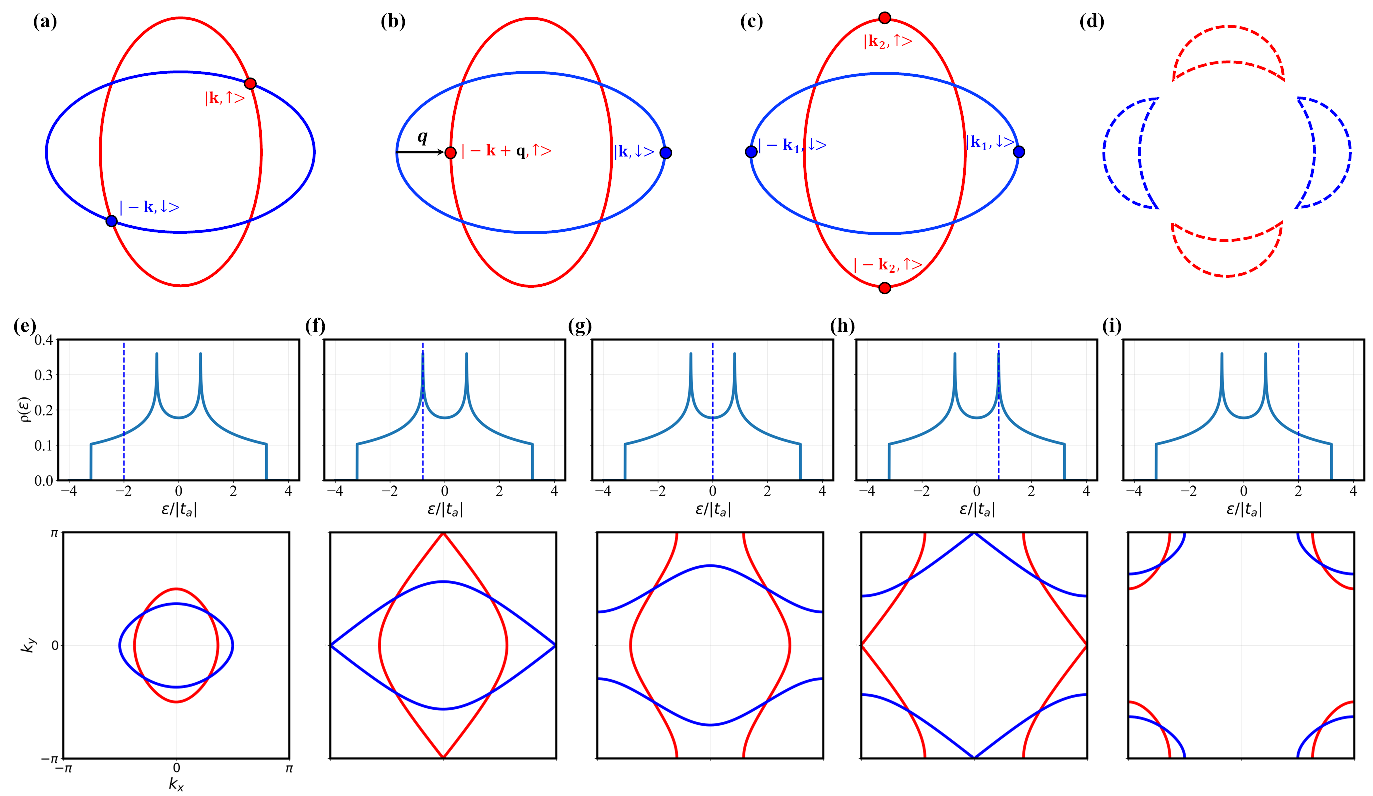}
\par\end{centering}
\caption{Demonstration of different pairing possibilities of $d_{x^2-y^2}$-wave NRSML in either continuum model (top panel) or 2D square lattice (bottom panel). (a) Spin-singlet $s$-wave paring. (b) Finite center-of-mass momentum pairing. (c) Spin-triplet $p$-wave paring. (d) Schematic representation of Bogoliubov Fermi surface, where the dashed lines  indicate regions with zero-energy Bogoliubov quasiparticle excitations. (e)-(g) Evolution of FSs (see Eq.\ref{eq:Feiguin2009-1} and Eq.\ref{eq:Feiguin2009-2}) with respect to chemical potential (black dashed line in the top panel). Color red and blue represents spin-up and -down channel, respectively}
\label{fig:s-wave-pairing} 
\end{figure*}

\section*{III. Superconductivity with nonrelativistic spin-momentum locked Fermi Surfaces}
As mentioned in previous section, the name ELC phase is inspired by the experimental observation of smectic phases in high-$T_c$ superconductors. Although the spin nematic phase has not been found at that time, theoretical predictions about the SC in spin nematic phase has been explored, as we show below.

Before entering detailed discussions, a simple inspection at the spin-momentum locked FSs shown in Fig.\ref{fig:s-wave-pairing}(a), allow us to anticipate three distinct types of Cooper pairings. The first is a zero-momentum spin-singlet pairing due to states $|\mathbf{k}, \uparrow>$ and $|-\mathbf{k}, \downarrow>$, as indicated in Fig.\ref{fig:s-wave-pairing}(a). In the absence of anisotropic spin splitting, this channel yields the conventional $s$-wave SC. Owing to its robustness against small perturbations, the $s$-wave character is expected to survive under weak anisotropic spin splitting. The second is a finite-momentum pairing, where the two mismatched FSs are connected by a finite momentum $\mathbf{q}$, as shown in Fig.\ref{fig:s-wave-pairing}(b). This state can be viewed as an unpolarized analogue of the Fulde-Ferrell-Larkin-Ovchinnikov (FFLO) state \citep{Fulde1964, Larkin1964}. For continuity, we still refer to it as a FFLO state, while noting that it may be also interpreted as a pair-density-wave (PDW) state within the context of ELC phases. The third is a zero-momentum spin-triplet pairing formed by states $|\mathbf{k}_1, \downarrow>$ and $|-\mathbf{k}_1, \downarrow>$ ($|\mathbf{k}_2, \uparrow>$ and $|-\mathbf{k}_2, \uparrow>$), as shown in Fig.\ref{fig:s-wave-pairing}(c). This spin-triplet pairing is unitary due to zero net magnetization \citep{Mazin2025}. Because both FFLO states and unitary spin-triplet pairings are inherently unconventional, their realization generally requires sufficiently strong anisotropic spin splitting to suppress conventional pairing tendency. Taken together, this simple inspection suggests that SC in AMs can closely resemble that found in either FM or collinear AFM systems.

\begin{figure}
\begin{centering}
\includegraphics[width=0.5\textwidth]{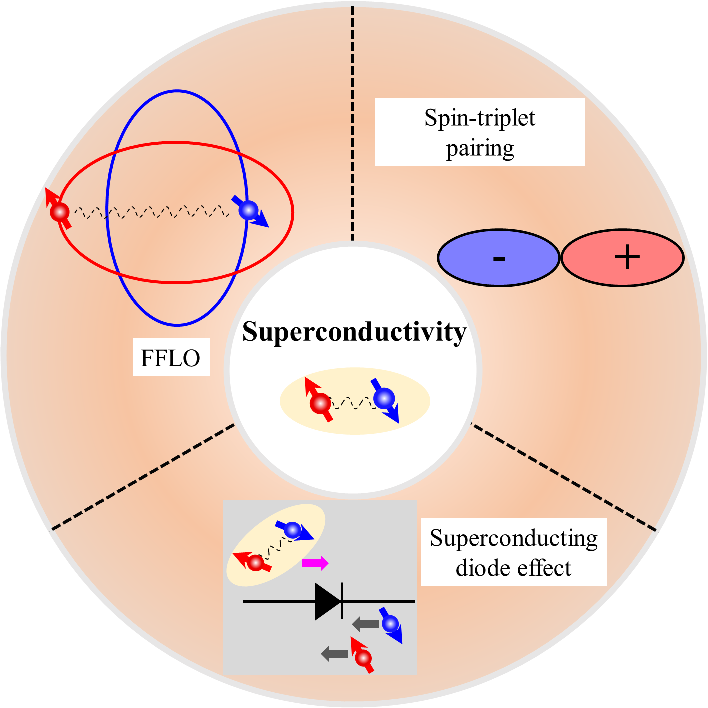}
\par\end{centering}
\caption{Emerging research landscape of superconductivity related to altermagnetism covered in this review.}
\label{fig:Fig-summary} 
\end{figure}

In the following, we focus on three key topics as summarized in Fig.\ref{fig:Fig-summary}: (1) FFLO states mediated by attractive interaction with $s$- and $d$-wave symmetry; (2) spin-triplet SC, which is closely associated with topological superconducting phases, and (3) superconducting diode effect (SDE). Our analysis in this section utilizes a \textit{single-band model}; while being an approximation in AMs, it successfully captures the essential features of NRSML. Moreover, it serves as a minimal framework for exploring the resulting exotic pairing phenomena that can be easily generalized to ultracold atom systems \citep{Wang2026}. In AMs, $s$-wave pairing may arise either from the intrinsic electron-phonon coupling or via the proximity effect through coupling with conventional SCs.

Obviously, the marriage between NRSML and SC is not limited to these three topics but can be also found in the classification of superconducting order parameters \citep{Feng2025, Maeda2025, Parshukov2025,  Heinsdorf2025},  Josephson effect \citep{Ouassou2023, ZhangS2024, Lu2024, Cheng2024-2, Fukaya2025, Zhao2025, Boruah2025, SunH2025, LiC2026, Zhao2026}, Andreev reflection \citep{Papaj2023, Sun2023, Das2024, Kazmin2025}, Majorana zero mode \citep{Li2023, Li2024, Ghorashi2024, Mondal2025, Hodge2025}, Floquet SC \citep{Pal2025, Fu2025, Fu2025-2} and many others \citep{Gill2024-1, Gill2024-2, Maiani2025, Monkman2025}. While we concentrate on the intrinsic regime where NRSML and SC coexist, hybrid systems combining conventional SC and AMs offer an alternative and experimentally feasible platform hosting NRSML, SOC, and SC simultaneously. A comprehensive discussion of such heterostructure can be found in the recent review by Fukaya \textit{et al.} \citep{Fukaya2025-2},  which surveys superconducting phenomena in unconventional magnetic systems, with particular emphasis on transport in superconducting junctions.

To describe the superconducting order parameter, we employ the standard Bogoliubov-de Gennes (BdG) mean-field approximation by default. This approach is reliable when interactions are not excessively strong. Accordingly, throughout our discussion of the three topics, we assume that the attractive interactions are of weak to medium strength relative to the Fermi energy. Other methods will be specified explicitly upon their introduction.

\subsection*{A. FFLO states with attractive pairing}

\subsubsection*{1, $s$-wave pairing potential}
In 2009, Feiguin \textit{et al.} \citep{Feiguin2009} considered the 
spin-dependent hopping in 2D square lattices subject to $s$-wave pairing instability, as described by the model Hamiltonian:
\begin{equation}
\hat{H}= \sum_{\mathbf{k}, \sigma} \varepsilon_{\sigma}(\mathbf{k}) \hat{c}^{\dagger}_{\mathbf{k} \sigma}  \hat{c}_{\mathbf{k} \sigma}  - U \sum_{i} \hat{n}_{i \uparrow} \hat{n}_{i \downarrow} 
, U > 0 \label{eq:Feiguin2009-3} 
\end{equation}
where
\begin{subequations}
\begin{align}
\varepsilon_{\uparrow}(\mathbf{k}) = -2t_a \cos k_x -2t_b \cos k_y - \mu 
, \label{eq:Feiguin2009-1} \\
\varepsilon_{\downarrow}(\mathbf{k}) = -2t_b \cos k_x  -2t_a \cos k_y - \mu 
, \label{eq:Feiguin2009-2} 
\end{align}
\end{subequations} 
where $t_a$ and $t_b$ are two NN hoppings determining the anisotropy of the two elliptical FSs.  Such a Hamiltonian exhibits $[C^{(s)}_{2\perp}||C^+_{4z}]$ symmetry with anisotropic spin splitting controlled by $t_b/t_a$ (without the loss of generality, $t_b/t_a \leq 1$ is taken). Since $[C^{(s)}_{2\perp}||\mathcal{P}]$ and $[C^{(s)}_{2\perp}||\pmb{\tau}]$ are absent, NRSML exists. At small filling factors, the two Fermi ellipses are centered at $\Gamma$ point as shown in Fig.\ref{fig:s-wave-pairing}(e). Expanding around the $\Gamma$ point, $\cos k_x \approx 1 - \frac{1}{2}k^2_x$, $\cos k_y \approx 1 - \frac{1}{2}k^2_y$, and defining 
\begin{equation}
Q^{ext}_0 \equiv  \frac{t_a + t_b}{2}, T^{ext}_{xy} \equiv \frac{t_a - t_b}{2}
\end{equation}
the single-particle dispersions Eq.\ref{eq:Feiguin2009-1} and Eq.\ref{eq:Feiguin2009-2} can be expressed uniformly as
\begin{equation}
\varepsilon_{\sigma}(\mathbf{k}) = Q^{ext}_0 \mathbf{k}^2 + \sigma T^{ext}_{xy}(k^2_x - k^2_y) -2(t_a+t_b) - \mu
\label{eq:Feiguin2009-dilute} 
\end{equation}
which has \textbf{E} monopole and \textbf{MT} quadrupole as defined in the previous section. Isotropic hopping occurs when $t_a = t_b$, thus $T^{ext}_{xy}$ measures the anisotropic spin splitting. 

The density of state (DOS) $\rho(\varepsilon)$ of Eq.\ref{eq:Feiguin2009-1} and Eq.\ref{eq:Feiguin2009-2} is given by:
\begin{equation}
\rho(\varepsilon) = \frac{2}{\pi^2 \sqrt{(4Q^{ext}_0)^2 - \varepsilon^2}} K(\sqrt{\frac{(4Q^{ext}_0)^2-(4T^{ext}_{xy})^2}{(4Q^{ext}_0)^2 - \varepsilon^2}})
\label{eq:Feiguin2009-DOS} 
\end{equation}
where $K(x) = \int^{\pi/2}_0 \frac{d\theta}{\sqrt{1-x^2 \sin^2(\theta)}}$ is the complete elliptic integral of the first kind. The band width is given by $[-4|Q^{ext}_0|, 4|Q^{ext}_0|]$ and there are two van Hove singularities at $\pm 4|T^{ext}_{xy}|$, see top panel of Fig.\ref{fig:s-wave-pairing}(e)-(i). The evolution of FSs with respect to $\mu$ is illustrated in Fig.\ref{fig:s-wave-pairing}(e)-(i). When $\mu$ is close to the bottom of the bands as shown in Fig.\ref{fig:s-wave-pairing}(e), both spin-up and -down FS are closed and Eq.\ref{eq:Feiguin2009-dilute} approximates well. With $\mu$ increasing, at $\mu = -4|T^{ext}_{xy}|$, both FSs undergo a Lifshitz transition from a closed loop to an open arc, and simultaneously a van Hove singularity emerges as shown in Fig.\ref{fig:s-wave-pairing}(f). Near this point, competing instabilities such as charge density wave (CDW) and spin density wave (SDW) will develop with repulsive interactions \citep{Parthenios2025, Rao2025}. At $\mu = 0$, the system is half-filled as demonstrated in  Fig.\ref{fig:s-wave-pairing}(g), and the perfect FS nesting holds as $\varepsilon_{\sigma}(\mathbf{k}+(\pi, \pi))  = - \varepsilon_{\sigma}(\mathbf{k})$. Therefore, an infinitesimal repulsive interaction will drive the system into an Ising AFM \citep{Xie2025}. Regardless of the sign of Hubbard interaction, it is found that the Hamiltonian is sign-problem-free for quantum Monte Carlo (QMC) method \citep{Xie2025}. As $\mu$ continues to increase, the open FSs evolve into closed ones centered at $(\pi, \pi)$ again, as shown in Fig.\ref{fig:s-wave-pairing}(h)-(i). Since Eq.\ref{eq:Feiguin2009-1} and Eq.\ref{eq:Feiguin2009-2} contain NN hoppings only, the sublattice particle-hole symmetry $\hat{c}_{i\downarrow} \rightarrow (-1)^{i_x + i_y} \hat{c}^{\dagger}_{i\downarrow}, \hat{c}^{\dagger}_{i\downarrow} \rightarrow (-1)^{i_x + i_y} \hat{c}_{i\downarrow}$ exists, which allows the mapping between positive and negative $U$.  

By fixing the interaction strength $U$ and applying the self-consistent BdG mean-field approximation, Feiguin \textit{et al.} mapped out the ground-state phase diagram as a function of the anisotropic ratio $t_b/t_a \in [0, 1]$ and electron filling (up to half filling) \citep{Feiguin2009}. For a dimensionless attractive interaction $U/t_a = 3.5$,  four distinct phases are identified: a conventional Bardeen-Cooper-Schrieffer (BCS) superfluid (SF) that dominates the phase diagram, a phase-separated region at intermediate $t_b/t_a$ and moderate filling, a nodal SF at small $t_b/t_a$ and moderate filling, and a normal metallic phase at small $t_b/t_a$ and high filling. The nodal SF arises from the vanishing of Bogoliubov quasiparticle excitation energy at specific momenta, as indicated in Fig.\ref{fig:s-wave-pairing}(d). These gapless excitations form what is referred to as Bogoliubov Fermi surface (BFS) \citep{Volovik1993}, due to their resemblance to conventional FSs. Crucially, the BCS SF remains remarkable robust against the $T_{xy}$ multipole, becoming unstable only when a sufficiently large $T^{ext}_{xy}$ is present. For instance, at a low filling of 0.2, the nodal SF emerges only after $T^{ext}_{xy}$ exceeds a threshold of approximately 0.45$t_a$. When the attractive interaction is increased further to $U/t_a = 4$, both phase-separated and normal phase vanish where BCS SF becomes the ground state. Such a fact matches the fact that strong attractive interaction favours conventional SC. Notably, the FFLO states, which is of particular interest, does not appear within the range of parameters explored in this study. What is more, the CDW ground state at half filling obtained by unbiased QMC simulation \citep{Xie2025} is also missed due to the ignorance of particle-hole channel in the mean-field approximation \citep{Feiguin2009}.

In 2014, Soto-Garrido \textit{et al.} \cite{Soto-Garrido2014} studied the possible FFLO states in a similar but continuum model to that of Feiguin \textit{et al.} \citep{Feiguin2009}. The single-particle Hamiltonian is:
\begin{equation}
\hat{H}_0 = \sum_{\mathbf{k} \alpha \beta} \hat{c}^{\dagger}_{\mathbf{k} \alpha}[\varepsilon({\mathbf{k}})(\sigma_0)_{\alpha \beta} + T^{ext}_{xy}(k^2_x - k^2_y)(\sigma_z)_{\alpha \beta}] \hat{c}_{\mathbf{k} \beta}
\label{eq:Soto-Garrido-2014-PRB-2.1}
\end{equation}
with the following interaction:
\begin{equation}
\hat{H}_{int} = - \frac{U}{N} \sum_{\mathbf{k}, \mathbf{k}',\mathbf{q}} \hat{c}^{\dagger}_{\mathbf{k}+\mathbf{q}/2 \uparrow}  \hat{c}^{\dagger}_{-\mathbf{k}+\mathbf{q}/2 \downarrow} \hat{c}_{-\mathbf{k}'+\mathbf{q}/2 \downarrow} \hat{c}_{\mathbf{k}'+\mathbf{q}/2 \uparrow}
\label{eq:Soto-Garrido-2014-PRB-2.5}
\end{equation}
Here $N$ is the number of unit cell and $T^{ext}_{xy} < 1$ is now a dimensionless parameter. The Fourier transformation of the pairing interaction in Eq.\ref{eq:Feiguin2009-3} takes the same form as Eq.\ref{eq:Soto-Garrido-2014-PRB-2.5}, therefore Eq.\ref{eq:Soto-Garrido-2014-PRB-2.1} and Eq.\ref{eq:Soto-Garrido-2014-PRB-2.5} can be understood as the dilute limit of the 2D square lattice model.

At finite temperature $T$, the opposite spin pairing susceptibility $\chi_{sc}(\pmb{Q}, i\nu_n)$  can be calculated by summing all the bubble diagrams in the particle-particle channel:
\begin{equation}
\begin{split}
\chi_{sc}(\pmb{Q}, i\nu_n) &= T \sum^{\infty}_{m = -\infty} \int \frac{d \mathbf{k}}{(2\pi)^2} G^0_{\uparrow}(\mathbf{k} + \frac{\pmb{Q}}{2}, i\omega_m + i\nu_n/2) \\
&\times G^0_{\downarrow}(-\mathbf{k} + \frac{\pmb{Q}}{2}, -i\omega_m + i\nu_n/2)  \\
\end{split}
\label{eq:Soto-Garrido-2014-PRB-3.1}
\end{equation} 
where $\pmb{Q}$ is the center-of-mass momentum of Cooper pairs, $\omega_m = (2m+1)\pi T$ and $\nu_n = 2n \pi T$ are fermionic and bosonic Matsubara frequencies, and 
\begin{equation}
G^0_{\alpha}(\mathbf{k}, i\omega_m) = \frac{1}{i\omega_m - \varepsilon_{\alpha}(\mathbf{k})}
\label{eq:Soto-Garrido-2014-PRB-3.2}
\end{equation}
is the free-fermion Green's function. After performing the Matsubara sum in Eq.\ref{eq:Soto-Garrido-2014-PRB-3.1} we yield

\begin{widetext}
\begin{equation}
\chi_{sc}(\pmb{Q}, i\nu_n) = \int \frac{d\mathbf{k}}{(2\pi)^2} \frac{1-n_F(\varepsilon_{\uparrow}(\mathbf{k}+\pmb{Q}/2))-n_F(\varepsilon_{\downarrow}(-\mathbf{k}+\pmb{Q}/2))}{\varepsilon_{\uparrow}(\mathbf{k}+\pmb{Q}/2)+\varepsilon_{\downarrow}(-\mathbf{k}+\pmb{Q}/2)-i\nu_n}
\label{eq:Soto-Garrido-2014-PRB-3.3}
\end{equation}
\end{widetext}

where $n_F(\varepsilon) = \frac{1}{1+e^{\varepsilon/T}}$ is the Fermi-Dirac distribution. At finite temperature, Eq.\ref{eq:Soto-Garrido-2014-PRB-3.3} in general has to be evaluated numerically. However, at zero temperature it is possible to obtain explicit analytic expressions for the pairing susceptibility at zero frequency, which gives us the well-known Thouless criterion for the onset of SC. Applying analytical continuation $i \nu_n = \nu + i 0^+$ and setting $\nu = 0$, we obtain the zero-temperature static expression:
\begin{widetext}
\begin{equation}
\chi_{sc}(\pmb{Q}) = \int \frac{d\mathbf{k}}{(2\pi)^2} \frac{1-\Theta(-\varepsilon_{\uparrow}(\mathbf{k}+\pmb{Q}/2))-\Theta(-\varepsilon_{\downarrow}(-\mathbf{k}+\pmb{Q}/2))}{\varepsilon_{\uparrow}(\mathbf{k}+\pmb{Q}/2)+\varepsilon_{\downarrow}(-\mathbf{k}+\pmb{Q}/2)}
\label{eq:Soto-Garrido-2014-PRB-3.5}
\end{equation}
\end{widetext}
where $\Theta(x)$ is the Heaviside step function. 

\begin{table*}[t]
	\caption{Comparison between different protocols in realizing FFLO phases in 2D.}
	\label{tab:four-FFLO}
	\begin{tabular}{c|c|c}
		\hline\hline
		FFLO type & Driven forces & Related multipole basis    \\
		\hline
		Conventional    & Zeeman field  & $M_{z}$       \\
		Rashba             & Rashba SOC + In-plane Zeeman field   & $Q_{z}$      \\
		Ising                  & Ising SOC + In-plane Zeeman field     & $Q_{3a/3b}$                       \\
		Altermagnetic   & $d$-wave SML  & $T_{xy}$       \\
		\hline\hline
	\end{tabular}
\end{table*}

To proceed, Soto-Garrido \textit{et al.} assumes that the integration over $\mathbf{k}$ can be approximated as:
\begin{equation}
\int \frac{d\mathbf{k}}{(2\pi)^2} \rightarrow N_F \int^{\omega_D}_{-\omega_D} d\zeta \int^{2\pi}_0 \frac{d\theta}{2\pi} 
\label{eq:Soto-Garrido-2014-PRB-3.8}
\end{equation}
where $\omega_D$ is an energy cutoff. Such an approximation implies that only a finite region near the FS contributes to pairing susceptibility, in line with the spirit of BCS theory. Using Eq.\ref{eq:Soto-Garrido-2014-PRB-3.8}, Eq.\ref{eq:Soto-Garrido-2014-PRB-3.5} can be integrated as:
\begin{equation}
\chi_{sc}(\pmb{Q}) = N_F \int^{2\pi}_0 \frac{d\theta}{2\pi} \mathrm{ln}\left| \frac{\omega_D}{T^{ext}_{xy} \cos(2\theta) - \frac{Q}{2}\cos(\theta - \phi)} \right|
\label{eq:Soto-Garrido-2014-PRB-3.7}
\end{equation}
where $Q$ (in units of the Fermi wave-vector $k_F$) and $\phi$ are the magnitude and the polar angle of $\pmb{Q}$. By choosing $\phi = n\pi/2$, i.e. $\pmb{Q}$ is along either $x$-axis or $y$-axis in Fig.\ref{fig:s-wave-pairing}(a), $\chi_{sc}(Q)$ is flat for $Q \leq 2T^{ext}_{xy}$ and decreases for $Q > 2T^{ext}_{xy}$. Therefore, there is no preference for a finite value of $Q$ in this situation, which exclude the possibility of a FFLO state. This result seems to be in accordance with that of Feiguin \textit{et al.} \citep{Feiguin2009} at low fillings. 

The recent discovery of $d$-wave AMs has revived interest in realizing FFLO states with nonrelativistic spin-momentum locked FSs. In 2024, Chakraborty \textit{et al.} \citep{Chakraborty2024-1} studied the 2D square lattice model with different $s$-wave pairing interaction strengths and different filling factors, and found the absence of FFLO state in the studied parameter range. At the same time, Zhang \textit{et al.} \citep{ZhangS2024} studied the 2D continuum model and obtained an analytical expression for the critical $\pmb{Q}$ ($\pmb{Q}_c$) at the large chemical potential limit. Using $T^{ext}_{xy}$ as the altermagnetic spin splitting,  the magnitude of $\pmb{Q}_c$ is expressed as:
\begin{equation}
Q_c(\phi) = \pm \left[ \sqrt{\frac{\mu}{1 + T(\phi)}} -   \sqrt{\frac{\mu}{1 - T(\phi)}}  \right]
\label{eq:ZhangS2024-NC}
\end{equation}
where $T(\phi) = T^{ext}_{xy} \cos(2\phi)$ and the chemical potential $\mu$ is measured in units of the Fermi energy, $\varepsilon_F$. 
Eq.\ref{eq:ZhangS2024-NC} gives an anisotropic $\phi$-dependent $\pmb{Q}_c$. With $\phi = \pi/4 + n \pi/2$, $Q_c = 0$, therefore no FFLO states are preferred. However, when $\pmb{Q}$  is along $x$-axis or $y$-axis, $\phi = n \pi/2$, $Q_c$ reaches the maximum value $ |\sqrt{\mu/(1 + T^{ext}_{xy})} -   \sqrt{\mu/(1 - T^{ext}_{xy})}|$, indicating the existence of a FFLO state. In this case, the value of the threshold $Q_c$ can be well-understood geometrically as the length of the arrow shown in Fig.\ref{fig:s-wave-pairing}(b), as pointed out recently by Liu \textit{et al.} \citep{Liu2025}. This finding by Zhang \textit{et al.} \citep{ZhangS2024} obviously disagrees with the results by Feiguin \textit{et al.} \citep{Feiguin2009}, Soto-Garrido \textit{et al.} \cite{Soto-Garrido2014}, and Chakraborty \textit{et al.} \citep{Chakraborty2024-1}. In 2025, Hong \textit{et al.} \citep{Hong2025} investigated the 2D square lattice model at a moderate interaction strength $U/t = 3$ and a fixed $\mu = -2t$, and demonstrated a phase transition from a BCS SF to the FFLO phase when $T^{ext}_{xy}$ exceeds a threshold of approximately 0.56t. Hu \textit{et al.} \citep{Hu2025-1} investigated the 2D continuum model numerically and confirmed the results of Hong \textit{et al.} \citep{Hong2025}. 

To clarify the above controversial results, most recently, Liu \textit{et al.} \citep{Liu2025} carried out an analytical study of the 2D continuum model. Without any further approximation, the ground state phase diagram has been obtained at both fixed chemical potential and fixed total particle number. In both cases, FFLO states do exist, in accordance with Zhang \textit{et al.} \citep{ZhangS2024}, Hong \textit{et al.} \citep{Hong2025}, and Hu \textit{et al.} \citep{Hu2025-1}. Liu \textit{et al.} \citep{Liu2025} clearly showed that the flatness of $\chi_{sc}(Q)$ in Eq.\ref{eq:Soto-Garrido-2014-PRB-3.7} is due to the approximation made in Eq.\ref{eq:Soto-Garrido-2014-PRB-3.8}. An appropriate way to remove the divergence in a continuum model is to renormalize the zero-range Hubbard $U$, then, $\chi_{sc}(Q)$ increases slowly as $Q$ increases below the threshold $Q_c$ given in Eq.\ref{eq:ZhangS2024-NC}.  Moreover, Liu \textit{et al.} \citep{Liu2025} can repeat the phase diagram of Hong \textit{et al.} \citep{Hong2025} and Hu \textit{et al.} \citep{Hu2025-1}. Finally, Liu \textit{et al.} \citep{Liu2025} demonstrated that a nodeless SF with topological BFS emerges when the chemical potential is fixed. 

At present, we could give an affirmative conclusion that the FFLO phase can be induced by $d$-wave altermagnetic spin splitting in two-dimensional spin-1/2 Fermi systems with attractive $s$-wave pairing interactions. 

In the above discussion, we did not distinguish FF and LO states. Sumita \textit{et al.}. \citep{Sumita2025} considered this issue by investigating three different microscopic models: (i) a two-sublattice model with AM order (see Eq.\ref{eq:s-d-model-2}), (ii) a continuum model (see Eq.\ref{eq:Soto-Garrido-2014-PRB-2.1}), and (iii) a conventional square lattice model with spin-dependent hoppings (see Eq.\ref{eq:Feiguin2009-1} and Eq.\ref{eq:Feiguin2009-2}). It is found that momentum doubling is important for stabilizing FF over LO state, which can be achieved by introducing sublattice dof (model (i)) or longer-range hoppings (model (iii)). This conclusion is pivotal for SDE to be discussed below. 

At this stage, it is instructive to compare FFLO phases realized here with other established protocols summarized in Tab.\ref{tab:four-FFLO}.  Conventional FFLO states are driven by Zeeman field, or equivalently by $M_z$ in the language of multipole basis. Such a Zeeman field may arise either from the Zeeman effect of an external magnetic field, or from an intrinsic exchange field in ferromagnetic metals \citep{Fulde1964, Larkin1964}. In the presence of an external magnetic field, however, orbital effects typically suppress SC before an FFLO phase emerges. Moreover, the FFLO phase breaks the continuous U($1$) rotational symmetry of the FSs. According to Nambu-Goldstone theorem \citep{Nambu1960, Goldstone1961, Goldstone1962}, this symmetry breaking leads to gapless modes associated with the fluctuations among equivalent $\mathbf{Q}$. In ferromagnetic metals, FFLO phases must further compete with spin-triplet SC, which is generally energetically favored \citep{Gorkov2001, Frigeri2004}. Owing to these limitations, although conventional FFLO phases has been proposed for over six decades, conclusive evidence remains elusive. In 2D noncentrosymmetric SCs, there exists Rashba (the related multipole basis is $Q_z$) or Ising SOC depending on whether the in-plane mirror ($\mathcal{M}_z$) symmetry is broken or preserved. In both cases, FFLO phases can be stabilized under in-plane magnetic fields, leading to Rashba-type \citep{Zheng2014} and Ising-type FFLO states \citep{LiuCX2017, Zhao2023, Wan2023}. In the Rashba case, the FFLO modulation originates from Zeeman effect-induced shifts of the FSs, whereas in the Ising case, the canonical momentum $\mathbf{k}$ is shifted through the orbital effect, described by the minimal coupling  $\mathbf{k} \rightarrow \mathbf{k} - e \pmb{A}$, with $\pmb{A}$ the vector potential. Since Zeeman effect-induced depairing can be largely avoided in these settings, the corresponding critical magnetic fields are largely enhanced. In contrast to all the above protocols, AM provides a field-free and SOC-free route to realize FFLO phases.

\subsubsection*{2, $d$-wave pairing potential}
\begin{figure*}
\begin{centering}
\includegraphics[width=0.90\textwidth]{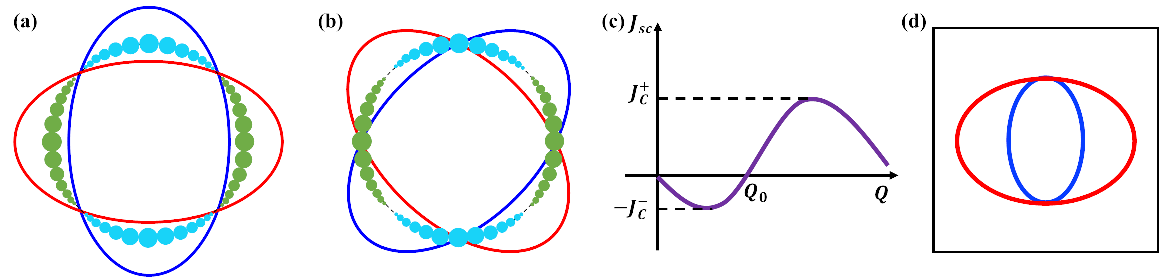}
\par\end{centering}
\caption{ Two possible orientation of $d$-wave SC and $d$-wave SML: (a) both SC and SML are $d_{x^2-y^2}$-wave; (b) SC is $d_{x^2-y^2}$-wave while SML is $d_{xy}$-wave. The colored lines and solid circles represent SML and SC components, respectively. (c) Schematic plot of $\pmb{J}_{sc}(\pmb{Q})$ for FF states in altermagnetic superconductors. (d) Schematic illustration of out-plane Zeeman field driven topological nodal-nodeless phase transition in altermagnetic SML. }
\label{fig:orientation} 
\end{figure*}

Having first considered the conventional $s$-wave pairing potential, we now turn our attention to the $d$-wave pairing potential. In 2D, both the $d$-wave SML and $d$-wave pairing potential can exhibit either $d_{x^2-y^2}$ or $d_{xy}$ symmetry. This lead to two possible configurations, depending on the relative orientation between NRSML and pairing potential. In the following, we fix the pairing potential to have $d_{x^2-y^2}$ symmetry. The two corresponding cases - with $d_{x^2-y^2}$-wave or $d_{xy}$-wave SML - are schematically illustrated in Fig.\ref{fig:orientation}. In Fig.\ref{fig:orientation}(a), the $d_{x^2-y^2}$-wave SML leads to pairing nodes that coincide with the intersection point between the two elliptic FSs. In contrast, Fig.\ref{fig:orientation}(b) shows that for the $d_{xy}$-wave SML, the strongest pairing occurs at the four interaction points of the two elliptic FSs. 

The $d$-wave pairing potential can arise from the nearest attractive interaction \citep{Sigrist1991}. Let us consider the extended attractive Hubbard model on 2D square lattice:
\begin{equation}
\hat{H}_{int} = -U \sum_{i} \hat{n}_{i \uparrow} \hat{n}_{i \downarrow} - \frac{V}{2} \sum_{\alpha \beta} \sum_{<i,j>} \hat{n}_{i \alpha} \hat{n}_{j\beta} 
\label{eq:extend-Hubbard}
\end{equation} 
By setting the distance of nearest neighbors to be unity, after Fourier transformation, we have
\begin{widetext}
\begin{equation}
\hat{H}_{int} =  -\frac{1}{2N}\sum_{\mathbf{k}, \mathbf{k}', \mathbf{q}} \sum_{\alpha \beta} V_{\alpha \beta}(\mathbf{k} - \mathbf{k}') \hat{c}^{\dagger}_{\mathbf{k}+\frac{\mathbf{q}}{2} \alpha}   \hat{c}^{\dagger}_{-\mathbf{k}+\frac{\mathbf{q}}{2} \beta} \hat{c}_{-\mathbf{k}'+\frac{\mathbf{q}}{2} \beta} \hat{c}_{\mathbf{k}'+\frac{\mathbf{q}}{2} \alpha}
\label{eq:general-interaction} 
\end{equation} 
\end{widetext}
with the two-body interaction $V_{\alpha \beta}(\mathbf{k} - \mathbf{k}')$ given by
\begin{widetext}
\begin{equation}
V_{\alpha\beta}(\mathbf{k} - \mathbf{k}') =U(1-\delta_{\alpha \beta}) + 2V[\cos(k_x -k'_x) + \cos(k_y - k'_y)] 
\end{equation}
\end{widetext}

Using the spherical harmonics decomposition on square lattices:
\begin{equation}
\begin{split}
g_s(\mathbf{k}) &= 1 \\
g_{es}(\mathbf{k}) &= \cos k_x + \cos k_y \\ 
g_{p \pm ip}(\mathbf{k}) &= \sin k_x \pm i\sin k_y \\
g_{d}(\mathbf{k}) &= \cos k_x - \cos k_y \\
\end{split}
\label{eq:form-factor}
\end{equation}
here "es" means extended $s$-wave, and noticing that 
\begin{equation}
2[\cos(k_x -k'_x) + \cos(k_y - k'_y)] = \sum_{m} g_{m}(\mathbf{k})g^*_{m}(\mathbf{k}') 
\end{equation}
where $m$ labels different square harmonics ($s$, $es$, $p \pm ip$, $d$), 
the extended attractive interaction Hamiltonian can then be decomposed into different pairing channels:
\begin{equation}
\begin{split}
\hat{H}_{int} &= - \frac{1}{2N}  \sum_{\mathbf{k}, \mathbf{k}', \mathbf{q}, m, \alpha, \beta} \lambda^{m}_{\alpha \beta} g_m(\mathbf{k}) g^*_m(\mathbf{k}') \\
& \times \hat{c}^{\dagger}_{\mathbf{k}+\frac{\mathbf{q}}{2} \alpha}   \hat{c}^{\dagger}_{-\mathbf{k}+\frac{\mathbf{q}}{2} \beta} \hat{c}_{-\mathbf{k}'+\frac{\mathbf{q}}{2} \beta} \hat{c}_{\mathbf{k}'+\frac{\mathbf{q}}{2} \alpha}  \\
\end{split}
\label{eq:square-harmonics-decom}
\end{equation}
with
\begin{equation}
\lambda^{s}_{\alpha \beta} = U(1-\delta_{\alpha \beta}), \lambda^{es/p \pm ip/d}_{\alpha \beta} = V
\label{eq:square-harmonics-decom-1}
\end{equation}
When $V = 0$, we have only $s$-wave pairing, which has been discussed in previous section. When $V \neq 0$, other partial-wave pairings become possible. In particular, the $d_{x^2-y^2}$-wave pairing potential mentioned at the beginning of this subsection, i.e., $g_{d}(\mathbf{k})$, appears in this square harmonics decomposition. Therefore, to isolate and highlight such unconventional superconductivity, we can set a small or even vanishing value for $U$. 

We can tailor Eq.\ref{eq:square-harmonics-decom-1} to select a particular paring. For example, if we choose 
\begin{equation}
\begin{split}
 \lambda^{es/p \pm ip/d}_{\uparrow \downarrow} &= \lambda^{es/p \pm ip/d}_{\downarrow \uparrow} = V  \\
 \lambda^{es/p \pm ip/d}_{\uparrow \uparrow} &= \lambda^{es/p \pm ip/d}_{\downarrow \downarrow} = 0 \\
\end{split}
\label{eq:square-harmonics-decom-singlet}
\end{equation}
then opposite-spin pairing potential are chosen which couple opposite spins at nearest neighbors. For extended $s$-wave and $d_{x^2-y^2}$-wave, this gives spin-singlet pairing, while for $p \pm ip$-wave, we have spin-triplet pairing.  Alternatively, we can choose 
\begin{equation}
\begin{split}
 \lambda^{es/p \pm ip/d}_{\uparrow \downarrow} &= \lambda^{es/p \pm ip/d}_{\downarrow \uparrow} = 0  \\
 \lambda^{es/p \pm ip/d}_{\uparrow \uparrow} &= \lambda^{es/p \pm ip/d}_{\downarrow \downarrow} = V \\
\end{split}
\label{eq:square-harmonics-decom-triplet}
\end{equation}
then only the $p \pm ip$-wave pairing are allowed due to Pauli exclusion principle. In the case that all the pairing potentials are present simultaneously (see Eq.\ref{eq:square-harmonics-decom-1}), all channels compete with one another, and the resulting superconducting state is determined by the channel that becomes unstable first as the temperature decreases.

In 2014, Soto-Garrido \textit{et al.} \cite{Soto-Garrido2014} also studied the $d_{x^2-y^2}$-wave pairing potential with $T_{xy}$ multipole in the continuum model, by setting $ \lambda^{d}_{\uparrow \downarrow} = \lambda^{d}_{\downarrow \uparrow} = V$ and suppressing all the other channels in Eq.\ref{eq:square-harmonics-decom}. As discussed in the above, such a choice corresponds to spin-singlet $d_{x^2-y^2}$-wave pairing. In this case, the pairing susceptibility can be expressed as:
\begin{widetext}
\begin{equation}
\chi_{sc}(\pmb{Q}) = N_F \int^{2\pi}_0 \frac{d\theta}{2\pi} \cos^2(2\theta) \mathrm{ln}\left| \frac{\omega_D}{T^{ext}_{xy} \cos(2\theta) - \frac{Q}{2}\cos(\theta - \phi)} \right|
\label{eq:Soto-Garrido-2014-PRB-3.7-d}
\end{equation}
\end{widetext}
By choosing $\phi = n\pi/2$, $\chi_{sc}(Q)$ increases for $Q \leq 2T^{ext}_{xy}$ and then decreases for $Q > 2T^{ext}_{xy}$. Therefore, one obtains a threshold $Q_c = 2T^{ext}_{xy}$ and the inhomogeneous FFLO state with centre-of-mass momentum $Q_c$ becomes the ground state.

The above result was recently explicitly examined by Chakraborty \textit{et al.} \citep{Chakraborty2024-1}, by numerically investigating a 2D square lattice model and by setting $\lambda^{es/d}_{\uparrow \downarrow} = \lambda^{es/d}_{\downarrow \uparrow} = V$ only in Eq.\ref{eq:square-harmonics-decom}. They found that when $V/Q^{ext}_0 \approx 2$ and the filling is around 0.6, the FFLO state emerges as the ground state within a narrow range of $T^{ext}_{xy}$, provided that  the modulation vector $\pmb{Q}$ is aligned along either $x$-axis or $y$-axis.

More recently, the ignoring of $\lambda^{p\pm ip}_{\uparrow \downarrow}$ and $\lambda^{p\pm ip}_{\downarrow \uparrow}$ channels in studies of FFLO states by Soto-Garrido \textit{et al.} \cite{Soto-Garrido2014} and Chakraborty \textit{et al.} \citep{Chakraborty2024-1} has been thrown into doubt. Hu \textit{et al.} \citep{Hu2026} studied two-body bound states in 2D square lattice using Eq.\ref{eq:square-harmonics-decom-singlet}. It is found that introducing $d_{x^2-y^2}$-wave SML generally drives the lowest-energy bound pair to acquire a finite $\pmb{Q}$, a sign of FFLO states in many-body systems. Once such finite-$\pmb{Q}$ bound states develop, all five pairing channels ($s$, $es$, $p \pm ip$, $d$) generally become mixed, even in the absence of SOC. Moreover, when $\pmb{Q}$ preserves mirror symmetry, an exotic singlet-triplet mixing arises, a characteristic feature of noncentrosymmetric SCs \citep{Bauer2012, Yip2014}. Complementary many-body results were obtained by Jasiewicz \textit{et al.} \citep{Jasiewicz2025} and Hu \textit{et al.} \citep{Hu2026-2}, who investigated the same lattice model at different fillings. In the absence of $T^{ext}_{xy}$,  $es$-wave ($d$-wave) superconducting state is the ground state at low (around half) filling. This is understandable from Eq.\ref{eq:form-factor}, at the low filling, $\mathbf{k} \approx 0$, the maximal weight comes from $g_{es}$; on the other hand, around half filling, the pairing mainly comes from states near $(\pi, 0)$ and $(0, \pi)$, then the maximal weight comes from $g_{d}$. FFLO states emerge with strong $T^{ext}_{xy}$ in both cases. Notably, at high filling, the resulting FFLO states  correspond to a mixture of spin-singlet and spin-triplet pairing. Taken together, these few-body and many-body studies underscore the importance of retaining all pairing channels in Eq.\ref{eq:square-harmonics-decom-singlet} when exploring FFLO states.

\subsection*{B. Spin-triplet superconductivity}
Let us now turn to the spin-triplet SC with $d$-wave SML. In 2014,  Gukelberger \textit{et al.} \citep{Gukelberger2014} revisited the lattice model originally proposed by Feiguin \textit{et al.} \citep{Feiguin2009}. Using diagrammatic QMC simulations to perform an unbiased sampling of the Feynman diagrammatic series, they uncovered a rich phase diagram. At the filling factor of 1.2 (equivalent to 0.8 due to particle-hole symmetry) and at low temperatures, four distinct phases were identified: a BCS SF, two different $p$-wave triplet SFs, and an incommensurate density wave. Notably, with on-site attractive interaction only, there is no direct interaction between identical particles. Therefore, at the first order, these two $p$-wave spin-triplet pairings are forbidden. Nevertheless, they can emerge as a second-order effect. These phases are stabilized only at sufficiently low temperatures, where competing instabilities are suppressed.  

In 2023, Zhu \textit{et al.} \cite{Zhu2023} considered the possibility of $p$-wave SC in 2D altermagnetic metals. It is well known that Rashba SOC can turn a $s$-wave SC into $p$-wave. Therefore, the following single-particle Hamiltonian with Rashba SOC is considered,
\begin{equation}
\begin{split}
\hat{H}_0 &= \sum_{\mathbf{k} \alpha \beta} h_{\alpha \beta}(\mathbf{k}) \hat{c}^{\dagger}_{\mathbf{k} \alpha}  \hat{c}_{\mathbf{k} \beta} \\ 
h(\mathbf{k}) &= -2Q^{ext}_0 g_{es}(\mathbf{k})\sigma_0 - 2T^{ext}_{xy}g_{d}(\mathbf{k})\sigma_z \\
&+ 2Q^{ext}_z(\sin k_y \sigma_x - \sin k_x \sigma_y) \\
\end{split} 
\end{equation}
where $Q^{ext}_z$ being the strength of Rashba SOC. For the interaction part, Zhu \textit{et al.} \cite{Zhu2023} adopted the following form: $\lambda^s_{\alpha \beta} = U(1-\delta_{\alpha \beta})$, $\lambda^{s/es/p \pm ip/d}_{\uparrow \uparrow} = \lambda^{s/es/p \pm ip/d}_{\downarrow \downarrow} = V$. In the absence of Rashba SOC, and under conditions of weak SML and low filling, increasing the interaction $V$ drives a transition from $s$-wave to $p$-wave SC. When Rashba SOC is included, $\mathcal{P}$ is broken, leading to a mixing of $s$-wave and $p$-wave components in the superconducting order parameter. 

In 2025, Hong \textit{et al.} \citep{Hong2025} extended the analysis by considering the full set of interaction potentials in Eq.\ref{eq:extend-Hubbard}, rather than limiting the model to spin-singlet or spin-triplet pairing alone. By fixing an intermediate $U$ and low filling, they found that $p + ip$-wave SC typically occurs when the nearest-neighbor interaction $V$ exceeds a critical threshold $V_c$. This threshold $V_c$ initially decreases rapidly with increasing $T^{ext}_{xy}$, and then saturates at large values of $T^{ext}_{xy}$.  Such spin-triplet pairing can support persistent currents with any spin polarization, including pure spin supercurrents realized in the charge counterflow regime, as mentioned by Monkman \textit{et al.} \citep{Monkman2025}.

\subsection*{C. Superconducting diode effect}
The superconducting diode effect (SDE) is one typical nonreciprocal transport phenomena in superconducting system \citep{Nagaosa2024}, where a dissipationless superconducting current flows in one direction while a normal current flow in the reverse direction with finite resistance. This unique property posit SDE a key element in superconducting quantum technology, like its traditional diode for dissipative semiconducting technologies. Nadeem \textit{et al.} \citep{Nadeem2023} recently summarized SDE across multiple dimensions, including the geometric structure of the diode device, underlying symmetries, the nature of SOC, the orientation of magnetic field or magnetization, direction of current flow, and the topological character of the constituent materials. 
Here we limit to the SDE originated from FF state, to be compatible with Tab.\ref{tab:four-FFLO}. FF state breaks both $\mathcal{T}$ and $\mathcal{P}$ and provides an oriented direction given by the centre-of-mass momentum $\pmb{Q}$. The critical current for superconducting along $\pmb{Q}$ differs from that in the opposite direction, leading to the diode effect. A solid experimental demonstration of SDE was reported by Ando \textit{et al.} \citep{Ando2020} in 2020 in artificially fabricated [Nb/V/Ta]$_n$ superlattice. The FF state in such system belongs to Rashba-type according to Tab.\ref{tab:four-FFLO} \citep{Daido2022, Yuan2022, He2022}. After that, more and more SDE are reported in Rashba-type systems \citep{Nadeem2023, Ma2025}. In 2022, the SDE based on Ising-type FF state was reported by Bauriedl \textit{et al.} \citep{Bauriedl2022} in few-layer NbSe$_2$. 

AM-type FF state offers an field-free approach towards SDE, which can be realized in both junction-free systems \citep{Chakraborty2025, Sim2025, Yang2025} and Josephson junctions \citep{Banerjee2024, Cheng2024}. Here we focus on the junction-free case. 

Close to the superconducting phase transition, the free energy density $f(\mathbf{r})$ as a functional of the superconducting order parameter $\Delta(\mathbf{r})$ reads \citep{Sim2025, Yuan2022}:
\begin{equation}
f(\mathbf{r}) = a_0 |\Delta(\mathbf{r})|^2 + a_2|\pmb{D} \Delta(\mathbf{r})|^2 + a_4 |\pmb{D}^2 \Delta(\mathbf{r})|^2 + \beta_1 |\Delta(\mathbf{r})|^4
\label{eq:Sim-2025-5}
\end{equation}
where $\pmb{D} = -i \pmb{\nabla} - 2e \pmb{A}$ is the covariant derivative. In Eq.\ref{eq:Sim-2025-5}, $a_0 = \frac{T-T_c}{T_c}$ is the reduced temperature.  

When restricted to FF order parameter $\Delta(\mathbf{r}) = \Delta e^{i \pmb{Q} \cdot \mathbf{r}}$, the corresponding free energy density takes the following forms:
\begin{equation}
f(\pmb{Q}-2e\pmb{A}, \Delta)  = \alpha(\pmb{Q}-2e\pmb{A}) \Delta^2 + \beta_1 \Delta^4
\label{eq:Yuan-2022-8-1}
\end{equation}
with $\alpha(\mathbf{q}) = a_0 + a_2 \mathbf{q}^2 + a_4 \mathbf{q}^4$. At the equilibrium, $\pmb{A} = 0$ and the Cooper pair momentum $\pmb{Q}_0$ is determined by minimizing $\alpha(\pmb{Q})$ over $\pmb{Q}$:
\begin{equation}
\frac{\partial \alpha}{\partial \pmb{Q}}|_{\pmb{Q}_0}  = 0 \mbox{ and }  \mathrm{Det}[\frac{\partial^2 \alpha}{\partial Q_i \partial Q_j }]|_{\pmb{Q}_0} >0
\label{eq:Yuan-2022-11}
\end{equation}

The introduction of $\pmb{A}$ allows us to calculate the supercurrent density $\pmb{J}_{sc}$ as follows
\begin{equation}
\pmb{J}_{sc} \equiv - \frac{\partial f}{\partial \pmb{A}}|_{\pmb{A} = 0} = 2e \frac{\partial }{\partial \pmb{Q}} f(\pmb{Q}, \Delta)
\label{eq:Yuan-2022-13}
\end{equation}
The minimization condition Eq.\ref{eq:Yuan-2022-11} constraint that the equilibrium state carries zero current, in accordance with the Bloch's theorem \citep{Bohm1949}. By connecting the system to an external source, one can pass a nonzero supercurrent $\pmb{J}_{sc}$ through the system. Such a current-carrying state has a Cooper pair momentum $\pmb{Q} \neq \pmb{Q}_0$ that is determined by $\pmb{J}_{sc}$ according to Eq.\ref{eq:Yuan-2022-13}. Minimizing the free energy $f(\pmb{Q}, \Delta)$ with respect to the gap magnitude $\Delta$ yields 
\begin{equation}
|\Delta|^2 = -\frac{\alpha(\pmb{Q})}{2\beta_1}
\end{equation}
when $\alpha(\pmb{Q}) < 0$. Then Eq.\ref{eq:Yuan-2022-13} becomes
\begin{equation}
\pmb{J}_{sc} = \frac{e}{\beta_1} |\alpha(\pmb{Q})| \frac{\partial \alpha(\pmb{Q})}{\partial \pmb{Q}}
\label{eq:Yuan-2022-14}
\end{equation}
Sim \textit{et al.} \citep{Sim2025} numerically calculated the coefficients based on 2D continuum model with $s$-wave spin-singlet pairing and obtained the $\pmb{J}_{sc}$ curve sketched in Fig.\ref{fig:orientation}(c), from which we can read the two critical currents $J^{\pm}_c$ along and against $\pmb{Q}_0$. Then the supercurrent diode coefficient $\eta \equiv \frac{J^{+}_c - |J^{-}_c|}{J^{+}_c + |J^{-}_c|}$ can be further calculated. 

Chakraborty \textit{et al.}\citep{Chakraborty2025} reported a nearly perfect SDE ($\eta$ can achieve 100\%) based on the fact FF state exists in $d$-wave spin-singlet SC under both Zeeman field and $d$-wave SML \citep{Chakraborty2024-1}. Such a large $\eta$ is ascribed to topological nodal-nodeless transition in this system \citep{Chakraborty2024-1}.  When the filling is low, the two FSs are closed as shown in Fig.\ref{fig:orientation}(a). The Zeeman field will expand the FS of spin up channel while reduce the FS of spin-down channel. Therefore, the system can undergo a nodal-nodeless transition, as shown in Fig.\ref{fig:orientation}(d). Near this topological transition, there is a close competition between BCS and FF state, which leads to a pretty small $-J^-_c$ when supurcurrent is along FF-BCS direction. Under such a circumstances, no matter how large  $J^+_c$ is, $\eta$ will be a large value and can even reach 100\%.

\section*{IV. Superconductivity in itinerant AM}
As emphasized in the previous sections, the single-band Hamiltonian employed to capture the NRSML in AMs is only an effective approximation. A more realistic description must incorporate the dual effects of strongly electron correlations associated with transition-metal cations and the anisotropic crystal fields generated by oriented ligands. Upon integrating out the ligand dof, effective models emerge in which the transition-metal ions reside on at least two sublattices characterized by anisotropic, spin-independent hoppings (see Eq.\ref{eq:s-d-model}). This intrinsic sublattice structure constitutes one defining feature of AMs. 

\begin{figure*}[t]
\centering
\includegraphics[width=0.95\textwidth]{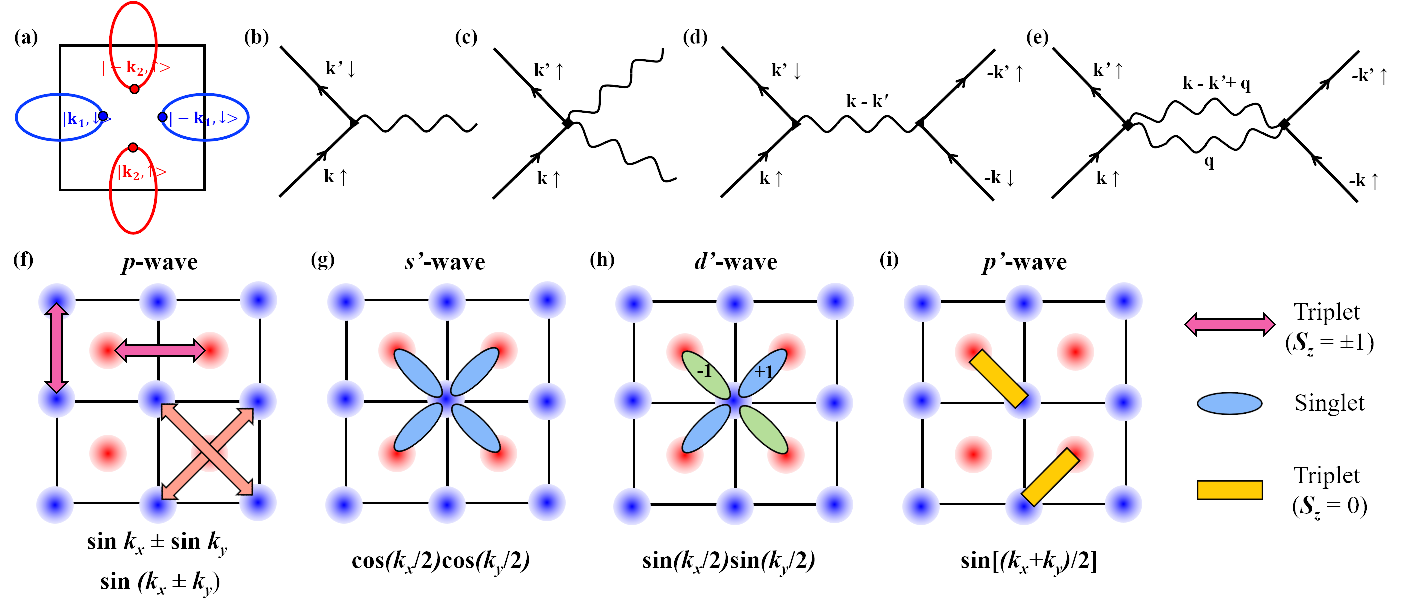}
\caption{(a) In a lattice model with sublattices, the FSs can exhibit $d_{x^2-y^2}$-wave SML while crossing the Brillouin zone boundary. (b)-(c) Feynman diagram for single-magnon process and double-magnon process of EMC. Straight/wavy line represents electron/magon. (d)-(e) Effective quasiparticle-quasiparticle interaction induced by EMC. Here spin up is chosen as an example. Note the free momentum $\mathbf{q}$ in the double-magnon-mediated quasiparticle-quasiparticle interaction.  (f-i) Real space configuration of four distinct pairing channels. The primes denote intra-unit-cell pairings whose Cooper pairs are formed by electrons from different sublattice. The bottom marks the corresponding form factor. (f) $p$-wave, where two parallel electrons come from the same sublattice. (g) $s'$-wave, where two antiparallel electrons come from different sublattices. (h) $d'$-wave, where two antiparallel electrons come from different sublattices. The "+1/-1" represents opposite pairing phase. (i) $p'$-wave, where two antiparallel electrons come from the same sublattice.}
\label{fig:EMC} 
\end{figure*}

The introduction of a sublattice dof enriches the structure of unconventional SC in AMs, as suggested by a growing body of theoretical works \citep{Brekke2023, Sumita2023, Bose2024, Maeland2024, Leraand2024, Chakraborty2025-2, Wu2025, Sumita2025,  Rasmussen2025, MaX2025, LuC2025, Zou2026, Xiao2026}. This can be understood from three complementary perspectives. First, the sublattice dof can substantially modify the topology of FSs. A prominent example is the $d_{x^2-y^2}$-wave SML shown in Fig. \ref{fig:EMC}(a), where the FSs cross the Brillouin zone boundary. Such FSs can be realized, for instance, in the low-filling regimes of the band structures shown in Fig. \ref{fig:projection}(g)-(h). Under suitable interaction conditions, spin-triplet pairing can emerge as the dominant pairing channel (see subsequent subsection A for an example). Second, the two-body interaction acquires a richer structure in band space. Since $\hat{c}_{\mathbf{k} a \alpha}$ is now a linear combination of band operators $\hat{d}_{\mathbf{k} u \alpha}$ (with $u$ the band index), a generic two-body interaction expressed in terms of 4 $c$-operators now expands into $(N_{sub})^4$ contributions (where $N_{sub}$ is the number of sublattice in the magnetic unit cell). Consequently, the interaction vertex $V_{u \alpha, v\beta}(\mathbf{k} - \mathbf{k}')$ becomes strongly modulated by the sublattice weights. Upon downfolding to the relevant band(s) at a given filling, one obtains an augmented effective interaction, which can significantly alter the preferred pairing (see the second example in Sect. V). 
This mechanism is particularly evident in Fig. \ref{fig:projection}(d) and Fig. \ref{fig:projection}(h), where FSs become both spin- and sublattice-pure under two distinct conditions. For the Hamiltonian $h_0(\mathbf{k})$ in Eq.\ref{eq:s-d-model-2}, the Bloch states along the boundary of the N\'{e}el AFM supercell Brillouin zone (cyan region in Fig. \ref{fig:projection}(d)) satisfying $cos k_x +  cos k_y = 0$ possess a purely sublattice character, independent of the inter-sublattice hopping $t$. Moreover, the sublattice composition exhibits an alternating pattern: if the boundary segment containing $(-\pi/2, \pi/2)$ is composed entirely of sublattice $A$, then the segment containing  $(\pi/2, \pi/2)$ is composed entirely of sublattice $B$, and vice versa (see the solid/dashed black line Fig. \ref{fig:projection}(d) for an example). Consequently, FSs located near these zone boundaries become nearly spin- and sublattice-pure. In contrast, the situation in Fig. \ref{fig:projection}(h) arises from a large exchange splitting $J_{sd}$S, which strongly suppresses the inter-sublattice hybridization, rendering the FSs nearly spin- and sublattice-pure throughout the entire Brillouin zone. As a result, conventional $s$-wave pairing is generally blocked, as pointed out by Chakraborty \textit{et al.} \citep{Chakraborty2025-2} and Rasmussen \textit{et al.} \citep{Rasmussen2025}. Such sublattice-selective pairing suppression is a characteristic of AM and distinguishes them from N\'{e}el AFMs.  
It is noted that quantum geometry effect may further modify the effective interaction in the presence of flat (or weakly dispersive) bands or band touching points\citep{Peotta2015, Shi2020, Qin2019}. Third, considering the zero-momentum, static superconducting order parameter $\Delta_{a \alpha, b \beta}(\mathbf{k}) \approx \langle \hat{c}_{-\mathbf{k} b \beta} \hat{c}_{\mathbf{k} a \alpha} \rangle$, the inclusion of two sublattices greatly enriches the pairing channels. By treating the sublattice index as a pseudospin dof, its combination with the physical spin expands the pairing space to 16 distinct channels, far exceeding the 4 channels available in a single-band system. This enlarged pairing space enables versatile unconventional superconducting phenomena, such as nonsymmorphic-symmetry-protected nodal line in odd-parity SCs \citep{Blount1985,Tou2005,Norman1995, Micklitz2009, Yanase2016, Kobayashi2016, Wang2016, Micklitz2017, Nomoto2017}, 
field-induced SC or pairing transitions in uranium-based ferromagnetic SCs \citep{Levy2005, Ran2019, Rosuel2023} and CeRh$_2$As$_2$ \citep{Khim2021, Nagaki2024}, as well as $\eta$-pairing in monolayer FeSe \citep{Ding2026, Hu2013}. A related example in AMs is discussed in subsequent subsection B.

\begin{table}[t]
	\caption{Magnon mediated superconductivity in itinerant FM, conventional AFM, and AM.}
	\label{tab:three-magnon}
	\begin{tabular}{c|c|c}
		\hline\hline
		Magnetic order & type of EMC & Pairing symmetry   \\
		\hline 
		FM    & Two-magnon  & Spin-triplet      \\
		AFM  & One-magnon   & Spin-singlet \\
		AM  & One(two)-magnon &                 Spin-singlet(triplet) \\
		\hline\hline
	\end{tabular}
\end{table} 

\subsection*{A. Superconductivity mediated by magnon}
Now we explicitly incorporate the static magnetic order into the system rather than approximating it via NRSML. Now the Hamiltonian contains three part, the free part of quasiparticles, the exchange coupling between localized magnetic moments, and the interaction between electrons and localized magnetic moments:
\begin{equation}
\hat{H} = \hat{H}_{0} + \hat{H}_{ex} + \hat{H}_{sd}
\label{eq:s-d+Heisenberg}
\end{equation}
where $\hat{H}_{ex}$ takes the form like Heisenberg model like Eq.\ref{eq:Heisenberg-AM}.

\begin{table*}[t]
	\caption{Spin-fluctuation mediated unconventional superconductivity near a quantum critical point.}
	\label{tab:three-paramagnon}
	\begin{tabular}{c|c|c|c}
		\hline\hline
		Magnetic order & Order momentum $\mathbf{Q}$ & Preferred pairing  & Typical materials  \\
		\hline 
		FM    & 0  & Spin-triplet    & U-based ferromagnetic SC  \\
		\multirow{2}{*}{AFM}
		  & $(\pi, \pi)$   & \multirow{2}{*}{Spin-singlet} & $d_{x^2-y^2}$-wave SC in cuprates and Ce-based materials \\
		  & $(\pi, 0)$    &                                                & $s_{\pm}$-wave SC in Fe-based materials  \\
		AM  & 0          &      Unitary  spin-triplet & /  \\
		\hline\hline
	\end{tabular}
\end{table*}

Magnon, the collective excitations of magnetic moments, can mediated unconventional SC in a way that phonon mediates conventional SC, where the electron-magnon coupling (EMC) characterized by $s-d$ interaction in $\hat{H}_{sd}$ plays the role of glue as electron-phonon coupling \citep{Akhiezer1959}. $\hat{H}_{sd}$ contains two components 
\begin{equation}
\hat{H}_{sd} = - \frac{J_{sd}}{2} \sum_{i} (\hat{\mathrm{S}}^+_i \hat{c}^{\dagger}_{i \downarrow} \hat{c}_{i \uparrow} + \hat{\mathrm{S}}^-_i \hat{c}^{\dagger}_{i \uparrow} \hat{c}_{i \downarrow}) + 
\hat{\mathrm{S}}^z_i (\hat{c}^{\dagger}_{i \uparrow} \hat{c}_{i \uparrow} - \hat{c}^{\dagger}_{i \downarrow} \hat{c}_{i \downarrow})
\label{eq:s-d-decomposition}
\end{equation} 
where $\hat{\mathrm{S}}^+$ ($\hat{\mathrm{S}}^-$) is the raising (lowering) operator of localized magnetic moment. Applying Holstein-Primakoff transformation to operator $\hat{\mathrm{S}}^{+/-/z}$, the low-energy excitation contains two types of terms: type i) $ \hat{H}_{sd, 1} =  -J_{sd} \sqrt{\frac{\mathrm{S}}{2}} \sum_i \hat{a}^{\dagger}_i \hat{c}^{\dagger}_{i \uparrow} \hat{c}_{i \downarrow}$, and ii) $ \hat{H}_{sd, 2} = \frac{J_{sd}}{2} \sum_i \hat{a}^{\dagger}_i \hat{a}_i \hat{c}^{\dagger}_{i \sigma} \hat{c}_{i \sigma}$. $ \hat{H}_{sd, 1}$  ($ \hat{H}_{sd, 2}$) contains one (two) magnon operator and is thus called one(two)-magnon process \citep{Maeland2024}. The momentum space versions of these two EMC are illustrated in Feynman diagrams Fig. \ref{fig:EMC}(b)-(c), where one-magnon process flips the electron spins while two-magnon process does not.  Fig. \ref{fig:EMC}(d)-(e) represent the corresponding effective quasiparticle-quasiparticle interaction mediated by these two EMC processes.

In a FM metal, the FSs are spin polarized as shown in Fig. \ref{fig:Fig-ELC}(b), obviously the two states with opposite momenta must have the same spin. In this case, only spin-polarized Cooper pairs are allowed due to double-magnon process, as shown in Fig. \ref{fig:EMC}(e). In a conventional AFM metal, the FSs are degenerated, this allows Cooper pairs with opposite spin and opposite momenta, which can be induced by single-magnon process, as illustrated in Fig. \ref{fig:EMC}(d). Both cases can occur in AM, but has strongly dependence on the shape of FSs. For FSs shown in Fig.\ref{fig:s-wave-pairing}(a), both one-magnon and two-magnon process can happen. Due to the limited number of magnons at low temperature, the one-magnon process always dominates and spin-singlet pairing is thus preferred, just as in conventional AFM (see Tab. \ref{tab:three-magnon}).

In 2023, Brekke \textit{et al.} \citep{Brekke2023} proposed a 2D model to microscopically study EMC mediated SC in AM with FSs illustrated in Fig. \ref{fig:EMC}(a). To describe the anisotropic exchange interaction in square lattice shown in Fig.\ref{fig:projection}(c), the following $\hat{H}_{ex}$ is applied:
\begin{equation}
\hat{H}_{ex} = \hat{H}_{AM} + \sum_{i} K(\hat{S}^z_i)^2
\label{eq:Brekke2023-PRB-4}
\end{equation}
where $\hat{H}_{AM}$ resembles Eq.\ref{eq:Heisenberg-AM} and the magnetic anisotropy $K$ gaps the Goldstone mode of Eq.\ref{eq:Heisenberg-AM} at the $\Gamma$ point and gives the preferred magnetization along $z$-axis. Brekke \textit{et al.} \citep{Brekke2023} derived an effective quasiparticle-quasiparticle interaction based on the Schrieffer-Wolff transformation \citep{Schrieffer1966}, from which a unitary spin-triplet $p$-wave superconducting states emerge, just like FM (see Tab. \ref{tab:three-magnon}).

\subsection*{B. Superconductivity mediated by spin fluctuations}
In the previous subsection, we assumed transverse fluctuations were weak and could be described by linear spin-wave theory. This approximation is generally robust for 3D metallic AMs far way from the AFM-PM phase transition. However, as the transition is approached, long-range AFM correlations vanish while both longitudinal and transverse fluctuations significantly intensify, rendering the spin-wave treatment insufficient. To account for these spin fluctuations, various theoretical frameworks have been developed, spanning the weak, intermediate, and strong coupling regimes \citep{Kohn1965, Berk1966, Leggett1975, Hirsch1985, Scalapino1986, Miyake1986, Monod1988, Schrieffer1988, Bickers1988, Bickers1989, Monthoux1991, Monthoux1992PRB, Monthoux1992PRL, Monthoux1993, Monthoux1994, Chubukov2002, Monthoux2007, Scalapino2012}. Specifically, when a superconducting dome emerges in the vicinity of an AFM-PM QCP, the SC state can be theoretically accessed starting from the PM.
At weakly to intermediate coupling, fermionic quasiparticles and collective spin fluctuations coexist, the effective quasiparticle-quasiparticle interaction induced by bosonic spin fluctuations is captured by the following action:
\begin{equation}
S_{eff} = -g^2 \int d\mathbf{r} dt \int d\mathbf{r}' dt' \chi(\mathbf{r} - \mathbf{r}', t-t') \mathbf{s}(\mathbf{r}, t) \cdot \mathbf{s}(\mathbf{r}', t')
\end{equation}
where $g$ is the exchange constant and $\chi(\mathbf{r} - \mathbf{r}', t-t')$ is the dynamic spin susceptibility. According to this picture, we can understand the property of spin-fluctuation mediated unconventional SC once $\chi(\mathbf{r} - \mathbf{r}', t-t')$ is given. To a first approximation, $\chi(\mathbf{r} - \mathbf{r}', t-t')$ could be replaced by the susceptibility of non-interacting electrons in crystalline, i.e. the Lindhard function. Although this approximation helps demonstrate how the effective interaction between quasiparticles can develop attractive regions in space and time, it typically leads to a $T_c$ so low that experimental observation would be practically impossible \citep{Kohn1965}. An improved representation of $\chi(\mathbf{r} - \mathbf{r}', t-t')$ is provided by the random phase approximation (RPA) \citep{Bohm1953} (see Sect. V for an example), which generalizes the Weiss-Stoner model by expression the magnetic susceptibility in terms of the Lindhard functions and the effective constant $g$. A further refinement is the fluctuation-exchange approximation (FLEX), which incorporates the self-consistent renormalization of both the quasiparticle Green's function and the susceptibility \citep{Bickers1988, Bickers1989}. An alternative way to determine $\chi(\mathbf{r} - \mathbf{r}', t-t')$ is to use input from experiments.

In 1990, based on the analysis of NMR, nuclear-quadrupole-resonance, and Knight-shift in cuprates, Millis \textit{et al.} proposed a phenomenological expression for the dynamical spin susceptibility \citep{Millis1990}:
\begin{equation}
\chi(\mathbf{q}, \omega) 
= \frac{\chi_{\mathbf{Q}_{AFM}}}{1+ \xi^2 (\mathbf{q} - \mathbf{Q}_{AFM})^2 - i\omega/\Gamma_{\mathbf{q}} }
\label{eq:Millis1990-PRB}
\end{equation}
where $\mathbf{Q}_{AFM} = (\pi, \pi)$, the AFM wave vector in cuprates, and $\chi_{\mathbf{Q}_{AFM}}$ is the static spin susceptibility at  $\mathbf{Q}_{AFM}$. $\xi$ is the magnetic correlation length, which diverges at the AFM-PM QCP. $\Gamma_{\mathbf{q}}$ encodes Landau damping by particle-hole excitations, which gives the character energy of spin fluctuations. Near QCP, $\chi(\mathbf{q}, \omega) $ can be strongly enhanced and potentially gives rise to high-$T_c$ SC. Later, Eq.\ref{eq:Millis1990-PRB} was generalized to FM case where $\mathbf{Q}_{FM} = 0$  \citep{Monthoux1999, Monthoux2001, Arita2000}. When combined with the linearized gap equation or the Eliashberg equations, it is found that AFM fluctuations induce spin-singlet SC while FM fluctuations give rise to spin-triplet SC, as summarized in Tab.\ref{tab:three-paramagnon}. 

The altermagnetic fluctuations may well enrich this paradigm. Since AMs host a $\mathbf{Q}_{AM} = 0$ order that does not expand the unit cell, it should give rise to spin-triplet SC, as illustrated in Tab.\ref{tab:three-paramagnon}. However, such a conclusion is near the QCP where $\xi$ is much larger than the period of AM order. When moving away from the QCP, $\xi$ is comparable to the AM unit cell, the staggered spin structure of AM should play a role and thus SC can have an internal structure \citep{Mazin2025}. Wu \textit{et al.}  investigated such an internal structure in detail and proposed the intra-unit-cell pairing \citep{Wu2025} in AMs. Wu \textit{et al.} \citep{Wu2025} started with Hamiltonian Eq.\ref{eq:s-d-model-2} in the magnetic unit cell
\begin{widetext}
\begin{equation}
h_0(\mathbf{k}) = [-2t'(\cos k_x + \cos k_y ) - \mu] \sigma_0 \tau_0 
-4t \cos(\frac{k_x}{2}) \cos(\frac{k_y}{2}) \sigma_0 \tau_x 
-2t' \delta (\cos k_x - \cos k_y) \sigma_0 \tau_z  
\label{eq:Wu2025-PRL-1}
\end{equation}
\end{widetext}
Within the parameter range $|t'| < t/2 $ (here we assume $t$ positive), $|\delta| > 1$, Eq.\ref{eq:Wu2025-PRL-1} features a quadratic band touching at $(\pi, \pi)$ point of the magnetic Brilloiuin zone when $\mu = 4t'$. As $\mu$ increasing from $4t'$, the FSs evolves from a small pocket around $(\pi, \pi)$ point to a large one close to the BZ boundary. 

To describe the AFM fluctuations, Wu \textit{et al.} \citep{Wu2025} takes the following form:
\begin{equation}
\chi(\mathbf{q}) = \frac{1}{r - J \cos(\frac{q_x}{2}) \cos(\frac{q_y}{2})}
\label{eq:Wu2025-PRL-chi}
\end{equation}
where $r$ measures the distance to the QCP and $J$ is the NN  exchange coupling. Clearly, the QCP is reached at $\mathbf{q} = 0$ when $r/J = 1$. By approaching QCP from PM side, we always have $r/J > 1$. 

With Eq.\ref{eq:Wu2025-PRL-1}, Eq.\ref{eq:Wu2025-PRL-chi}, and $\hat{h}_{sd}$ from Eq.\ref{eq:s-d-model-2}, Wu \textit{et al.} \citep{Wu2025} obtained the AM-fluctuation mediated interaction between the electrons:
\begin{widetext}
\begin{equation}
S_{eff} = - (\frac{J_{sd} \mathrm{S}}{2})^2 \int d\mathbf{k} d\mathbf{p} d\mathbf{q}  \Psi^{\dagger}_{\mathbf{k}}(\sigma_z \tau_z) \Psi_{\mathbf{k}+\mathbf{q}}\chi(\mathbf{q})
\Psi^{\dagger}_{\mathbf{p}}(\sigma_z \tau_z) \Psi_{\mathbf{p}-\mathbf{q}}
\label{eq:Wu2025-PRL-4}
\end{equation}
\end{widetext}

Keeping the particle-particle channel and selecting the uniform SC (zero centre-of-mass momentum), it is found that without SOC, Eq.\ref{eq:Wu2025-PRL-4} is approximated to 
\begin{equation}
S_c = - \frac{(J_{sd}\mathrm{S})^2}{16} \sum^6_{i = 1}  \int d\mathbf{k} d\mathbf{p} 
(\Psi^{\dagger}_{\mathbf{k}} \Gamma^i \Psi^*_{-\mathbf{k}})\chi(\mathbf{k}-\mathbf{p})(\Psi^{T}_{-\mathbf{p}} \Gamma^i \Psi_{\mathbf{p}})
\label{eq:Wu2025-PRL-5}
\end{equation}
where $\Gamma^1 = \sigma_0 \tau_0$, $\Gamma^2 = \sigma_z \tau_z$, $\Gamma^3 = \sigma_0 \tau_z$, $\Gamma^4 = \sigma_z \tau_0$, $\Gamma^5 = \sigma_y \tau_x$, and $\Gamma^6 = \sigma_x \tau_x$. 
By solving the linearized gap equation at $t = 1$, $t' = -0.05$, and $\delta = 5$,  Wu \textit{et al.} identified four distinct pairings, as demonstrated in Fig. \ref{fig:EMC}(f)-(i): one inter-unit-cell pairing (Fig. \ref{fig:EMC}(f)) and three intra-unit-cell pairing (Fig. \ref{fig:EMC}(g)-(i)), characterized as a function of $r/J$ and $\mu$. When $r/J$ is close to 1, $p$-wave pairing dominants in a wide range of $\mu$, such a result is consistent with Tab.\ref{tab:three-paramagnon}. What's interesting is when $r/J$ is larger than 1, the AM correlation length is shorter and hence the staggered magnetic order manifests, the intra-unit-cell pairings emerge and system can be $s'$-, $p'$-, and $d'$-wave pairing with decreased $\mu$. If the FSs are centred at $(\pi, \pi)$ point, it is the $d'$-wave pairing that dominates, this is reasonable because the $d'$-wave pairing form factor $|\sin(\frac{k_x}{2})\sin(\frac{k_y}{2})|$ is the largest at $(\pi, \pi)$. When FSs are centred at $\Gamma$ point, the $s'$-wave pairing form factor $|\cos(\frac{k_x}{2})\cos(\frac{k_y}{2})|$ has the large magnitude, so $s'$-wave now is the leading pairing. For intermediate $\mu$, the FSs is large and close to the BZ boundary, disfavoring both the $s'$- and $d'$-wave states, which have nodes either on the BZ boundaries or along $k_{x/y} = 0$. As a result, the $p'$-wave becomes the leading instability. Notably, the trends observed for intra-unit-cell pairings are consistent with those obtained in the single-band model discussed at the end of Sect. III-A-2.

\section*{V. Superconductivity originates from repulsive interaction}
In this section, we consider the repulsive Hubbard model and treat competing orders on an equal footing. On the 2D square lattice with NN hopping $t$ only, the model exhibits strong AFM correlations at half filling. Upon charge doping, long-range AFM is suppressed, but numerous large-scale numerical studies indicate that the ground state is not a simple homogeneous superconductor. Instead, the system tends to develop intertwined spin and charge stripe orders, with superconducting correlations that are either subdominant or coexist with these inhomogeneous states. While short-range $d_{x^2-y^2}$-wave pairing correlations are clearly enhanced relative to other channels, growing evidence suggests that the pure $t-U$ Hubbard model does not generically exhibit uniform $d$-wave SC as the ground state in the intermediate to strong coupling regime at the thermodynamic limit with finite doping \citep{Zheng2017, Qin2020, Qin2022}.

A qualitatively different behavior emerges when longer-range hoppings are included, most notably the NNN hopping $t'$. The parameter $t'$ alters the band structure and FS topology, weakens the perfect nesting, and modifies the competition among ordered phases. In particular, negative $t'$ (relevant for hole-doped cuprates) tends to frustrate stripe formation and enhance spin fluctuations near momentum $(\pi, \pi)$. Consequently, many numerical studies find that a finite $t'$ substantially enlarges the regime where robust $d$-wave SC becomes the leading instability \citep{Jiang2019, Xu2024, Zhang2025}. 

As discussed in previous sections, AMs intrinsically exhibit NNN hoppings, making the $t-t'-U$ Fermi-Hubbard model a natural minimal description. Since NNN hopping is anisotropic in AMs, AMs provide a new platform for exploring the interplay between magnetism and SC within the anisotropic $t-t'-\delta-U$ Fermi-Hubbard model, as discussed subsequently. It is noted that, the state-of-art many-body computational methods used in studying 2D isotropic $t-t'-U$ Hubbard model, such as constrained-path auxiliary-field QMC \citep{Zhang1997}, density matrix renormalization group \citep{White1992}, infinite projected entangle pair states \citep{Verstraete2004, Jordan2008}, density matrix embedding theory \citep{Knizia2012},  remain at an early stage of development for AMs. 

\begin{figure*}[t]
\centering
\includegraphics[width=0.90\textwidth]{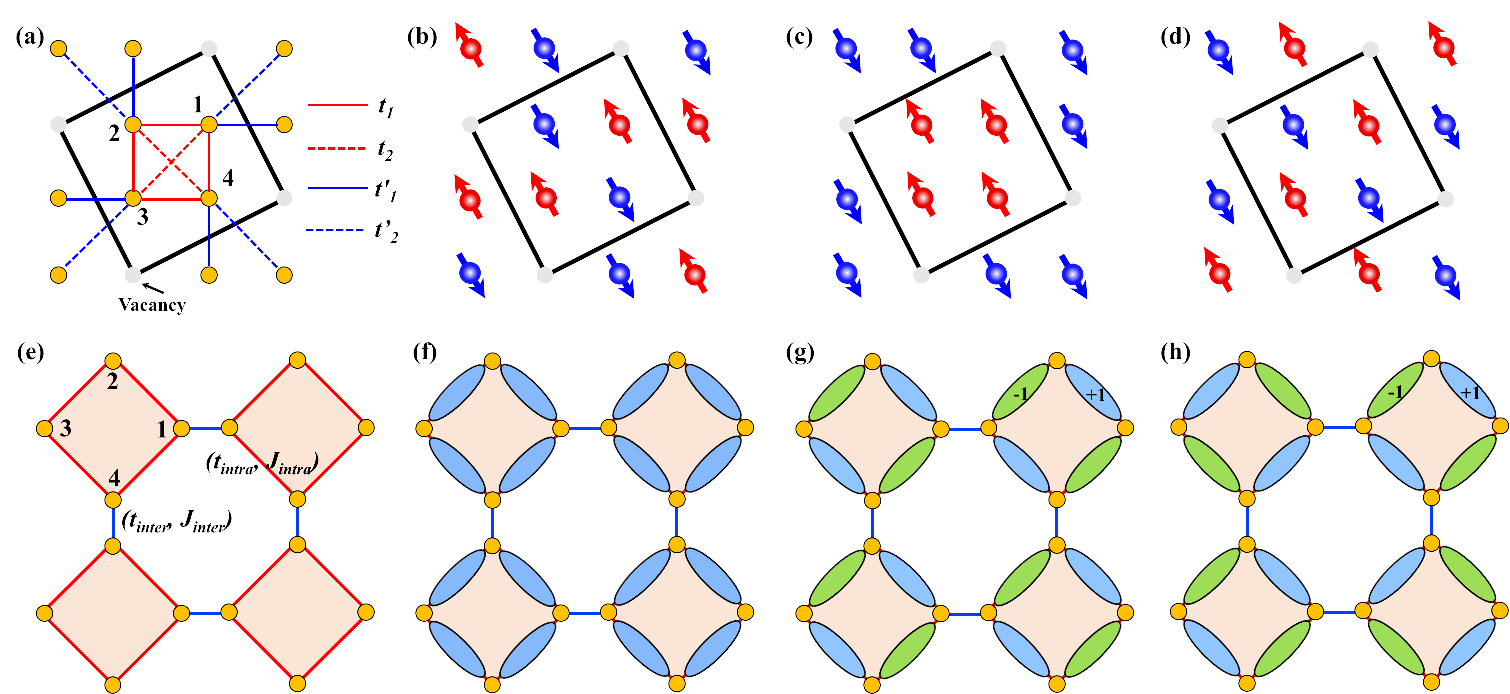}
\caption{(a) Schematic view of a $\sqrt{5} \times \sqrt{5}$ vacancy-ordered square lattice. The red/blue lines indicate intra-block/inter-block hoppings, and the solid/dashed lines indicate NN/NNN hoppings, respectively. (b)-(d) show the real space perspective of N\'{e}el-AM (sublattice 1,3 have opposite magnetization than sublattice 2, 4, and inter-unit-cell wave vector is $0$), BS-AFM (all four sublattices have the same magnetization, and inter-unit-cell wave vector is $(\pi, \pi)$), and N\'{e}el-AFM (sublattice 1,3 have opposite magnetization than sublattice 2, 4, and inter-unit-cell wave vector is $(\pi, \pi)$) orders. (e) Schematic view of the decorated square lattice model, where the unit cell contains four sublattices. (f-h) Real space configuration of three distinct spin-singlet pairing channels. (f) Uniform $s$-wave with the same phase and pairing strength on each inter-block bond. (g) Uniform $d$-wave with the uniform strength but alternating phase of the pairing within the block. (h) $d$-wave PDW with a $d$-wave pattern of pairing within the block, but now with a wave vector of $(\pi, \pi)$.}
\label{fig:diamond} 
\end{figure*} 

In 2025, motivated by a $\sqrt{5} \times \sqrt{5}$ vacancy-ordered square lattice in iron chalcogenide SCs \citep{Fang2011}, Ma \textit{et al.} \citep{MaX2025} considered Eq.\ref{eq:Naka2019-1} for a lattice shown in Fig. \ref{fig:diamond}(a), there are four sublattices ($a, b$ =1, 2, 3, 4) in the unit cell (black line in Fig. \ref{fig:diamond}(a)) which form a block, and the hoppings up to NNN are labelled as $t_{1/2}$ (intra-block), $t'_{1/2}$ (inter-block) as shown in Fig. \ref{fig:diamond}(a), which preserves the $I\frac{4}{m}$ symmetry. In the large-$U$ limit, the hoppings lead to AFM exchanges between NN and NNN with exchange coupling strength $J_{1/2} \approx \frac{4t^2_{1/2}}{U}$, $J'_{1/2} \approx \frac{4(t^{'}_{1/2})^2}{U}$, resulting in an extended Heisenberg model. The ratio $\alpha \equiv t_2/t_1$ ($\alpha' \equiv t'_2/t'_1$ ) therefore measures the intra-block (inter-block) frustration. By fixing $\alpha$ = 0.2, at half-filling and $U = |8t_1|$ (large-$U$ limit), This model has three magnetic orders: N\'{e}el-AM (see Fig. \ref{fig:diamond}(b)), Block spin-AFM (BS-AFM, see Fig. \ref{fig:diamond}(c)), and N\'{e}el-AFM (see Fig. \ref{fig:diamond}(d)) depending on the value of $\alpha'$ and $\beta \equiv t'_1/t_1$. When $\alpha'$ is large and  $\beta$ is small, $J_1$ and $J'_2$ dominate intra- and inter-block exchange couplings, therefore, the N\'{e}el-AM phase is the ground state.

In order to investigate the SC properties,  Ma \textit{et al.} \citep{MaX2025} suppresses the long-range magnetic order by choosing a smaller $U$ near metal-insulator transition and introducing charge doping. By combined RPA and the linearized gap equations, Ma \textit{et al.} \citep{MaX2025} obtained a phase digram for different pairings. Deep inside the N\'{e}el-AM region, it is found that the suppression of N\'{e}el-AM order leads to AM fluctuation characterized by $\mathbf{Q} = 0$ and a strong peak at $\Gamma$ for $\chi^{S, N}(\mathbf{q})$ (intra-block N\'{e}el channel spin susceptibility). When considering leading pairings, the $E_u$ channel (with nodal lines) is dominantly larger than the $A_g$ and $B_g$ channel,  which can further mix to form fully gapped $p \pm ip$ SCs. The story near the (N\'{e}el-AM)-(BS-AFM) QCP is totally different, where the BS-AFM-type spin fluctuation dominates with peak at $(\pi, \pi)$. Therefore, such a spin fluctuation prefers spin-singlet $A_{1g}$ pairing. Obviously, the results obtained from Ma \textit{et al.} \citep{MaX2025} agree with Tab.\ref{tab:three-paramagnon}. 

In 2024, Bose \textit{et al.} \citep{Bose2024} considered a large-$U$ limit on a decorated square lattice model (see Fig. \ref{fig:diamond}(e)), where $\hat{H}_{int}$ is expressed within a multi-sublattice $t-J$ model:
\begin{equation}
\hat{H}_{int} = \sum_{i,j, a, b} J_{ia,jb} \hat{\mathbf{s}}_{i a} \cdot \hat{\mathbf{s}}_{j b}  
\label{eq:Bose-2024-PRB-1}
\end{equation}
Such a model can be regarded as a simplified version of Fig. \ref{fig:diamond}(a) \citep{Ma2025}, the four sites within a unit cell form a block, and the intra(inter)-block hoppings and exchange coupling strengths are $t_{intra}$ ($t_{inter}$) and $J_{intra}$ ($J_{inter}$), respectively, as illustrated in Fig. \ref{fig:diamond}(e). $J_{intra}$ is set to be positive, indicating inter-unit-cell AFM coupling. Obviously, the N\'{e}el-AM order is favored when  $J_{inter} < 0$ (see Fig. \ref{fig:diamond}(b)). It is noteworthy that while $J_{inter} > 0$ arises naturally from a Hubbard model, the case $J_{inter} < 0$ can only airse from additional Hund's coupling with multiorbitals at the same site. 

By solving Eq.\ref{eq:Bose-2024-PRB-1} with mean-field theory, when $J_{inter}/J_{intra} = -0.5$ and $t_{inter}/t_{intra} = 0.5$, according to the filling (up to half-filling) of the model, there are versatile orders including uniform $s$- (see Fig. \ref{fig:diamond}(f)), $d$-wave SCs (see Fig. \ref{fig:diamond}(g)), strong and weak N\'{e}el-AM phase (characterized by large and small magnetizations), weak N\'{e}el-AFM, band insulator, and potential $d$-wave PDW (see Fig. \ref{fig:diamond}(h)). The uniform $s$-wave SC occurs with filling smaller than 1/2, which can be understood from the $t_{inter} \ll t_{intra}$ limit where the following intra-block orbitals are a better description of $\hat{H}_0$:
\begin{subequations}
\begin{align}
\hat{s}_{i} &= \frac{1}{2} (\hat{c}_{i 1}+\hat{c}_{i 2}+\hat{c}_{i 3}+\hat{c}_{i 4}) 
,\label{eq:Bose-2024-PRB-3} \\
\hat{X}_{i} &= \frac{1}{\sqrt{2}} (\hat{c}_{i 1}-\hat{c}_{i 3}) 
,\label{eq:Bose-2024-PRB-5} \\
\hat{Y}_{i} &= \frac{1}{\sqrt{2}} (\hat{c}_{i 2}-\hat{c}_{i 4}) 
,\label{eq:Bose-2024-PRB-6} \\
\hat{d}_{i} &= \frac{1}{2} (\hat{c}_{i 1}-\hat{c}_{i 2}+\hat{c}_{i 3}-\hat{c}_{i 4}) 
, \label{eq:Bose-2024-PRB-4}
\end{align}
\end{subequations}
which corresponds to a $s$ orbital, a pair of degenerate $p$ orbitals $X, Y$, and a $d$ orbital with on-site energy $-2t_{intra}$, $0$, and $2t_{intra}$ respectively. These intra-block orbitals will form corresponding bands, which for $t_{inter} \ll t_{intra}$ are nonoverlapping. With increase filling, we go from filling the $s$-orbital band (filling up to 1/2), to the $X,Y$-orbital bands (filling up to 3/2), and eventually the $d$-orbital band (filling up to 2), with gapped band insulators appearing in between when some of these bands are filled. By plugging the inverse of Eq.\ref{eq:Bose-2024-PRB-3}-Eq.\ref{eq:Bose-2024-PRB-4} into $\hat{H}$, and discarding high-energy terms, we obtain effective model within inter-block orbitals. 

When only $s$-orbital is filled (up to 1/2), $\hat{H}_0$ of Eq.\ref{eq:Bose-2024-PRB-1} can be downfolded to an effective $s$-orbital with NN hoppings $t_{inter}/4$ in 2D square lattice:
\begin{equation}
\hat{H}^{s}_{eff, 0} = - \frac{t_{inter}}{4}\sum_{i, j, \sigma} (\hat{s}^{\dagger}_{i \sigma} \hat{s}_{j \sigma} + h.c.) - \mu\sum_{i} \hat{n}^{s}_{i} 
\label{eq:Bose-2024-PRB-8-9}
\end{equation} 
At the same time, $\hat{H}_{int}$ of Eq.\ref{eq:Bose-2024-PRB-1} can be downfolded as follows:
\begin{equation}
\hat{H}^{s}_{eff, int} = -\frac{3J_{inter}}{8} \sum_{i} \hat{n}^{s}_{i \uparrow}\hat{n}^{s}_{i \uparrow} + \frac{J_{inter}}{16} \sum_{i, j} \hat{\mathbf{s}}^{s}_{i} \cdot \hat{\mathbf{s}}^{s}_{j}
\label{eq:Bose-2024-PRB-13}
\end{equation}
where $\hat{\mathbf{s}}^{s}_{i}$ is the $s$-orbital electron spin operator. The intracell exchange couplings thus lead to an effective attractive on-site interaction with $U_{eff} = -\frac{3J_{inter}}{8}$, which favors $s$-wave spin-singlet pairing. 

What is intriguing is that at half-filling (i.e. half of the $p$-band is filled) and with strong $J_{intra}$, it is found that strong N\'{e}el-AM is the ground state. When the filling slightly deviates from half-filling, the strong N\'{e}el-AM order is suppressed and there is a regime where SC and N\'{e}el-AM coexist in the mean-field theory. In this coexistence phase, the N\'{e}el-AM order mixes the real space $d$-wave spin-singlet and $s$-wave triplet pairing with $S_z = 0$ on the intra-unit cell. Incorporating longer-range repulsion interactions, such phase coexistence may organize into more structured pattern, such as stripes \citep{Bose2024}.

The discussion in this section and in Sect. IV-B rely on tuning of AMs through external stimuli such as external magnetic fields, pressure, and chemical doping to realize unconventional SC. Since crystalline symmetry is pivotal in protecting NRSML, chemical doping inevitably break this symmetry in 3D bulk AMs such as MnTe \citep{Krempasky2024} and CrSb \citep{Reimers2024, Zeng2024, Lu2025, Zhou2025}. This limitation is substantially mitigated in quasi-2D AMs, such as KV$_2$Se$_2$O \citep{Jiang2025} and RbV$_2$Te$_2$O \citep{ZhangF2025}, where disorder in the charge-reservoir layers has only a weak influence on the AM layers. Recently Sun \textit{et al.} \citep{Sun2026} have reported a superconducting transition temperature 16.3 K in the altermagnetic candidate Na$_{2-x}$V$_2$Se$_2$O. Pressure, by contrast, provides a much cleaner tuning parameter for driving quantum phase transition, although accompanying structural transitions are often unavoidable. In this context, first-principles calculations play an important role in predicting candidate systems and potential phase transitions. Notably, layered ternary iron-selenides under pressure have recently predicted to host coexisting AM and SC phases \citep{LiZ2025}.

\section*{V. Discussion and outlook}
Thus far, we have highlighted the two central themes: (1) the heuristic development of novel AM order, and (2) the ensuing exotic SC in AM systems. In the first theme, we reviewed the connections among ELC phases, multipole basis, and AM, emphasizing nonrelativistic SML as the unifying principle, as illustrated in Fig.\ref{fig:Fig-1}. In the second theme, we examined the influence of AM order and fluctuations on SC, with particular emphasis on the similarities and distinctions between AM-, Rashba-, Ising-type, and conventional FFLO phases (see Tab.\ref{tab:four-FFLO}), as well as among AM, FM, and conventional AFM superconductors (see Tab.\ref{tab:three-magnon} and Tab.\ref{tab:three-paramagnon}). 

In this final section, we adopt a broader perspective on SC, stepping beyond the technical details to reflect on open problems that are currently attracting growing attentions. We also outline two promising research directions, each linked to fundamental challenges in the physics of AMs and SC. Although not intended to provide an exhaustive outlook, these perspectives highlight the rapidly evolving nature of the field and highlight opportunities where significant breakthroughs may emerge in the near future. 

\subsection*{A. $p$-wave (unconventional) magnetism}
While significant attention has been devoted to $d$-wave SML and AMs, their  counterpart - $p$-wave SML and $p$-wave magnet - represent another important unconventional magnetism. Analogous to the $d$-wave SML, $p$-wave SML can either originate from PI in the spin channel \citep{Varma2006, Wu2007} or be realized in certain AFMs \citep{Hayami2020-PRB-2, Hayami2022, Hellenes2023, Mitscherling2026, Song2025, Yamada2025, Chakraborty2025-NC, Yu2025, Brekke2024, Ezawa2024, Zhuang2025, Sim2026, Lin2026, Huang2025, Zhu2025, LiuD2025, Zhang2026}. Symmetry analysis on spin space groups elucidates that collinear magnets only support even-parity SML like $s$-wave in FMs, $d$-wave in AMs \citep{Smejkal2022-1}. Consequently, the search for  $p$-wave SML in pristine magnetic materials must focus on non-collinear coplanar or noncoplanar systems.

In 2023, Hellenes \textit{et al.} \citep{Hellenes2023} revealed that  $[C^{(s)}_{2\perp}||\boldsymbol{\tau}]$ in noncentrosymmetric non-collinear coplanar magnets can protect $p$-wave SML. It is important to note that in non-collinear coplanar magnets, the combination of  $C^{(s)}_{2\perp}$ and $\mathcal{T}$ gives $\varepsilon(\bf{k}, \sigma_{\perp}, \sigma_{\parallel}) = \varepsilon(-\bf{k}, -\sigma_{\perp}, \sigma_{\parallel})$. Therefore, the effective $\mathcal{P}$ present in collinear magnets is absent in non-collinear coplanar magnets. This breaking of  $\mathcal{P}$ implies $\varepsilon(\bf{k}, \sigma_{\perp}, \sigma_{\parallel}) \neq \varepsilon(-\bf{k}, \sigma_{\perp}, \sigma_{\parallel})$. Taken together, these symmetry considerations lead to $p$-wave SML, as illustrated in Fig.\ref{fig:Fig-ELC}(c). This insight has far-reaching implications, when SC is taken into account. The absence of $\mathcal{P}$ naturally leads to non-centrosymmetric SC \citep{Bauer2012}, among which $p$-wave SC is especially sought after \citep{Nagae2025, Sukhachov2025, Sun2025}. If such a symmetry condition are realized experimentally, they may offer a viable pathway to realizing topological SCs - highly desirable for their potential in fault-tolerant quantum computation. Therefore, exploring $p$-wave SML and $p$-wave magnets not only deepens our understanding of magnetism, but also paves the avenue toward technologically transformative superconducting phases.

\subsection*{B. Multipole fluctuation mediated superconductivity and multipole expansion of Cooper pairs}
Beyond specific material realizations, there is significant potential to develop new conceptual frameworks. One particular promising direction involves extending multipole basis fluctuation mediated SC \citep{Kozii2015, Sumita2020, Mineev2024, Kirikoshi2024} to that with extra dof. Multipole basis fluctuation is a generalization of $\textbf{E}$ monopole and $\textbf{M}$ dipole fluctuation. One straightforward hint is Tab.\ref{tab:four-FFLO}, where a direct correspondence between multipole basis and FFLO phases is obvious.
Actually, Sect. III can be thought as $\textbf{MT}$ quadrupole $T_{xy}$ fluctuation induced SC. When $T_{xy}$ is zero, the SC is conventional; nevertheless, FFLO phases emerges with large $T_{xy}$. From this perspective, every multipole basis should bestow certain type of pairing. In 2020, Sumita \textit{et al.} \citep{Sumita2020} exhausted all ferroic multipole basis-fluctuation-mediated SC. In this framework, Eq.\ref{eq:Hayami-2018-PRB-33} is understood as the addition of four Hermitian multiple basis operators:
\begin{equation}
\hat{Q} = \sum_{\mathbf{k} \alpha \beta} \Lambda_{\alpha \beta}(\mathbf{k}) \hat{c}^{\dagger}_{\mathbf{k} \alpha} \hat{c}_{\mathbf{k} \beta}
\label{eq:Sumita-2020-PRR-1}
\end{equation}
where $\Lambda_{\alpha \beta}(\mathbf{k})$ can be $\varepsilon^{E}(\mathbf{k})\delta_{\alpha \beta}, \varepsilon^{O}(\mathbf{k})\delta_{\alpha \beta}, f^{E}_{\alpha \beta}(\mathbf{k}), f^{O}_{\alpha \beta}(\mathbf{k})$, see Eq.\ref{eq:Hayami-2018-PRB-34}-Eq.\ref{eq:Hayami-2018-PRB-37} (in Sumita \textit{et al.} \citep{Sumita2020}, they are called even-parity electric, odd-parity magnetic, even-parity magnetic, and odd-parity electric multipole).  With electronic correlations or electron-phonon attractive interaction, there is multipole-fluctuated interaction:
\begin{equation}
\hat{H}_{int} = \frac{1}{2N} \sum_{\mathbf{q}} V_{\mathbf{q}} \hat{Q}(\mathbf{q})\hat{Q}(-\mathbf{q})
\label{eq:Sumita-2020-PRR-5}
\end{equation}
where $\hat{Q}(\mathbf{q}) = \hat{Q}^{\dagger}(-\mathbf{q})$ is the Fourier transformation of the order parameter in real space:
\begin{equation}
\hat{Q}(\mathbf{q}) = \frac{1}{2} \sum_{\mathbf{k} \alpha \beta}[\Lambda_{\alpha \beta}(\mathbf{k}+\mathbf{q}) +\Lambda_{\alpha \beta}(\mathbf{k}) ]\hat{c}^{\dagger}_{\mathbf{k}+\mathbf{q} \alpha} \hat{c}_{\mathbf{k} \beta}
\label{eq:Sumita-2020-PRR-6}
\end{equation}
$\hat{Q}(\mathbf{q} = 0)$ corresponds to $\hat{Q}$ introduced in Eq.\ref{eq:Sumita-2020-PRR-1}.

Restricting the effective interaction to pairing channels with zero centre-of-mass momentum, the following reduced Hamiltonian can be obtained:
\begin{equation}
\hat{H}_{int} = \frac{1}{N} \sum_{\mathbf{k}, \mathbf{k}'} \sum_{\alpha \beta \gamma \delta} V_{\alpha \beta \gamma \delta}(\mathbf{k}, \mathbf{k}') \hat{c}^{\dagger}_{\mathbf{k} \alpha} \hat{c}^{\dagger}_{-\mathbf{k} \beta} \hat{c}_{-\mathbf{k}' \gamma} \hat{c}_{\mathbf{k}' \delta}
\label{eq:Sumita-2020-PRR-7}
\end{equation}
where the pairing interaction vertex is given by:
\begin{widetext}
\begin{equation}
V_{\alpha \beta \gamma \delta}(\mathbf{k}, \mathbf{k}')  = \frac{1}{8}[
V_{\mathbf{k} - \mathbf{k}'} (\Lambda_{\alpha \delta}(\mathbf{k}) + \Lambda_{\alpha \delta}(\mathbf{k}')) (\Lambda_{\beta \gamma}(-\mathbf{k}) + \Lambda_{\beta \gamma}(-\mathbf{k}'))
-V_{\mathbf{k} + \mathbf{k}'} (\Lambda_{\alpha \gamma}(\mathbf{k}) + \Lambda_{\alpha \gamma}(-\mathbf{k}')) (\Lambda_{\beta \delta}(-\mathbf{k}) + \Lambda_{\beta \delta}(\mathbf{k}'))
]
\label{eq:Sumita-2020-PRR-8}
\end{equation}
\end{widetext}
By inserting in different multipole basis expressed in Eq.\ref{eq:Hayami-2018-PRB-34}-Eq.\ref{eq:Hayami-2018-PRB-37} and Eq.\ref{eq:Hayami-2018-PRB-21}-Eq.\ref{eq:Hayami-2018-PRB-24}, different pairing orders can be obtained, such as spin-triplet SC by FM fluctuation and $d$-wave SC by AFM fluctuation (see the first two rows of Tab.\ref{tab:three-paramagnon}). Here we choose $\textbf{E}$ monopole, i.e. $\Lambda_{\alpha \beta}(\mathbf{k}) = \varepsilon^E(\mathbf{k}) \delta_{\alpha \beta} = \delta_{\alpha \beta} $, Eq.\ref{eq:Sumita-2020-PRR-8} is the reduced to $\frac{1}{2}(V_{\mathbf{k} - \mathbf{k}'} \delta_{\alpha \delta} \delta_{\beta \gamma} + V_{\mathbf{k} + \mathbf{k}'}\delta_{\alpha \gamma} \delta_{\beta \delta})$, which is nothing but $V_{\alpha \beta}(\mathbf{k} - \mathbf{k}')$ in Eq.\ref{eq:general-interaction}. Therefore, SC originating from the attractive Hubbard model and the extend attractive Hubbard model can be regarded as being mediated by $\textbf{E}$ monopole.
However, such a generalization does not consider additional internal dof like valley, sublattice. It is thus natural to extend this framework to systems with valley dof so the multipole associated with Ising-type FFLO states (see the third line of Tab.\ref{tab:four-FFLO}) can be systematically identified, as well as to systems with sublattice dof, as in AMs. 

Another extension is to generalize the language of multipole expansions to Cooper pairs. As discussed earlier, ELC phases can also describe Cooper pairs. Given the close alignment between the  language of ELC phases and multipole expansions, it is therefore natural to apply multipole expansion language to Cooper pairs, particularly in multi-orbital systems \citep{Nomoto2016, Yatsushiro2021, Kirikoshi2024}.

\begin{dataavailability}
The data is not available, because this is a review paper. The data cannot be made publicly available upon publication because they are owned by a third party and the terms of use prevent public distribution. The data that support the findings of this study are available upon reasonable request from the authors.
\end{dataavailability}

\begin{acknowledgments}
We thank H. Kusunose, H. Zhai and J. Wang for helpful discussions. We also thank J. Sinova for valuable comments. This research was supported by the Australian Research Council's (ARC)
Discovery Program, Grants Nos. DP240101590 (H.H.) and DP240100248
(X.-J.L.). 
\end{acknowledgments}

\end{document}